\documentclass[pdflatex]{sn-jnl}
\usepackage[utf8]{inputenc}
\usepackage{amsfonts}
\usepackage{amsmath}

\usepackage{pdflscape}
\usepackage{geometry}

\newgeometry{
  a4paper,
  landscape,
  margin=1in
}

\usepackage{array}
\usepackage{makecell}
\newlength\savedwidth
\newcolumntype{?}{!{\vrule width2pt}}
\newcommand\thickhline{\noalign{\global\savedwidth\arrayrulewidth\global\arrayrulewidth 2pt}
\hline
\noalign{\global\arrayrulewidth\savedwidth}}

\DeclareMathAlphabet\mathbfcal{OMS}{cmsy}{b}{n}

\usepackage[table]{xcolor}

\usepackage{array}
\usepackage{makecell}
\usepackage{multirow}

\newcolumntype{+}{!{\vrule width 2pt}}

\title{CLOVE: Travelling Salesman's approach to hyperbolic embeddings of complex networks with communities}

\author[1,2]{\fnm{Sámuel G.} \sur{Balogh}}\email{balogh@hal.elte.hu}

\author[2]{\fnm{Bendegúz} \sur{Sulyok}}\email{bendeguz.sulyok@ttk.elte.hu}

\author[2]{\fnm{Tamás}  \sur{Vicsek}}\email{vicsek@hal.elte.hu}
\author*[2,3]{\fnm{Gergely} \sur{Palla}}\email{gergely.palla@emk.semmelweis.hu}

\affil[1]{\orgname{
National Laboratory for Health Security, Alfr{\'e}d R{\'e}nyi Institute of Mathematics}, \orgaddress{\street{Re{\'a}ltanoda utca 13-15}, \city{Budapest}, \postcode{1053}, \country{Hungary}}}

\affil[2]{\orgdiv{Department of Biological Physics}, \orgname{Eötvös Lor{\'a}nd University}, \orgaddress{\street{P{\'a}zm{\'a}ny P.\ stny.\ 1/A}, \city{Budapest}, \postcode{H-1117}, \country{Hungary}}}

\affil[3]{\orgdiv{Health Services Management Training Centre}, \orgname{Semmelweis University}, \orgaddress{\street{K{\'u}tv{\"o}lgyi {\'u}t 2.}, \city{Budapest}, \postcode{H-1125}, \country{Hungary}}}

\date{\today}

\abstract{
The embedding of complex networks into metric spaces has become a research topic of high interest with a wide variety of proposed methods. Low dimensional hyperbolic spaces offer a natural co-domain for embeddings allowing a roughly uniform spatial distribution of the nodes even for scale-free networks and the efficient navigability and estimation of linking probabilities. According to recent results, the communities of a complex network after optimization can be naturally mapped into well-defined angular sectors of the hyperbolic space.  Here we introduce CLOVE, an embedding method exploiting this property based on iterative arrangement of the communities in a hierarchical manner, down to individual nodes. A crucial step in the process is finding the optimal angular order of the communities at a given level of the hierarchy, which is solved based on the Travelling Salesman Problem. Since CLOVE outperforms most of the alternative methods regarding different embedding quality measures and is computationally very efficient, it can be very useful in related down-stream machine learning tasks such as AI based pattern recognition.
}


\begin{document}

\maketitle

\section*{Introduction}
The network approach for describing and analysing complex systems has become ubiquitous in the last two decades \cite{Laci_revmod,Dorog_book,Newman_Barabasi_Watts,Jari_Holme_Phys_Rep,Vespignani_book}, building on the fundamental concept of representing the interactions between the constituents of the studied system by a graph. A general approach for augmenting the network reflecting the structure of the web of connections (that serve as a sort of a skeleton for a complex system) is to apply network embedding techniques \cite{Goyal_Ferrara_embedding_survey,Radicchi_compare_embeddings,Embedding_book_Yang,Bianconi_zoo_guide_to_embedding}. These methods are aimed at finding an optimal arrangement of the network in a metric space, thereby associating coordinates to the nodes of the network based on the network topology. These coordinates can be useful from several different aspects, e.g., they enable the prediction of missing links, can help navigation over the network, or may serve as input for further machine learning tasks such as node classification, community finding, etc. 

Although the majority of network embedding techniques operate in Euclidean spaces (see e.g., a recent review in Ref.\cite{Bianconi_zoo_guide_to_embedding}), hyperbolic methods offer an alternative approach with unique advantages\cite{Boguna_Krioukov_review}. Probably most important is that while Euclidean algorithms usually embed in high dimensions, hyperbolic approaches can yield good quality embeddings already in 2 dimensions. The intuitive reason behind this is that the exponential growth of the volume as a function of the radius for spheres in hyperbolic spaces allows more "freedom" in node placement compared to the case of Euclidean spaces, where the volume is increasing only like a power-law \cite{hyperGeomBasics}. The majority of hyperbolic embedding methods work in the native representation of the hyperbolic space, which in 2 dimensions, is often referred to as the native disk. In this representation, the radial coordinates are usually strongly coupled with the node degree, where the high degree nodes tend to be placed closer to the centre of the native disk, while the low degree nodes occupy the disk periphery. (A brief description of the native disk and hyperbolic geometry is given in Methods).

Several different hyperbolic embedding algorithms have been proposed in the literature, starting from the optimisation of the likelihood with respect to hyperbolic network models \cite{Boguna_Krioukov_Internet_2010,EPSO_HyperMap},  through the dimension reduction of a non-linear Laplacian matrix \cite{Alanis-Lobato_LE_embedding,Alanis-Lobato_liekly_LE_emb}, the dimension reduction of a Lorentz-matrix using the hiperboloid model \cite{Hydra,our_dir_embedding} and the family of coalescent embeddings (applying dimension reduction to different pre-weighted matrices encapsulating the network structure) \cite{linkWeights_coalescentEmbedding} to various mixed approaches combining both dimension reduction and local optimisation \cite{S1H2_Mercator,dmercator,our_opt_embedding} and neural network based embeddings. In Ref.~\cite{cannistraci_minimum_curvilinearity}, the authors introduced a method called  Minimum Curvilinear Automaton (MCA) that uses the minimum spanning tree to obtain hyperbolic embeddings of complex networks.

Hyperbolic embeddings are closely coupled with the modular structure of networks \cite{analogyBetweenHypEmbAndComms,modularity_of_RHG_2021,balogh2023maximally, Balogh_intracomod_2024}. On the one hand, graphs generated by geometric network models operating explicitly in hyperbolic spaces have been shown to exhibit a highly pronounced modular nature, wherein communities (corresponding to densely connected modules in the networks) occupy tightly localized domains within the geometric space and share an asymptotically negligible fraction of inter-connections between one another~\cite{modularity_of_RHG_2021,balogh2023maximally, Balogh_intracomod_2024}. On the other hand, this separability of the network modules in the metric space can be also considered to be a fundamental prerequisite for high-quality hyperbolic embeddings, suggesting a deep connection between the embeddings and the community structure of complex networks~\cite{analogyBetweenHypEmbAndComms}.
Indeed, when embedding a given network, we essentially mean to provide an $f_{E}$ mapping function of the form 
$f_{E}:\ V\to\mathbb{R}^{d}$ equipped with a metric,  
where $V$ denotes the set of nodes in the network and $d$ is the dimension of the embedding space. 
In parallel, partitioning the same network is equal to constructing a $f_{P}: \ V\to \mathbb{N}$ mapping, which can be regarded to some extent as a coarsened version of its embedding~\cite{analogyBetweenHypEmbAndComms}. Additionally, in Ref.~\cite{generalized_mod_matrix}, the authors show that the embedding technique relying on the Laplacian Eigenmap is merely a specific instance of a broader trace maximization problem associated with the generalized modularity matrix.

Notably, the emergence of this formal analogy between embedding and partitioning gives rise to a variety of intriguing implications; e.g., one can reasonably assess the quality of hyperbolic embeddings by quantifying the extent to which nodes within the same community have similar angular coordinates in the embedding space (angular coherence of the communities). As expected, state-of-the-art hyperbolic embedding methods such as the coalescent embedding~\cite{linkWeights_coalescentEmbedding}, or the $D$-Mercator~\cite{dmercator} perform 
excellently in this respect, as shown through specific quality measures capturing the communities' angular coherence in Refs.~\cite{dmercator, Cannistraci_ASI}. Additionally, in Ref.~\cite{Lizotte_bigue}, the authors introduce BIGUE (Bayesian Inference of a Graph’s Unknown Embedding), an efficient Markov chain Monte Carlo algorithm that uses a set of cluster (community) based transformations to improve the exploration of the posterior distribution. 

Perhaps, an even more explicit manifestation of the previous analogy emerges, when the hyperbolic embeddings of a given network are constructed based on the information encoded in its community structure~\cite{commSector_hypEmbBasedOnComms_2016, commSector_hypEmbBasedOnComms_2019}. Herein, the authors introduce a family of embedding methods that rely on the iterative assignment of the network communities and their respective sub-communities to distinct angular sectors on the native disk. It is important to note, however, that the crux of the aforementioned procedure lies in the reasonable arrangement of communities, a task that unfortunately lacks a well-principled systematic solution scheme. Although a computationally very fast greedy-like methodology has been proposed in Ref.~\cite{commSector_hypEmbBasedOnComms_2019} under the name of Hyperbolic Mapping based on the hierarchical Community Structure (HMCS) method, our empirical findings show its diminished efficiency under specific circumstances. Driven by this incompleteness, in the present paper we propose a novel hyperbolic embedding method built upon the modular structure of networks, where the arrangement problem of the found communities is solved according to the renowned Travelling Salesman Problem\cite{ancient_tsp_robinson,ancient_tsp_verblunsky,ancient_tsp_dantzig, ancient_tsp_karp} (TSP). Originally, the TSP focuses on finding the minimum weight Hamiltonian path, which, in this context, can directly be used to determine the angular order of (sub-)communities on the native disk. Since the angular arrangement is optimised according to a well-known route finding problem borrowed from the domain of computer science, we abbreviate our method as CLOVE, standing for Cluster Level Optimised Vertex Embedding. The key concept of  this method involves detecting the communities in the network, constructing a weighted super-graph from them, and subsequently employing approximate algorithms for the Travelling Salesman Problem in order to identify the minimum-weighted cycle of the communities. This sequence of instructions is then iterated hierarchically, encompassing increasingly smaller scales, until reaching a point, where no further community structure can be uncovered.

On the one hand, since the TSP has to be solved only on relatively small networks, the method is fast, capable of embedding networks having millions of nodes under just a few hours. On the other hand, due to the repeated optimisation, the quality of the obtained embedding is high according to various different measures. In the upcoming sections, we compare the performance of CLOVE with various state of the art embedding algorithms from the aspect of both the computation time and the quality of the end result.

\section*{Results}

\subsection*{Embedding networks into hyperbolic space via the Travelling Salesman Problem}

The Travelling Salesman Problem (TSP) is one of the most well-known and extensively studied optimization problem in computer science and mathematics~\cite{ancient_tsp_robinson, ancient_tsp_verblunsky,ancient_tsp_dantzig, ancient_tsp_karp}. It deals with the issue of finding the shortest possible route that a salesman can take to visit a given set of cities and return to the starting point, visiting each city only once (tour). The problem can effectively be modelled as a graph, wherein the nodes represent the cities to be visited by the salesman, whereas the edges of the graph correspond to the paths along which the salesman may travel. Each edge connecting two cities in the graph is assigned a weight being equivalent to the distance or cost of travelling between the two cities. In addition, provided that the resulting graph is fully connected, i.e. all pairwise distances are known in advance, the TSP can eventually be reformulated as the task of finding the shortest Hamiltonian cycle
in the graph.

In our approach, the first step is the identification of the communities in the network and the definition of weighted links between them based on their level of connectivity. Notably, the pre-weighting scheme we apply satisfies the triangle inequality, endowing the assigned weights with the role of virtual distance measures encapsulating the hyperbolic proximity between the detected communities (see Supplementary Information for more details). Consequently, this metric property ensures the seamless adaptation of the TSP to unveil the optimal angular arrangement of the modules in the native disk. As a next step of the algorithm, sub-modules are identified separately within each community that are arranged locally, again with the help of the TSP. This iteration is continued in a hierarchical manner, always dividing the communities at a given level into smaller smaller parts, defining weighted links between the found sub-modules and optimising the angular arrangement of the sub-modules within the original community via the TSP. After settling the angular coordinates in the above manner, the radial coordinate $r$ of the nodes are determined based on the node degree $k$, following a simple relation between $r$ and $k$ established in multiple hyperbolic network models\cite{PSO,hyperGeomBasics,S1H2_Mercator} and used in various other embedding methods\cite{EPSO_HyperMap,our_opt_embedding,commSector_hypEmbBasedOnComms_2016,commSector_hypEmbBasedOnComms_2019} (the details are described in Methods).

\begin{center}
\begin{figure}[hbt!]
\centering
    \includegraphics[width=0.95\textwidth]{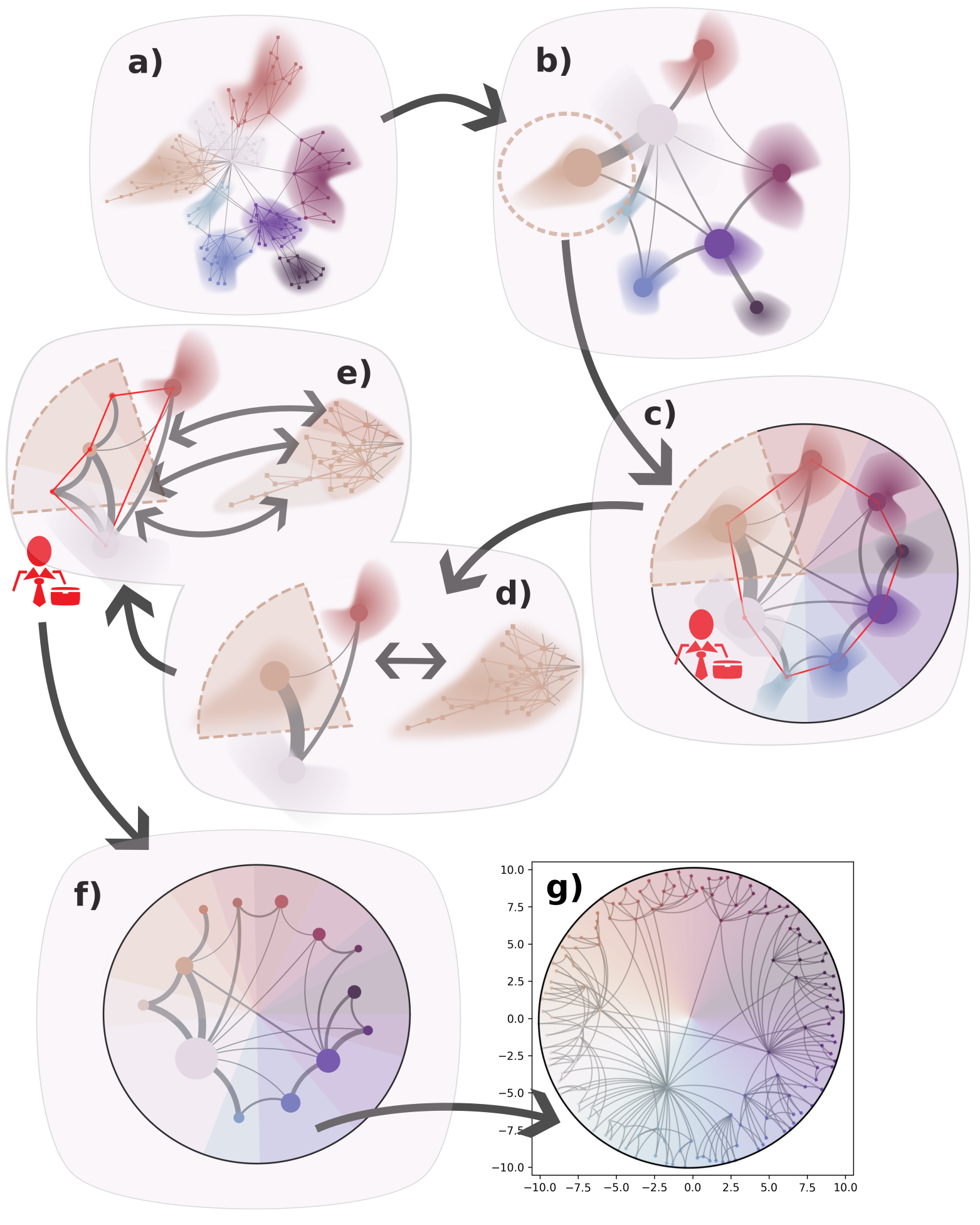}
\caption{\textbf{Illustration of the CLOVE algorithm.} a) A network with the detected communities indicated by the different colours. b) The weighted network between the communities. c) Optimal arrangement of the communities on the native disk according to the solution of the TSP on the weighted network in b). d) Zooming into one of the modules with the two neighbouring communities also shown. e) Sub-modules in the previous community and their optimal arrangement based on the TSP, taking into account also the neighbours from the top-level. f) Optimal arrangement of the sub-modules at the second level based on the local TSPs.  g) The network embedded into the native disk.
    \label{fig:algorithm_illustration}}
\end{figure}
\end{center}

An illustrating flow-chart of our algorithm is presented in Fig.\ref{fig:algorithm_illustration}., where the communities detected in the original network are marked by the different colours in Fig.\ref{fig:algorithm_illustration}a. This is followed by the definition of a weighted, complete graph between the found modules (shown in Fig.\ref{fig:algorithm_illustration}b), where the strength of a given connection roughly quantifies the extent of surprise that would be associated to it if the graph would have been generated by the configuration model and the resulting weights satisfy the triangle inequality (see Methods and the Supplementary Information for details). By solving the TSP on this weighted graph and taking the found shortest Hamiltonian path, we can arrange the communities on the native disk representation of the 2 dimensional hyperbolic space (Fig.\ref{fig:algorithm_illustration}c), where each community occupies an angular range proportional to its size, measured in the number of member nodes. 

In the next stage, we iterate over the communities, locating and arranging sub-modules within each of them. These sub-modules are found by simply applying the same community finding method as in the case of the original network but now only on the sub-graph of the given community (detached from the original network). Similarly to the top-level communities, we arrange the sub-modules based on the TSP, however this time the weighted graph between the sub-modules also includes two neighbouring communities from the top-level as indicated by Figs.\ref{fig:algorithm_illustration}d-e. The reason behind this is that these provide ''anchors'' for the sub-modules, allowing an arrangement that is using information coming from the surrounding of the original module. The angular range of the sub-modules is again proportional to their size. 

The above procedure is repeated in a hierarchical manner over each sub-module (and the even smaller sub-modules found within). When reaching to the point where the community finding method does not break the sub-module to further smaller communities, one can either  use a simple heuristic for the arrangement of the nodes within the sub-module (detailed in Methods) or treat the individual nodes as if they were the communities to be arranged on the next level below (and use again the TSP as in the case of the higher levels in the community hierarchy). 

Apart from the method for arranging the nodes on the lowest level in the hierarchy, our framework also allows flexibility in the choice of the community finding method (we use Leiden~\cite{leiden}, corresponding to a fast method that guarantees well-connected communities) and the method for solving the TSP (we apply the Christofides, sometimes referred to as Christofides–Serdyukov~\cite{tsp_solution_christofides,tsp_solution_serdyukov} algorithm with a threshold accepting boosting scheme~\cite{tsp_solution_ta}). A fully detailed description of our embedding algorithm is given in Methods.

Remarkably, our method not only proves scalable, allowing for the embedding of networks even with millions of nodes in a reasonable amount of time, but as we demonstrate below, it outperforms many state-of-the-art methods in several different quality scores.

\subsection*{Comparison with current state-of-the-art methods}

We tested CLOVE on several networks representing the web of connections in real complex systems. The size of these networks spanned from $N=10^{3}$ nodes to $N=2.7\cdot 10^6$ nodes and the studied systems belonged to various different domains, including social, biological and technological networks alike. We compared the performance of our approach with different state-of-the-art hyperbolic embedding methods according to multiple quality scores. These include the mapping accuracy\cite{mappingAccuracyAsSPLcorr}, MA, measuring the correlation between the shortest path distance and the geometric distance in the embedding space, the edge prediction precision, EPP, and the area under the receiver operating characteristic curve, AUR, in graph reconstruction\cite{AUROCmeaningInLinkPred,descriptionOfMeasuresOfGraRecAndLinkPred}, the greedy routing success rate\cite{Boguna_2009_nat_phys}, GR, corresponding to the fraction of successful paths when navigating according to the node coordinates in the network, the greedy routing score\cite{linkWeights_coalescentEmbedding}, GS, taking into account also the length of the paths during greedy routing and the greedy routing efficiency\cite{Carlo_Nat_coms_hyp_congruency}, GE, comparing the geometric distances and the projected greedy routing paths (the precise definition for all of these measures is provided in Methods). The alternative embedding methods serving as a baseline for comparison were the hyperbolic non-centered minimum curvilinear embedding (ncMCE) \cite{linkWeights_coalescentEmbedding}, relying on the dimension reduction of a weighted matrix encoding the distance relations, Mercator\cite{S1H2_Mercator}, combining the dimension reduction of the Laplacian matrix with a local optimisation with respect to the random hyperbolic graph, and the HMCS method\cite{commSector_hypEmbBasedOnComms_2019}, taking advantage of the hierarchical community structure of networks in a similar fashion to our approach, however, arranging the modules and sub-modules in a simple greedy fashion.

\begin{table}[hbt!]
\begin{tabular}{?l?c|c|c|c|c|c|c|c?}
    \thickhline
    & \makecell{MA} & \makecell{EPP} & \makecell{AUC} & \makecell{GR} & \makecell{GS} & \makecell{GE} & \makecell{Running\\ Time \\ (min.)}  & \makecell{Peak \\ Mem. \\ (GB)} \\
    \thickhline
    \makecell{CLOVE\\(default)} & \cellcolor[RGB]{96.42983343465943, 162.3210600851139, 45.43509339818203}$0.365$ & \cellcolor[RGB]{39.555148087280834, 100.67202136881365, 25.116873281532808}\textcolor{white}{$0.485$} & \cellcolor[RGB]{41.58478489939489, 103.12895014137277, 25.544165241977872}\textcolor{white}{$0.962$} & \cellcolor[RGB]{41.7156646413554, 103.2873835132197, 25.571718871864295}\textcolor{white}{$0.505$} & \cellcolor[RGB]{40.829961483183276, 102.2152165322745, 25.385255049091214}\textcolor{white}{$0.426$} & \cellcolor[RGB]{60.03006903336279, 125.45745198775495, 29.427382954392165}$0.180$ & \cellcolor[RGB]{247.12008672220384, 246.53966756488535, 246.8398843703949}$0.496$ & \cellcolor[RGB]{248.34805788938036, 241.83244475737536, 245.20258948082622}$0.426$ \\ \hline
    \makecell{CLOVE\\ (with SA)}  & \cellcolor[RGB]{97.06006955582002, 162.85045842688882, 45.83844451572482}$0.364$ & \cellcolor[RGB]{39.0, 100.0, 25.0}\textcolor{white}{$\mathbf{0.487}$} & \cellcolor[RGB]{41.6373728044765, 103.19260918436629, 25.55523637988979}\textcolor{white}{$0.962$} & \cellcolor[RGB]{40.89490576911658, 102.29383329945692, 25.398927530340334}\textcolor{white}{$\mathbf{0.507}$} & \cellcolor[RGB]{40.036134350506956, 101.2542678979821, 25.218133547475148}\textcolor{white}{$\mathbf{0.428}$} & \cellcolor[RGB]{59.0387094916268, 124.25738517407454, 29.218675682447746}$\mathbf{0.181}$ & \cellcolor[RGB]{247.12101482718944, 246.53610982910715, 246.83864689708074}$0.500$  & \cellcolor[RGB]{248.43338471289567, 241.50535860056652, 245.08882038280575}$0.453$ \\ \hline
    \makecell{CLOVE\\ (Louvain)}  & \cellcolor[RGB]{98.1714114457387, 163.78398561442052, 46.54970332527277}$0.362$ & \cellcolor[RGB]{39.63711694182546, 100.77124682431503, 25.13412988248957}\textcolor{white}{$0.485$} & \cellcolor[RGB]{41.67071382918305, 103.23296937216895, 25.562255542985906}\textcolor{white}{$0.962$} & \cellcolor[RGB]{43.53675445018305, 105.49186065022158, 25.955106200038536}\textcolor{white}{$0.500$} & \cellcolor[RGB]{43.00525989452214, 104.84847250389522, 25.843212609373083}\textcolor{white}{$0.421$} & \cellcolor[RGB]{61.854536475092246, 127.66601783826957, 29.811481363177315}$0.178$ & \cellcolor[RGB]{247.13224696164167, 246.49305331370692, 246.8236707178111}$0.546$ & \cellcolor[RGB]{248.38684931469348, 241.683744293675, 245.15086758040871}$0.438$ \\ \thickhline
\makecell{ncMCE \\(hyperbolic)} & \cellcolor[RGB]{115.09562108626892, 178.00032171246588, 57.381197495212106}$0.328$ & \cellcolor[RGB]{194.26670791014928, 229.463786047891, 150.51600837719667}$0.173$ & \cellcolor[RGB]{44.86653549431641, 107.10159559838303, 26.235060104066612}\textcolor{white}{$0.946$} & \cellcolor[RGB]{200.47369730151672, 232.16247708761597, 160.5011652241791}$0.168$ & \cellcolor[RGB]{197.50646697198982, 230.8723769443434, 155.72779469407055}$0.147$ & \cellcolor[RGB]{201.05683300226076, 232.41601434880903, 161.4392530905934}$0.066$ & \cellcolor[RGB]{248.43270192195942, 241.50797596582225, 245.0897307707208}$5.918$  & \cellcolor[RGB]{142,1,82}\textcolor{white}{$9.475$} \\ \hline    
    \makecell{ \\Mercator} & \cellcolor[RGB]{39.0, 100.0, 25.0}\textcolor{white}{$\mathbf{0.506}$} & \cellcolor[RGB]{53.78151030392455, 117.89340721001392, 28.111896906089378}\textcolor{white}{$0.449$} & \cellcolor[RGB]{39.0, 100.0, 25.0}\textcolor{white}{$\mathbf{0.976}$} & \cellcolor[RGB]{116.44197215016649, 179.13125660613986, 58.242862176106556}$0.329$ & \cellcolor[RGB]{103.3544990440166, 168.13777919697395, 49.86687938817063}$0.299$ & \cellcolor[RGB]{129.73985085119702, 189.77849967533842, 68.31666155671219}$0.119$ & \cellcolor[RGB]{142,1,82}\textcolor{white}{$123.921$}  & \cellcolor[RGB]{237.13319929274917, 169.17850291806297, 209.04530362531384}$4.176$ \\ \hline
    \makecell{ \\HMCS}  & \cellcolor[RGB]{113.16808828596413, 176.38119416020987, 56.14757650301705}$0.331$ & \cellcolor[RGB]{159.03466076930826, 208.79442892042817, 103.7787998786363}$0.237$ & \cellcolor[RGB]{42.66700785104561, 104.43900950389732, 25.772001652851706}\textcolor{white}{$0.957$} & \cellcolor[RGB]{131.45485110426753, 190.89174545364733, 70.39271449463965}$0.300$ & \cellcolor[RGB]{124.67790396919997, 186.04943933412798, 63.513858540287984}$0.262$ & \cellcolor[RGB]{132.76205210519396, 191.74027943670484, 71.97511570628741}$0.117$ & \cellcolor[RGB]{247.05313905797476, 246.79630027776346, 246.92914792270034}$\mathbf{0.220}$  & \cellcolor[RGB]{248.12565969608795, 242.68497116499623, 245.49912040521608}$\mathbf{0.356}$ \\
    \thickhline
\end{tabular}
\caption{ {\bf Average quality scores for small and medium sized networks.} We show the results for the mapping accuracy, MA, the edge prediction precision, EPP, the area under the receiver operating characteristic curve, AUC, the greedy routing score, GR, the greedy success rate, GS and the greedy routing efficiency, GE, averaged over 10 networks with size ranging between $N=1000$ and $N=20,000$ nodes. Beside the quality scores, we also display the running time in seconds and the peak memory usage in GB. In the top part the table we list the scores obtained for CLOVE with default settings, for CLOVE with 
simulated annealing optimisation during the solution of the TSP problem  and  for CLOVE with Louvain communities. For comparison, in the bottom part of the table we give the results for hyperbolic ncMCE, Mercator and HMCS.
}
\label{table:small_nets}
\end{table}

In Table \ref{table:small_nets}. we show the quality scores averaged over 10 networks falling into the size range between $N=1000$ and $N=20000$. (In Tables S2-S11 in the Supplementary Information we also display the results for the individual networks one by one). In addition to the quality scores, Table \ref{table:small_nets}. also provides the running time and the peak memory usage during the different processes. According to the results for the different algorithms, Mercator achieved far the best mapping accuracy score and the best AUC value, whereas CLOVE with an additional simulated annealing during the solution of the TSP turned out to be the best according to the edge prediction precision, the greedy routing score, the greedy success rate and the greedy routing efficiency. We note that all CLOVE versions outperformed both HMCS and hyperbolic ncMCE according to all quality scores, and also Mercator regarding the greedy routing based scores (GR, GS and GE). 

In terms of the time consumption, not surprisingly HMCS turned out to be the best, followed by the different CLOVE implementations. 
Furthermore, all CLOVE versions ran roughly 10 times faster than hyperbolic ncMCE in our experiments, and more than 200 times faster than Mercator. Finally, HMCS has the lowest peak memory usage, where the results for the different CLOVE versions are not far behind, and are considerably smaller compared to the memory usage of hyperbolic ncMCE and Mercator.

\begin{table}[hbt!]
\begin{tabular}{?l?c|c|c|c|c|c|c|c?}
    \thickhline
    & \makecell{MA} & \makecell{EPP} & \makecell{AUC} & \makecell{GR} & \makecell{GS} & \makecell{GE} & \makecell{Running \\ Time \\ (min.)}  & \makecell{Peak \\ Mem. \\ (GB)} \\
    \thickhline
    \makecell{CLOVE\\(default)} & \cellcolor[RGB]{212.0225908320109, 237.18373514435257, 179.07982003410447} $\mathbf{0.278}$ & \cellcolor[RGB]{182.52483574923767, 224.04243724073322, 132.21427485434035}$0.405$ & \cellcolor[RGB]{45.917409083431785, 108.37370573257532, 26.456296649143532} \textcolor{white}{$\mathbf{0.964}$} & \cellcolor[RGB]{206.1697210677327, 234.63900915988378, 169.66433389157}$0.304$ & \cellcolor[RGB]{224.60099502609322, 242.652606533084, 199.31464417241082}$0.223$ & \cellcolor[RGB]{240.32243259835434, 246.2144038351005, 231.68087478445995}$\mathbf{0.079}$ & \cellcolor[RGB]{249.7635777983912, 236.40628510616705, 243.31522960214505}$663.259$ & \cellcolor[RGB]{249.03084349632263, 239.21509993076324, 244.29220867156982}$\mathbf{2.708}$ \\ \hline
    \makecell{CLOVE\\ (with SA)} & \cellcolor[RGB]{212.54155519961492, 237.40937182591952, 179.91467575590224}$0.276$ & \cellcolor[RGB]{181.46276001887298, 223.35301966137368, 130.92860423337254}$\mathbf{0.409}$ & \cellcolor[RGB]{45.91338323872956, 108.36883234161999, 26.455449102890434} \textcolor{white}{$\mathbf{0.964}$} & \cellcolor[RGB]{205.66379193735457, 234.41903997276285, 168.85044789922256}$\mathbf{0.306}$ & \cellcolor[RGB]{224.27237373328643, 242.50972771012454, 198.7859925274608}$\mathbf{0.225}$ & \cellcolor[RGB]{240.2815657663593, 246.20959597251286, 231.58712146400077}$\mathbf{0.079}$ & \cellcolor[RGB]{249.88756818694756, 235.93098861670097, 243.1499090840699}$693.016$ & \cellcolor[RGB]{249.3723382609231, 237.9060366664614, 243.83688231876917}$3.163$ \\ \hline
    \makecell{CLOVE\\ (Louvain)} & \cellcolor[RGB]{212.19516337507676, 237.25876668481598, 179.35743673381913}$0.277$ & \cellcolor[RGB]{182.40991671027416, 223.96784067158146, 132.07516233348977}$0.406$ & \cellcolor[RGB]{46.19018754776987, 108.70391124203721, 26.51372369426734} \textcolor{white}{$0.962$} & \cellcolor[RGB]{209.07714392172758, 235.90310605292504, 174.34149239582266}$0.291$ & \cellcolor[RGB]{226.68726613523276, 243.55968092836207, 202.67081943493966}$0.214$ & \cellcolor[RGB]{240.56630144635812, 246.24309428780683, 232.2403386122333}$0.076$ & \cellcolor[RGB]{249.70520126995171, 236.6300617985184, 243.3930649733977}$649.248$  & \cellcolor[RGB]{249.385768362454, 237.85455461059297, 243.818975516728}$3.181$ \\
    \thickhline
   \makecell{ \\ HMCS} & \cellcolor[RGB]{213.6854304347977, 237.90670888469467, 181.75482287337022}$0.271$ & \cellcolor[RGB]{237.81026439794607, 245.91885463505247, 225.9176653835233}$0.108$ & \cellcolor[RGB]{47.4864999040496, 110.2731314627969, 26.786631558747285} \textcolor{white}{$0.955$} & \cellcolor[RGB]{240.79442511874205, 246.26993236691084, 232.76368115476117}$0.073$ & \cellcolor[RGB]{241.9731764290571, 246.40860899165378, 235.46787533724864}$0.059$ & \cellcolor[RGB]{245.20031766573786, 246.7882726665574, 242.87131699786917}$0.021$ & \cellcolor[RGB]{247.37108819994927, 245.57749523352777, 246.50521573340097}$\mathbf{89.061}$  & \cellcolor[RGB]{249.33495128154755, 238.0493534207344, 243.88673162460327}$3.113$ \\
    \thickhline
\end{tabular}
\caption{ {\bf Average quality scores for large networks.} We display the measured average scores for the mapping accuracy, MA, the edge prediction precision, EPP, the area under the receiver operating characteristic curve, AUC, the greedy routing score, GR, the greedy success rate, GS  and the greedy routing efficiency, GE, averaged over 17 networks with size ranging between $N=20000$ and $N=2.7\cdot 10^6$ nodes. Beside the quality scores, we also display the running time and the peak memory usage in GB. In the top part the table we list the scores obtained for CLOVE with default settings, for CLOVE with simulated annealing optimisation during the solution of the TSP problem and  for CLOVE with Louvain communities. For comparison, in the bottom row we give the results for HMCS.
}
\label{table:large_nets}
\end{table}

In Table.\ref{table:large_nets}. we provide the average values for the studied embedding quality scores in large networks, corresponding to systems where the number of nodes varies between $N=2\cdot 10^{3}$ and $N=1.3\cdot 10^6$. The same quality indicators for the individual networks are listed in a similar manner in tables S12-S25 in the Supplementary Information. An important difference compared to the case of smaller networks is that since the scores are defined as various sums over node pairs, their exact evaluation becomes unfeasible, and therefore, we relied on sampling from all possible node pairs when calculating the quality measures. In Sect.S4 in the Supplementary Information we examine the relation between the exact quality score values and their estimates based on sampling in smaller systems, arriving to the conclusion that sampling offers a reasonably precise estimate of the exact values already at relatively low frequency values. A further difference compared to Table.\ref{table:small_nets}. is that due to the larger resource requirements in terms of computation time or memory, Mercator and the hyperbolic ncMCE method were not applied to the larger networks.

According to the results shown in Table.\ref{table:large_nets}., CLOVE significantly outperforms HMCS according to all quality scores at the cost of having a roughly 8 times as large computation time. While CLOVE with extra simulated annealing seems to be the best among the different CLOVE versions in Table.\ref{table:large_nets}., when examining the detailed list of results for the individual networks in tables S12-S25 in the Supplementary Information, it becomes clear that in certain systems it is the default version or the one relying on Louvain communities that achieves the best result. Nevertheless, aside from MA and AUC, clear gap between the scores of CLOVE and that of HMCS is always present.

\newcolumntype{?}{!{\vrule width2pt}}

\newcolumntype{C}[1]{>{\centering\arraybackslash}m{#1}}

\begin{figure}[hbt]
    \centering
    \includegraphics[width=\linewidth]{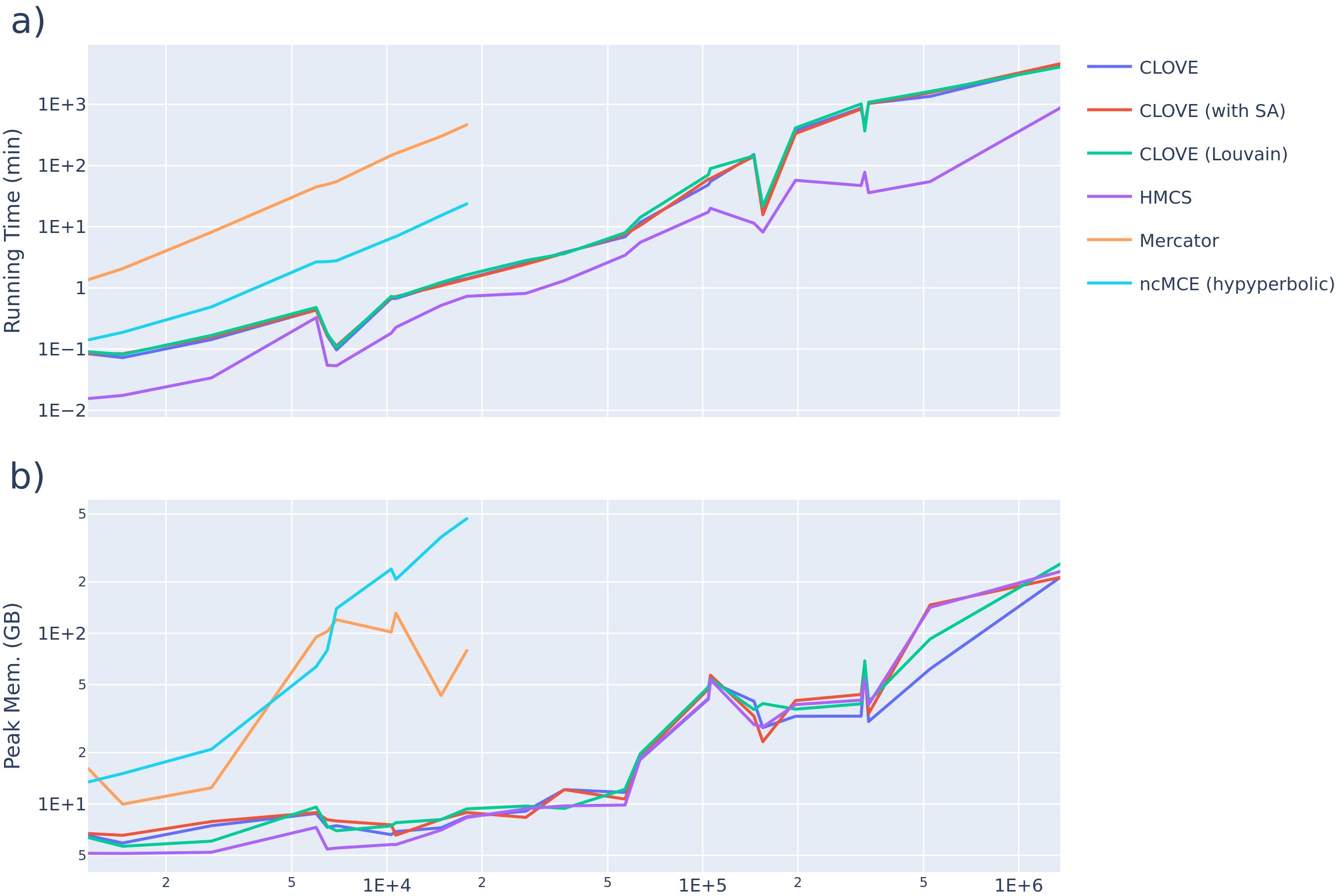}
    \caption{\textbf{Resource usage of the studied algorithms as a function of the size of the embedded networks.} We plot the average running time in panel a) and the peak memory usage in panel b), with the colour code of the different methods given in the legends.}
    \label{fig:resource_usage}
\end{figure}

In Fig.\ref{fig:resource_usage}. we show the computational resource usage of the studied embedding methods as a function of the network size (measured in the number of nodes). Naturally, all of the curves show an overall increasing tendency, however, they are not strictly monotonic, indicating that besides the size, also the structure of the network can have a strong effect on the amount of computational resources needed for the embedding. The comparison between the different curves leads to a conclusion that is consistent with the previous results shown in Tables.\ref{table:small_nets}-\ref{table:large_nets}.: As expected, among the studied methods HMCS is the fastest followed by our different CLOVE implementations. The time curves for the hyperbolic ncMCE and Mercator seem to be steeper compared to the previous approaches and these methods run slower by at least one order of magnitude at the upper size limit of smaller networks ($N=20,000$ in our study). In parallel, the peak memory usage (Fig.\ref{fig:resource_usage}b) displays two bundles of curves, where the CLOVE implementations and HMCS show very similar memory needs, which are considerably more moderate compared to those of Mercator and hyperbolic ncMCE.

\subsection*{Hyperbolic maps of real networks with ground-truth modules}

In this section we demonstrate that the embeddings generated by our approach can provide intuitive node arrangements in the native disk for different complex systems. For a small fraction of the networks we analysed, "ground truth" modules and/or additional node labels were also available beside the network topology. Although our method is agnostic with respect to any extra node labels and calculates the coordinates solely based on the network structure, still, the organisation of the obtained layouts is meaningful also in the light of these additional features.

\subsubsection*{The network of tennis matches between ATP players}

ATP stands for the Association of Tennis Professionals, which serves as the governing body for men's professional tennis. It is responsible for overseeing and managing various aspects of this sport, including the organization of tournaments and the establishment of player rankings. Related to that, here we examine a tennis dataset accessible at~\cite{tennis_atp}, with a central question in mind; Can the two-dimensional hyperbolic space efficiently host the network representing the matches between ATP tennis players?

\begin{figure}[hbt!]
\includegraphics[width=1.\textwidth]{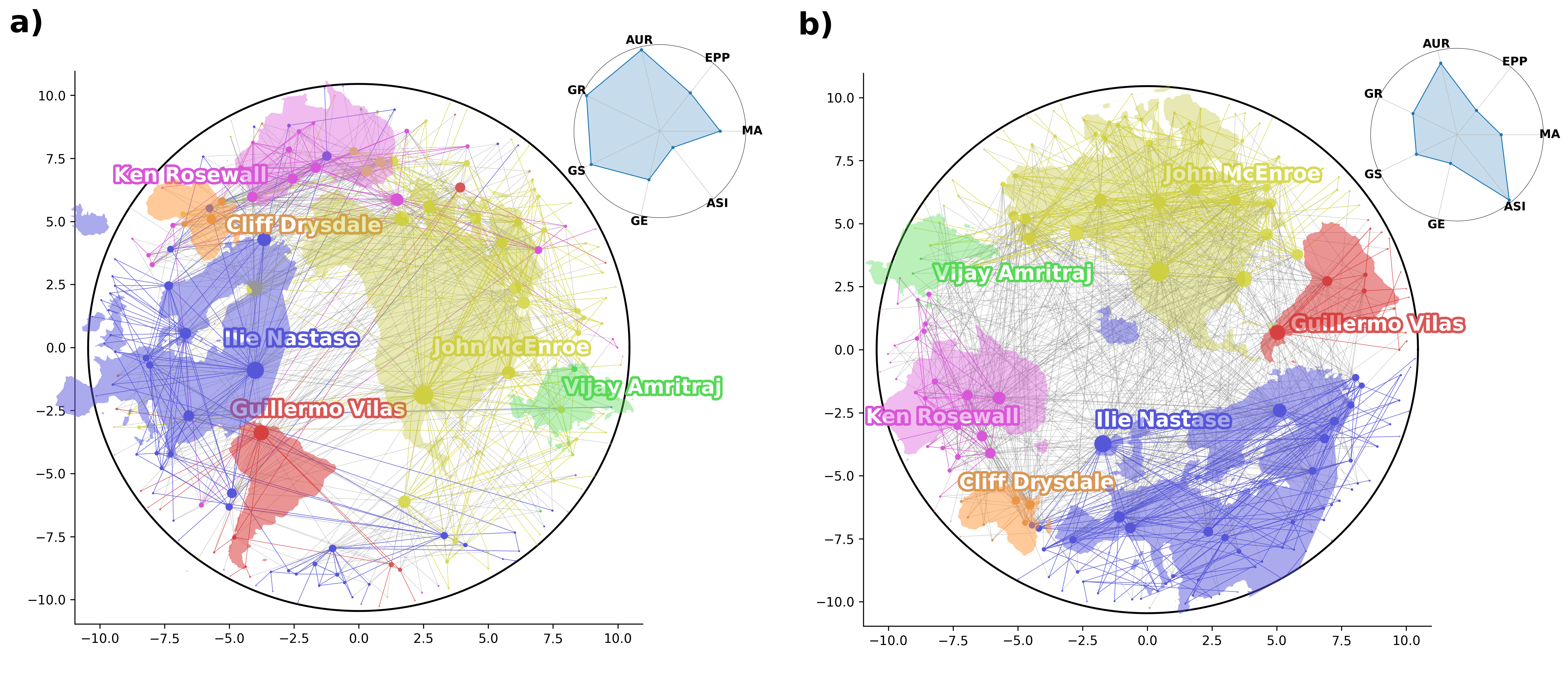}
\caption{\textbf{The ATP tennis network embedded into the two-dimensional hyperbolic space using two different strategies.} a) The hyperbolic layout obtained by running the CLOVE method in its default settings 
alongside with the associated metric scores displayed in a radar chart at the top-right corner. We show the results for the mapping accuracy, MA, the edge prediction precision, EPP, the area under the receiver operating characteristic curve, AUC, the greedy routing score, GR, the greedy success rate, GS and the greedy routing efficiency, GE. b) Embedding the tennis network by relying on a regional dendrogram comprising ground-truth information about the ethnicities of the players. In a similar fashion to panel a), the same metric scores are presented again in a radar chart at the top-right corner. In both panels, the network nodes are colored based on the continent to which the corresponding players belong, with the continents outlined and positioned according to the angular coordinates of their respective players. In panel a) the higher metric scores and fewer edge crossings suggest that using CLOVE with the default settings, as shown generally yields better embedding quality.
\label{fig:tennis_network}}
\end{figure}

In order to investigate this question in detail, we first build the network by considering the matches between the top-ranked ATP players who competed against each other during the period from 1969 to 1989 and participated in at least 7 official matches. Subsequently, we apply the CLOVE algorithm with various parameter settings to map this network to the native disk representation of the two-dimensional hyperbolic space. Our approach consists of two rather different embedding strategies. In the first case, we run our algorithm with its default settings, where communities are identified and arranged in a nested fashion using a fast community detection method (e.g. Louvain or Leiden) applied across increasingly finer scales. The resulting hyperbolic layout is displayed in Fig.\ \ref{fig:tennis_network}a, along with the angular sectors where players from distinct continents are predominantly clustered. Moreover, in Fig.\ \ref{fig:tennis_network}a we also indicate the position of a prominent tennis player for each continent.

In our second embedding approach, the identification of network modules to be positioned on the native disk is not dictated by the output of a pre-defined community detection method. Instead, we rely on a two-level dendrogram that incorporates ground-truth information regarding the ethnicities of the players. The first level pertains to the nationalities of the players, while the second level maps nations to continents, thus forming a complete dendrogram of communities. This regional dendrogram is passed to the embedding algorithm, which then arranges the communities accordingly, again based on the TSP. We show the obtained hyperbolic layout in Fig.\ref{fig:tennis_network}b, where in a similar fashion to Fig.\ref{fig:tennis_network}a, both the angular sectors corresponding to the continents and the same top-tier players for each continents are highlighted.

Overall, by observing the quality scores displayed at the top-right corner of the panels in Fig.\ref{fig:tennis_network}, we can deduce that the embedding quality is superior in the first scenario, i.e., when the modules to be arranged on the disk are derived from a community detection method, rather than being constructed based on the regional dendrogram. This phenomenon can roughly be explained by the presence of intercontinental links in the ATP network. More specifically, when modules are defined based on regional information, these intercontinental links can become excessively long, as different continents may be positioned far apart on the native disk, eventually leading to a sub-optimal embedding. Contrarily, when modules to be arranged by the algorithm are derived from a community detection method, the majority of links tend to fall within the same angular sector. This spatial concentration of the links results in shorter average link lengths, which in turn enhances the overall quality of the embedding. This explanation is perfectly corroborated by the observation of fewer link crossings in the embedding shown in Fig.\ref{fig:tennis_network}a.

\subsubsection*{The air transportation network}

\begin{figure}[hbt!]
\centering
    \includegraphics[width=\textwidth]{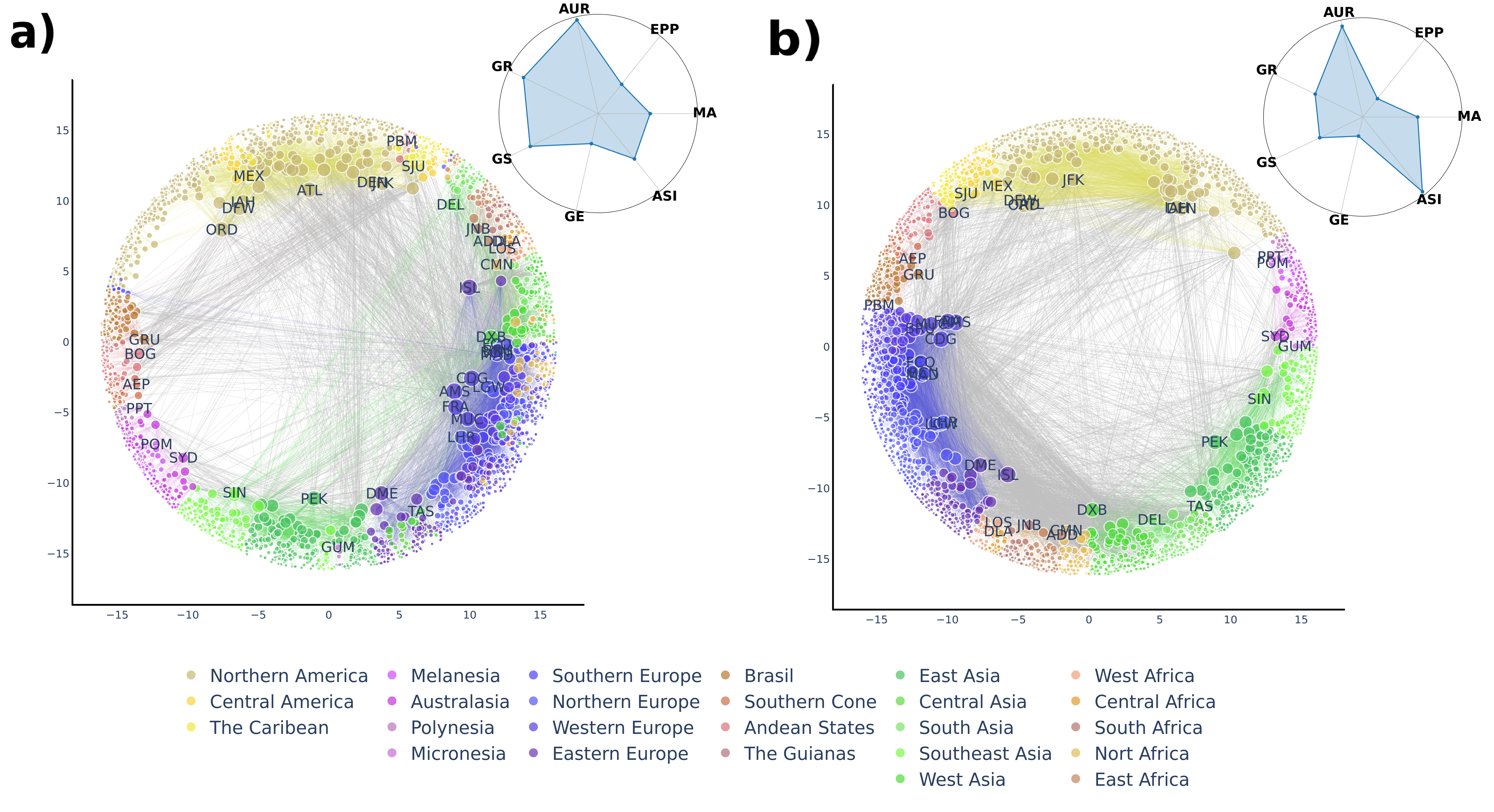}
    \caption{{\bf Embedding of the air transportation network.} The major geographical regions such as continents and sub-continents are colour coded and the size of the nodes indicates the degree. The most important airports are marked by their IATA code and the radar plots in the insets show the different quality scores of the embedding. a) The embedding obtained with CLOVE at default settings. b) The embedding with CLOVE using a dendrogram corresponding to the hierarchy of geographic locations. In both panels, the radar charts positioned in the top-right corners show the qualities of the embeddings using the same metric scores as depicted in Fig. 2. The large similarity between the panels indicates that CLOVE with default settings in panel a) found an arrangement very close to the ground truth categorisation of geographical regions solely based on the network structure. This is accompanied by a clear separation of continents in terms of angular coordinates, despite the embedding being completely agnostic to geographical information.}
    \label{fig:airports_figure}
\end{figure}

The OpenFlights database\cite{openflights} provides detailed information on regular commercial flights between major airports worldwide, containing more than 3,000 airports and roughly 67,000 flights, defining a transportation network of crucial importance. Similarly to the ATP tennis network, in our study of this system  we applied CLOVE both with default settings (results shown in Fig.\ref{fig:airports_figure}a) and with a pre-defined dengrogram of geographical regions (results shown in Fig.\ref{fig:airports_figure}b). The seemingly large similarity between the two layouts in Fig.\ref{fig:airports_figure}. indicates that our algorithm was able to find a natural arrangement for the nodes even when it was completely unaware of the ground truth geographical categorisation of the airports and calculated the embedding coordinates solely based on the network structure.  

\begin{figure}[hbt!]
\centering
    \includegraphics[width=0.93\textwidth]{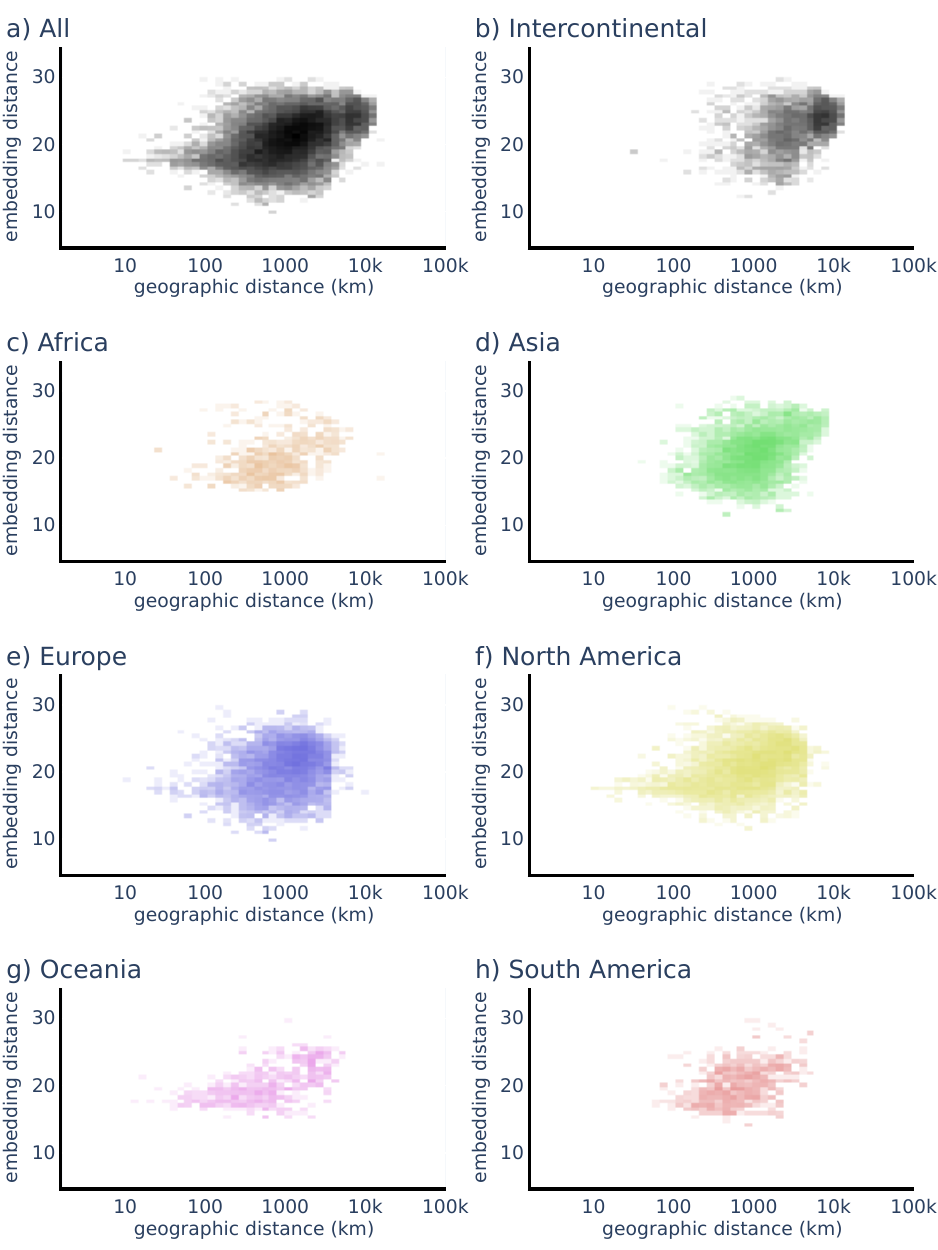}
    \caption{{\bf Embedding distance and geodesic distance in the air transportation network.} We plot the distance measured on the hyperbolic disk (the embedding distance) for connected airport pairs as a function of the geodesic distance on the globe, measured in kilometers. The panels depict heat-maps corresponding to different large geographical regions as indicated in the panel titles. The fact that intercontinental connections tend to be longer than continental ones also in the embedding space reinforces that the embedding obtained solely based on the network structure captures essential features of the original system.}
\label{fig:openflight_geodesic_vs_embedding_distance}
\end{figure}

Additionally, in Fig.\ref{fig:openflight_geodesic_vs_embedding_distance}a we plot the embedding distance (measured on the native disk) as a function of the real-world geodesic distance a given flight covers between two airports. The intercontinental flights (Fig.\ref{fig:openflight_geodesic_vs_embedding_distance}b) tend to travel the largest distance in both the real world and in the embedded space. In turn, the flights within a given continent (Figs.\ref{fig:openflight_geodesic_vs_embedding_distance}c-h) are usually shorter, again according to both distance measures. This shows that in spite of the difference in the curvature of the underlying geometry and the fact that the embedding is completely unaware of the true flight distances (i.e., it is inputted an unweighted network), still our algorithm is finding an arrangement of the airports on the native disk which is coherent with the real world geographical positioning of the airports. This is also supported by a Pearson correlation coefficient of 0.40 between the embedding distance and the geodesic distance.

In summary, as demonstrated by the examples of ATP and Openflights networks, CLOVE performs notably well, whether using its default settings (see Fig.\ref{fig:tennis_network}a and Fig.\ref{fig:airports_figure}a) or a pre-defined dendrogram of geographical regions (see Fig.\ref{fig:tennis_network}b and Fig.\ref{fig:airports_figure}b). However, the former strategy is generally better than the latter, as evidenced by the reduced number of long-range interconnections in Fig.\ref{fig:tennis_network}a and Fig.\ref{fig:airports_figure}a corresponding to the default versions of CLOVE. This superiority is further reflected by the fact that running the method with its default settings almost always yields higher metric scores (shown in the upper right corner of panels Fig. \ref{fig:tennis_network}a-b and Fig. \ref{fig:airports_figure}a-b). Nonetheless, it is important to note an exception, specifically the ASI score, which measures the angular coherence of communities. In general, a high ASI value indicates well-separated communities in terms of angular coordinates, thus reaching its maximal value when the arrangement is explicitly constructed based on the ground-truth dendrogram of communities. This observation is supported by the radar charts illustrated in Fig. \ref{fig:tennis_network}b and Fig. \ref{fig:airports_figure}b. For a more detailed description of ASI and the other metric scores employed in our analysis, please refer to the Methods section.

\section*{Discussion}
A prevalent and very essential feature of numerous complex systems -- observed in either nature or society--, lies in the presence of an inherent hierarchical structure that governs the relationships among their constituent components~\cite{zafeiris_Vicsek_hier, Sole_hier_PNAS, Balogh_mesh}. 
Gaining access to these nested hierarchical structures can be beneficial from various aspects; for instance, it can streamline the design of efficient search protocols among the constituents~\cite{opt_hier_search}, facilitate optimal decision-making~\cite{zafeiris_Vicsek_hier}, and even economize the costs associated with reliable information transfer~\cite{guimera2001communication_opt_hier}. 

In this study, we utilised these distinctive architectures to effectively address the hyperbolic embedding of complex networks. Specifically, we introduced a method called CLOVE, which accomplishes the mapping of networks into the two-dimensional hyperbolic space through a series of optimization tasks performed in a hierarchical manner. When dealing with a given network, the CLOVE method involves two fundamental steps; initially, it begins by reasonably partitioning the network into smaller interconnected entities, followed by determining their optimal arrangement within the hyperbolic disk. While advanced community finding methods such as the Leiden method can effectively achieve a sensible partitioning of the network into smaller units, finding the optimal arrangement of these sub-modules on the hyperbolic disk remains a highly challenging task. The CLOVE method brings significant progress in addressing this challenge by leveraging the Travelling Salesman Problem~\cite{ancient_tsp_robinson, ancient_tsp_verblunsky, ancient_tsp_karp, ancient_tsp_dantzig} --an extensively studied problem in computer science -- to optimize the arrangement of communities and their respective sub-communities in the hyperbolic disk. While the MCA method described in Ref.~\cite{cannistraci_minimum_curvilinearity} employs a somewhat related minimum spanning tree based approach, to the best of our knowledge, this study is the first to explicitly use the TSP for solving the embedding of complex networks. CLOVE introduces a whole new family of embedding techniques, providing a highly efficient alternative framework to well-established methods such as likelihood optimization and spectral-based embeddings.

The TSP is undoubtedly one of the best-known combinatorial optimisation problems, with applications ranging from DNA sequencing~\cite{DNA_sequencing_with_TSP}, aerospace engineering~\cite{tsp_aerospace}, the analysis of crystals' structures~\cite{tsp_crystal}, to the planning of telescope movement in astronomy~\cite{WJCook_book_on_TSP,TSP_in_astronomy}. Additionally, it has proven to be highly effective in defining and measuring the geometric separability (both linear and nonlinear) of mesoscale patterns in multidimensional datasets~\cite{carlo_tsp}. In this paper, we introduced a novel application in complex network theory, facilitating the rapid optimization of node arrangements in 
the two-dimensional hyperbolic space.

Even though the complexity of the chosen heuristic approximation method is $\mathcal{O}(C^3)$ in the number of "cities", $C$, since we do not run the TSP on all the nodes at once, instead only on modules appearing together on a given level in a given branch of the module-hierarchy, the embedding of networks with millions of nodes can be accomplished in less than 50 hours. Although this falls behind the running time of very fast methods like HMCS\cite{commSector_hypEmbBasedOnComms_2019}, in our opinion CLOVE provides a favorable balance between speed and accuracy. On average, CLOVE outperformed HMCS according to all studied quality indicators and its embedding quality is comparable, and in many cases, even superior to state-of-the-art methods, such as Mercator\cite{S1H2_Mercator}.

A slight limitation of our method is its ability to embed networks solely into the two-dimensional hyperbolic space. Recent progress in complex network theory has both generalised the fundamental hyperbolic network models to higher dimensions\cite{RHG_d_dim_mathematics,RHG_d_dim_krioukov,dPSO} and has also introduced higher dimensional hyperbolic embeddings\cite{our_dir_embedding,dmercator}. Nonetheless, extending CLOVE to higher dimensions poses a non-trivial task, offering an intriguing challenge for future research, although it falls beyond the scope of the present paper.

In conclusion, owing to the scalable nature of CLOVE, it becomes feasible to map even very large networks into hyperbolic space within a reasonable amount of time, moreover with a high level of reliability. This remarkable efficiency of the CLOVE method undoubtedly represents a significant step towards the creation of hyperbolic maps for a wide range of real-world complex networks.

\section*{Acknowledgements}
S. G. B. acknowledge support from the National Laboratory for Health Security, RRF-2.3.1-21-2022-00006. This research was also partially supported by the National Research, Development and Innovation Office – NKFIH, grants No: OTKA-SNN 139598 and K128780.

\section*{Author contributions}

G. P. and S. G. B. developed the concept of the study, S. G. B. proposed and developed the application of the Travelling Salesman Problem (TSP) in the embedding method, S. G. B. and B. S. implemented the embedding methods, B. S.  pre-processed the network data, tested the embedding methods and performed the analyses, B. S. and S. G. B. prepared the figures, B. S. prepared the tables, G. P. and S. G. B. wrote the paper, G. P., S. G. B. and T. V. contributed to the interpretation of the results. All authors reviewed the manuscript.

\section*{Competing interests}
The authors declare no competing interests.

\section*{Code availability}
The Python implementation of CLOVE will be made available in a Github repository upon publication.

\section*{Data availability}
All data used in this work is publicly available from the KONECT project at \url{http://www.orgnet.com}. The access links the individual network data sets are provided in the Supplementary Information.

\section*{Methods}
\subsection*{Networks in the native disk representation of the hyperbolic space}

A common approach to the study of hyperbolic network geometry is the use of the native representation  of the two-dimensional hyperbolic space \cite{hyperGeomBasics}, where the hyperbolic plane of constant curvature $K<0$ is represented by a disk of infinite radius in the Euclidean plane. The advantage of this representation compared to the famous Poincaré disk model is that the radial coordinate $r$ of a point (defined as its Euclidean distance from the disk centre) is equal to its actual hyperbolic distance from the disk centre. In addition, the Euclidean angles between hyperbolic lines are also equal to their hyperbolic counterparts.

The hyperbolic distance between two points can be measured along the connecting geodesic, which is either a hyperbola, or -- if the disk centre falls on the Euclidean line connecting the two points -- the corresponding diameter of the disk. The hyperbolic distance $x$ between two points at polar coordinates $(r,\theta)$ and $(r',\theta')$ can be calculated from the hyperbolic law of cosines written as
\begin{equation}
    \mathrm{cosh}(\zeta x)=\mathrm{cosh}(\zeta r)\,\mathrm{cosh}(\zeta r')-\mathrm{sinh}(\zeta r)\,\mathrm{sinh}(\zeta r')\,\mathrm{cos}(\Delta\theta),
    \label{eq:hypDist}
\end{equation}
where $\zeta=\sqrt{-K}$ and $\Delta\theta=\pi-|\pi-|\theta-\theta'||$ is the angle between the examined points. According to Ref. \cite{hyperGeomBasics}, for $2\cdot\sqrt{e^{-2\zeta r}+e^{-2\zeta r'}}<\Delta\theta$ and sufficiently large $\zeta r$ and $\zeta r'$, the hyperbolic distance can be approximated as
\begin{equation}
    x\approx r+r'+\frac{2}{\zeta}\cdot\ln\left(\frac{\Delta\theta}{2}\right).
    \label{eq:hypDistApprox}
\end{equation}

When generating random graphs via geometric network models operating in the native disk, or embedding networks into the native disk, there seems to be an intimate relation between the node degree and the radial position. Hyperbolic network models are centered around the the idea of placing nodes on the native disk (in a uniform or close to uniform fashion) and drawing links with a probability depending
on the metric distance. In general, such models can be regarded as a particular case of a broader hidden variable framework~\cite{caldarelli_first_peaked,caldarelli_general_sf_recipe,boguna_general_hv,Garlaschelli_Entropy,gen_treshold_scirep}, where the hidden variables of the nodes are associated with the coordinates of the nodes in the hyperbolic space, whereas the connection probability between pair of nodes depends specifically on their respective distances.

One of the best-known hyperbolic network models is given by Popularity-Similarity Optimisation (PSO) model\cite{PSO}. In case of the PSO model (where new nodes are added to the network one be one with logarithmically increasing radial coordinate and random angular coordinate), a rather intuitive analogy was drawn between the coordinates and plausibly important features of the nodes such as the popularity and similarity that govern the network growth. In this picture a small angular distance indicates a high similarity between a node pair, whereas the popularity (the degree) of the nodes is controlled by their radial coordinate, with hubs appearing closer to the disk centre and low degree nodes occupying the disk periphery. 

More specifically, in the PSO-model the expected degree of node $i$ at time point $t$ in the network generation is $\bar{k_i}(t)\sim \exp(r_{it}-r_{tt})$ where $r_{tt}=\frac{2}{\zeta}\ln t$ is the radial coordinate of the newly appearing node at $t$ (with $\zeta=\sqrt{-K}$ originating from the hyperbolic curvature $K$, usually assumed to be $\zeta=1$) and $r_{it}$ is the actual radial coordinate of node $i$ that was shifted from its original $r_{ii}$ value as $r_{it}=\beta r_{ii}+(1-\beta)r_{tt}$, where $\beta\in (0,1]$ corresponds to the popularity fading parameter\cite{PSO}. Related to this, when assuming that a network was generated by the PSO-model, the maximum likelihood estimate for the radial coordinate be given as \cite{HyperMap}
\begin{subequations}
\begin{align}
   r_{ii}^* &=\frac{2}{\zeta}\ln i^*, \label{eq:opt_fading_radial} \\
    r_{iN}^* &= \beta r_{ii}^* +(1-\beta)r_{NN}^*, 
    \label{eq:opt_radial}
\end{align}
\end{subequations}
where the optimal ordering of the nodes given by $i^*$ is following the node degrees, with the largest degree node in the network obtaining $i^*=1$, second largest degree node receiving $i^*=2$, etc., and equation (\ref{eq:opt_fading_radial}) corresponds to the initial radial coordinate of node $i^*$, whereas equation (\ref{eq:opt_radial}) takes into account also the outward drift due to the popularity fading.

A similarly close relation occurs between the node degree and the radial coordinate in the random hyperbolic graph (RHG) model\cite{hyperGeomBasics}, also known as the  $\mathbb{S}^1/\mathbb{H}^2$ model\cite{S1H2_Mercator}. In this static approach nodes are given uniform random angular coordinates and a hidden degree variable $\kappa$ sampled from a power-law distribution. Node pairs are connected according to a probability that is decreasing as a function of the angular distance but also takes into account the product of the hidden variables, resulting in a scale-free network where the degree decay exponent is the same as for the hidden variable distribution and the expected degree of node $i$ is given by $\kappa_i$. When mapping the network onto the native disk, the radial coordinate is defined as $r_i=R_0-2\ln(\kappa_i)$ where $R_0$ is a constant depending on the model parameters. Hence, similarly to the PSO-model, the hubs are placed close to the disk centre, the low-degree nodes are located towards the periphery and there is an overall logarithmic dependence between the degree and the radial coordinate.

Numerous hyperbolic embedding methods take advantage of the above intrinsic connection between the radial coordinate and the node degree. For example, Hypermap\cite{HyperMap}, one of the first hyperbolic embedding methods, is based on likelihood maximisation with respect to a generalised version of the PSO model, where the optimisation shuffles only the angular coordinates with the radial coordinates being assigned according to the degree. Another well-known hyperbolic embedding approach is provided by the family of coalescent embeddings\cite{linkWeights_coalescentEmbedding}, where the angular coordinates are inferred using dimension reduction techniques on weighted matrices representing the distance relations between the nodes, however the radial coordinates are again distributed according to the PSO model, based on the degree. This choice for setting the radial coordinates was left unchanged  when the coalescent embedding approach was combined with local angular optimisation of the node positions\cite{our_opt_embedding}. The radial arrangement of the nodes is according to the PSO model also in the case of Laplacian Eigenmaps\cite{Alanis-Lobato_LE_embedding}, where the angular coordinates are obtained from the non-linear dimension reduction of Laplacian matrices. The RHG model can also be used for inferring the radial coordinates based on the node degree, as was shown in the case of the Mercator embedding method\cite{S1H2_Mercator,dmercator}. Nevertheless, the radial coordinates assigned based on the PSO model or based on the RHG model are very similar, since both depend logarithmically on the node degree. The only major difference between these two options is that all nodes obtain a unique radial coordinate according to the PSO model, whereas it is allowed for multiple nodes to have the same radial coordinate in the RHG model.

\subsection*{Detailed description of the CLOVE method}

Let us consider the task of embedding an arbitrary undirected  (and not necessarily connected) network  consisting of $N$ number of nodes and $E$ number of edges into the two-dimensional native disk representation of the hyperbolic space. We employ a hierarchical multi-level arrangement of the communities within the native disk by leveraging information about the connectedness of these communities and their respective sub-communities across different scales of the network. We denote the communities at a given hierarchy level $l$ by $t^{(l)}_m$, where the lower index $m$ is running from $0$ to the total number of communities at the given level. 

\begin{enumerate}
    \item{\textbf{Arranging the communities at the topmost $l=0$ level}}  
    \begin{itemize}
        \item[a)] \textit{Detecting communities:} We can identify the top-level communities $t_m^{(0)}$ by using any arbitrary non-overlapping community finding algorithm. Here, the Leiden method\cite{leiden} is adopted as the default approach for community detection, which is an advanced technique based on modularity maximisiation. Nevertheless, other built-in options, such as the Louvain method, are also available in the provided code. (A brief description of both the Leiden and the Louvain approaches, as well as the concept of modularity is provided in the Supplementary Information). If an entire hierarchical dendrogram of the communities is accessible, e.g., as it might be the case for the Louvain algorithm\cite{Louvain}, in this step we use the partition at the topmost level ($l=0$) of the dendrogram.
        \item[b)]\textit{Defining a weighted network between the communities:} We construct the proximity graph of the communities, i.e., build up a complete weighted super-graph, whose nodes correspond to the communities $t_m^{(0)}$ found earlier in step 1a).  The edge weight between any pair of super-nodes $i$ and $j$ is defined as 
        \begin{equation}
            W_{ij}=f\left(\frac{2E_lC_{ij}}{K_iK_j}\right)+1,
        \label{eq:preweighting_method_description}
        \end{equation}
        where $E_l=E_0$ is the number of edges, $K_i$ and $K_j$ denotes the number of intra-community links within the communities $t_i^{(0)}$ and $t_j^{(0)}$, respectively, and $C_{ij}$ stands for the number of inter-connections between $t_i^{(0)}$ and $t_j^{(0)}$. Note that although the function $f$ defined in Eq. (\ref{eq:preweighting_method_description}) can be any arbitrary decreasing function of its argument, taking values on the unit interval, we use an exponentially decaying form $f(x)=e^{-x}$ by default. In the Supplementary Information, we demonstrate that adopting this choice for the weights between modules guarantees compliance with the triangle inequality, thereby justifying the utilization of TSP in the later steps.  
        \item[c)]\textit{Approximate solution for the TSP:} We look for the minimal-weight Hamiltonian cycle of the super-nodes (communities) in the proximity graph defined in 1b). This corresponds to solving the TSP on the proximity graph, and the obtained solution represents the inferred angular order of communities. We use the Christofides method supplemented with a threshold accepting boost~\cite{tsp_solution_ta}  for solving the TSP by default, however, further possible choices are also available in the provided code, including e.g., the greedy method, simulated annealing, or the threshold accepting method solely. Note that the latter two metaheuristic algorithms can also be applied in combination with the greedy or Christofides method, providing therefore, an option of boosting that may enhance the quality of the final embedding in particular cases. (We give a brief summary of the implemented TSP solvers in the Supplementary Information). 
        \item[d)]\textit{Circular arrangement of the communities:}
        We arrange the communities on the native disk such that subsequent communities become adjacent on the disk. Each community is allocated a circular sector, the size of which is proportional to the number of nodes within that community. More precisely, the community $t_i^{(0)}$ in the minimal-weight order is assigned to the angular interval
        \begin{equation}
            \left[\Phi_{i,{\rm start}}^{(0)},\Phi_{i,{\rm end}}^{(0)} \right) =\left[\frac{2\pi}{N}\sum\limits^{i-1}_{j=1}n_j^{(0)},\ \frac{2\pi}{N}\sum\limits^{i}_{j=1}n_j^{(0)}\right) \label{eq:angular_range}
        \end{equation}
        where $n_m^{(0)}$ denotes the number of nodes in community $t_m^{(0)}$.
            
    \end{itemize}
   \item{\textbf{Arranging the communities at level $l+1>0$}}\\ For convenience, the current level is considered to be level $l+1$, whereas the previous level (immediately above the hierarchy) is assumed to be level $l$.
   \begin{itemize}
       \item[a)] {\it Detecting sub-communities:} For each community at the previous level, $l$, we run the same community finding algorithm as in 1a) on the sub-graph spanning between the community members (detached from the rest of the network). Let us focus on the sub-modules found this way within community $t_{i}^{(l)}$ from the previous level, and let us denote these sub-modules as $t_{i1}^{(l+1)},t_{i2}^{(l+1)},\dots t_{ik}^{(l+1)}$ for convenience.
       \item[b)] {\it Defining a weighted network between the sub-communities:} For each group of sub-modules found within a specific larger community from the previous level, we define a separate weighted network, similarly to step 1b). However, an important difference is that this time we also include two extra nodes in this complete graph, corresponding to the neighbouring communities from the previous level. These serve as ''anchors'' for a more optimal arrangement of the sub-modules. Specifically, for the sub-modules $t_{i1}^{(l+1)},t_{i2}^{(l+1)},\dots t_{ik}^{(l+1)}$ listed in 2a), we include the left and right neighbouring communities of $t_{i}^{(l)}$ according to the angular arrangement of the communities in level $l$. The link weights are defined again by using (\ref{eq:preweighting_method_description}).
       \item[c)]\textit{Approximate solution for the TSP:} For each separate weighted complete graph defined in 2b) we solve the TSP using the same heuristic as in 1c), receiving a Hamiltonian cycle over the sub-modules and the two extra neighbouring communities from the previous level. 
       \item[d)]\textit{Arrangement of the sub-communities:} Naturally, the sub-modules located within  $t_{i}^{(l)}$ must be placed inside the angular range $ \left[\Phi_{i,{\rm start}}^{(l)},\Phi_{i,{\rm end}}^{(l)} \right)$ associated with $t_{i}^{(l)}$. Any sub-module $t_{ik}^{(l+1)}$ receives a circular sector having a central angle of $\frac{2\pi}{N} n_{ik}^{(l+1)}$ (with $n_{ik}^{(l+1)}$ denoting the number of nodes in $t_{ik}^{(l+1)}$), and the order of the sub-modules is determined by the Hamiltonian cycle received in 2c). Under optimal circumstances, the ''anchoring'' super-nodes (communities from level $l$) are neighbours in the Hamiltonian cycle, and we can apply a cyclic permutation bringing the ''anchor'' placed aside $t_{i}^{(l)}$ at $\Phi_{i,{\rm start}}^{(l)}$ to the beginning of the cycle and the ''anchor'' placed aside $t_{i}^{(l)}$ at $\Phi_{i,{\rm end}}^{(l)}$ to the end of the cycle. Based on the cycle obtained, now aligned with the ''anchor'' positions, the angular range of $t_{ik}^{(l+1)}$ can be given as
        \begin{equation}
            \left[\Phi_{ik,{\rm start}}^{(l+1)},\Phi_{ik,{\rm end}}^{(l+1)} \right) =\left[\Phi_{i,{\rm start}}^{(l)}+\frac{2\pi}{N}\sum\limits^{k-1}_{j=1}n_{ij}^{(l+1)},\ \Phi_{i,{\rm start}}^{(l)}+\frac{2\pi}{N}\sum\limits^{k}_{j=1}n_{ij}^{(l+1)}\right) \label{eq:sub_angular_range}
        \end{equation}
        In case the ''anchors'' are not neighbours in the Hamiltonian cycle received in 2c), we look for a cyclic permutation where the left "anchor" is in the correct position, specifically, at the beginning of the Hamiltonian cycle. Starting from this left anchor node and moving to the right, we maintain the order obtained in step 2c) until we reach the right "anchor." Ideally, this right "anchor" should be the rightmost element in the order. To meet this condition, we initiate the reversal of the remaining part of the cycle from this right ''anchor'' node, and concatenate it with the preceding unchanged segment. With this concatenation in the ''middle'', we can guarantee that the longest directionally correct sub-sequences of Hamiltonian cycle are preserved. 
   \end{itemize}
\item  \textbf{Iteration and stopping criterion for the angular arrangement of the communities}\\
After the completion of the angular arrangement of the communities at any level $l$, we proceed to the next level as described in 2. However, if for any communities $t_{i}^{(l)}$ the community finding algorithm returns no sub-modules in 2a), meaning that $t_{i}^{(l)}$ is already so small and compact that it is not worth dividing into sub-communities, we do not carry out steps 2b-d, and leave $t_{i}^{(l)}$ as it is. Although $t_{i}^{(l)}$ can still act as an anchor for the sub-modules of neighbouring communities, the angular arrangement procedure is locally stopped at $t_{i}^{(l)}$. Naturally, for other communities at the same level the algorithm will carry on and may discover contained sub-modules, where we position these according to steps 2. 

When the recursive discovery of contained sub-communities is stopped locally everywhere, we have reached the stage where it is not worth dividing further any of the modules at the lowest level in any branch of the community hierarchy. (Naturally, the maximal depth of the branches can vary.) In order to fully specify the angular coordinates of the individual nodes, we can now move on to the next phase in the algorithm, described in step 4.

\item \textbf{Angular arrangement of individual nodes within communities}\\
There are several options for arranging the members of a given community (assumed to be on the possible lowest level in the corresponding branch of the community hierarchy). In all cases the node positions are distributed in a uniform regular fashion inside the considered sub-module, where the angular distance between neighbouring nodes is always $\frac{2\pi}{N}$.
\begin{itemize}
    \item[a)] Probably the most natural choice is to apply the same principles as in the case of the sub-modules, outlined in step 2. Here we basically replace the sub-modules  $t_{i1}^{(l+1)},t_{i2}^{(l+1)},\dots t_{ik}^{(l+1)}$ by the individual community members, but otherwise carry out exactly the same steps from 2b to 2d. Although this is likely to provide the best quality local arrangement among the other options, it is also computationally the most demanding. 
    \item[b)] Another very simple choice is to distribute the members randomly among the available angular positions. This is the fastest option, albeit also with the lowest quality.
    \item[c)] A further heuristic solution we propose is based on the node degrees. If the number of community members is odd, the member with the largest degree will occupy the central position and the node with second largest degree will be its left or right neighbour (chosen at random). If the number of members is even, the first two nodes according to the degrees will occupy the two central positions (again, in random order). The further nodes are added in the order of their degree, always occupying a position next to the already occupied positions either from the left or from the right. We decide about inserting to the left or to the right based on the number of connections between the given node and the already inserted nodes on the right or on the left. (In the case we observe an equal number of connections to the right and to the left, we choose randomly). 
    This method yields usually better quality arrangements compared to random positions and it is faster compared to option a).
\end{itemize}
By default we use option c), however, the code we provide allows both a) and b) as well.

\item \textbf{Radial arrangement of the nodes}\\
The radial coordinates are defined solely based on the node degree, independently from the angular coordinates. For simplicity, we use the radial coordinates predicted based on the PSO model and apply Eqs.(\ref{eq:opt_radial}-\ref{eq:opt_radial}) for assigning $r_i$, where the node indices are distributed according to the order dictated by the node degrees, as explained in Section Networks in the native disk representation of the hyperbolic space. The parameter $\beta$ necessary for calculating the coordinates is obtained by fitting the tail of the degree distribution of the embedded network with a power-law decaying function and applying the well-known relation $\beta =\frac{1}{\gamma-1}$ between the degree decay exponent $\gamma$ and the popularity fading parameter.

\end{enumerate}

\subsubsection*{Additional parameters of the CLOVE method}

\begin{itemize}
    \item \textbf{Number of ''anchor'' nodes} 
    \newline
    Originally, CLOVE uses $z=2$ number of ''anchor'' nodes in steps 2b)-d) by default. However, the implementation we provide allows to handle neighbors of higher orders as well. In such cases, for each sub-community, we include $z=2l, \ l\in\mathbb{N}^{+}, \ l>1$ number of neighbouring communities from the preceding level, hence exploiting a more global information about the connectedness of the communities in the arrangement step.
    \newline
    \item \textbf{Decomposition of isolated nodes and components}
    \begin{itemize}
        \item \textbf{Embedding networks with multiple components}
        \newline
        Despite the difficulty that most embedding algorithms have in dealing with networks comprising multiple connected components, the CLOVE algorithm can handle this type of networks in a natural manner. If the network we need to embed is not fully interconnected, the default approach for CLOVE is to start by optimizing the position of the different components on the hyperbolic disk instead of the top-level communities. Subsequently, the algorithm proceed conventionally by detecting sub-communities inside these distinct components using a predefined community finding method. Notably, the default application of the Leiden algorithm ensures the preserved 
        connectivity of these identified sub-communities~\cite{leiden}. This embedding option of the algorithm is referred to as the decomposition of connected components, which is governed by a Boolean variable in the provided code. Conversely, if the decomposition of connected components is disabled, the algorithm can still effectively manage multiple components. In such cases, instead of seeking the optimal arrangement of the separate components at the highest level, the algorithm directly arranges the communities themselves consistently across all scales.
        \item \textbf{Decomposition of nodes with degree $k=0$} 
        \newline
        If the network contains isolated nodes, CLOVE can embed these isolated nodes separately by detaching them from the rest of the network. When this feature is enabled, random angular coordinates are allocated to the isolated nodes, while the remaining portion of the network is embedded using the standard procedure outlined in steps 1-5 above. The assignement of the radial coordinates are not affected. It is important to note that this option is primarily designed to improve runtime efficiency; nevertheless, it may also result in enhanced accuracy in particular cases. We refer to this feature of the algorithm as the "decomposition of $k0$ nodes" controlled by a boolean variable in the provided code. 
        \item \textbf{Decomposition of nodes with degree $k=1$}
        \newline
        Similarly to the decomposition of isolated nodes, CLOVE allows the decomposition of nodes with degree $k=1$ as well. Upon enabling this feature, controlled again by a boolean variable, the algorithm starts by detaching nodes with only one degree from the rest of the network. First, the remaining part of the network is embedded, then detached nodes with only one degree receive the same angular coordinates as their single neighbor. In case two nodes are only connected to each other, hence having been detached during the decomposition procedure, they both receive the same uniformly sampled random angular coordinate. The assignement of the radial coordinates are not affected here either. 
    \end{itemize}
    \item \textbf{The sizes of angular sectors corresponding to the communities}
    \newline
    During the arrangement of the communities in steps 1d) and 2d), the CLOVE method allocates an angular sector to each community with the central angle being proportional to the number of nodes it contains, as demonstrated by Eq.(\ref{eq:angular_range}) and Eq.(\ref{eq:sub_angular_range}). However, in the code we provide, there is also an option to allocate circular sectors to each community in such a way that their central angle is proportional to the sum of node degrees within the community. This method, also utilized in Ref.~\cite{commSector_hypEmbBasedOnComms_2019}, enhances the flexibility of the algorithm.
\end{itemize}

\subsection*{Embedding quality metrics}

Broadly speaking, embedding quality metrics are scalar values ranging from 0 to 1, used to quantify how well an embedding of a given network fits into the two-dimensional hyperbolic space. In order to reasonably assess the quality of our resulting embeddings and to make meaningful comparisons with other state-of-the-art methods, we systematically tracked various such metrics for each and every studied embedding algorithm. The results of these assessments are presented in the main text of the manuscript, while the subsequent section provides a comprehensive list and detailed explanations for each metric score employed in our analysis.

\subsubsection*{Mapping accuracy}
Mapping accuracy (MA) assesses the relationship between geodesic distances and topological shortest paths in an embedded network by determining the Spearman's rank correlation between the two:

\begin{equation}
\text{MA} = \frac{cov[R(GD), R(TP)]}{\theta_{R(GD)} \theta_{R(TD)}}.
\end{equation}
Here, $GD$ and $TD$ represent the lists of geodesic distances and topological distances for vertex pairs, respectively, while $R(GD)$ and $R(TD)$ denote the corresponding ranks of these lists.

\subsubsection*{Edge prediction AUROC}
The Edge Prediction AUROC is a measure that evaluates how well an embedding reflects the anticipated pattern of positioning connected vertices closer together than unconnected ones. It involves computing the area under the ROC curve, where predicted scores are determined by the inverse of vertex distances. Positive ground truth classes are represented by existing edges, and negative ground truth classes are represented by non-existing edges. The ROC curve depicts the true positive rate (TPR) against the false positive rate (FPR), showcasing the performance of a binary classifier across different acceptance thresholds. An AUROC score of $0.5$ is expected for a random predictor.

\subsubsection*{Edge predicition AUPRC}
Edge Prediction AUPRC is an alternative metric that assesses the same behavior as AUROC, but employs a different approach. In this case, the area under the Precision-Recall curve is computed for the same predictions and ground truth occurrences.

\subsubsection*{Greedy routing success rate}

The Greedy Routing Success Rate (GR) is an embedding metric that evaluates the efficiency of Greedy Routing paths in reaching their target vertex. This is determined by simply counting the number of successful greedy paths and dividing this sum by the total number of directed vertex pairs. More precisely, the GR score is defined as
\begin{equation}
    \text{GR} = \frac{1}{|V|(|V|-1) / 2 - 2 |E|} \sum_{\substack{\forall u \in V \\ \forall v \in \Bar{N}(u)}} \rho(u, v),
    \label{eq:GR}
\end{equation}
where $\rho(u, v)$ counts the number of successful greedy paths between vertices $u,v$,
$\Bar{N}(u)$ is the complement of the neighbourhood of the vertex $u$, i.e. the set of vertices $v \in V$ that are not adjacent to $u$ (including $u$ itself). 
By excluding adjacent vertex pairs in Eq.(\ref{eq:GR}), we eliminate a constant offset from the definition of the GR metric. This is due to the fact that such pairs consistently represent successful greedy paths in embeddings that forbid the assignment of identical coordinates to more than one vertex.

\subsubsection*{Greedy routing score}

The Greedy Routing Score (GS) is an extension of the previously discussed Greedy Routing Success Rate, introducing weights to provide a more comprehensive metric for evaluating the embedding quality. In this refined approach, successful paths are assigned weights determined by the ratio of the topological shortest path length between the source and target vertices to the number of visited vertices along the greedy path. This weighting scheme ensures that the contribution to the score is diminished for successful greedy paths that are significantly longer than the topological shortest path. Mathematically, the GS can be expressed as
\begin{equation}
    \text{GS} = \frac{1}{|V|(|V|-1) / 2 - 2 |E|} \sum_{\substack{\forall u \in V \\ \forall v \in \Bar{N}(u)}} \frac{TSPL(u, v)}{GPL(u, v)},
    \label{eq:grs_metric}
\end{equation}
where $TSPL(u, v)$ is the length of the topological shortest path between vertices $u$ and $v$, and $GPL(u, v)$ is the length of the greedy path starting from vertex $u$ and ending in $v$. If a path is unsuccessful, $GPL(u, v)$ is set to infinity, thus having zero contribution to the GS in Eq.(\ref{eq:grs_metric}).

\subsubsection*{Greedy routing efficiency}

The Greedy Routing Efficiency metric evaluates the relationship between geodesic distances and projected greedy paths, as given by the formula:
\begin{equation}
    \text{GE} = \frac{1}{|V|(|V|-1) / 2 - 2 |E|} \sum_{\substack{\forall u \in V \\ \forall v \in \Bar{N}(u)}} \frac{GD(u, v)}{PGPL(u, v)},
\end{equation}
where $GD(u, v)$ is the geodesic distance between vertices $u$ and $v$, and $PGPL(u, v)$ is the projected greedy path length between $u$ and $v$, i.e. the sum of the lengths traveled along the greedy path.

\subsubsection*{Angular separation index}

The common characteristics of the above metrics is that they depend only on the topology of the graph and the coordinates of the nodes. However, the quality of an embedding can also be quantified via the angular coherence of the communities capturing the extent to which nodes within the same community have similar angular coordinates in the embedding space. A possible quantity that measures this tendency is given by the angular separation index (ASI)~\cite{Cannistraci_ASI}, which contrarily to the previous metrics depends on the communities of the network as well. The key idea behind this metric is to compare the number of "mistakes" in the angular arrangement —i.e. the number $o_i$ of nodes from other communities mistakenly placed between the boundaries of the given module $i$—summed over all the $q$ communities of the network with the highest total number of mistakes obtained with the same clustering of the nodes when the angular coordinates are shuffled at random. More precisely, the ASI can be written as follows
\begin{equation}
\text{ASI} =1-\frac{\sum\limits^q_{i=1}o_i}{\max\limits_{r} \left ( \sum\limits^q_{i=1}o^{(r)}_i \right )}
\label{eq:asi}
\end{equation}
where the max function in the denominator of Eq.(\ref{eq:asi}) is taken over a fixed number of random shuffles. By default, we use 1000 samples, i.e. consider $r=1,2,\dots,1000$ in Eq.(\ref{eq:asi}) as suggested in Ref.\cite{Cannistraci_ASI}.

\bibliographystyle{sn-nature}
\bibliography{references} 

\newpage

\begin{center}
\LARGE{\bf SUPPLEMENTARY INFORMATION}
\end{center}

\renewcommand{\thefigure}{S\arabic{figure}}
\renewcommand{\thetable}{S\arabic{table}}
\renewcommand{\theequation}{S\arabic{equation}}
\renewcommand{\thesection}{S\arabic{section}}

\setcounter{section}{0}
\setcounter{figure}{0}
\setcounter{equation}{0}
\setcounter{table}{0}

\section{Finding communities in complex networks}

As pointed out in the main text of this article, community detection plays a pivotal role in the formulation of the CLOVE method. Despite the plethora of diverse methods available in the literature, each adopting distinct approaches~\cite{Newman_modularity_original, Clauset_comm_in_large_nets, Louvain, leiden, Infomap, label_prop}, our primary focus revolves around techniques based on modularity optimization. The subsequent sections provide a brief overview of the concept of modularity and discusses two prominent modularity optimization methods, namely the Louvain~\cite{Louvain} and Leiden~\cite{leiden}, both of which are incorporated as selectable options in the code we provide.

\subsection{Modularity}
Modularity, often denoted as $Q$, is a widely used measure for evaluating the strength of communities in complex networks~\cite{Newman_modularity_original}. Fundamentally, this measure entails comparing a given network partition to a random baseline by assessing the difference between the observed fraction of intra-community links and its expected value in the random null model. Typically, the null model is represented by the configuration model, and $Q$ in this case can simply be written as
\begin{equation}
Q=\sum\limits_{c=1}^{n}\frac{l_c}{E}-\left ( \frac{\sum\limits_{i \in c}k_{i}}{2E} \right )^2,
\label{eq:Q_def}
\end{equation}
where the summation runs over the communities, $l_{c}$ stands for the number of intra-community links in  module $c$, $k_i$ is the degree of node $i$, and $E$ denotes the total number of links in the network~\cite{Newman_modularity_original}. The quality measure above can take any values in the range of $Q\in \left[-\frac{1}{2},1\right]$, where greater values of $Q$ are considered as a convincing sign of a strong community structure, whereas lower modularity values usually imply the lack of modules in the studied network~\cite{Clauset_comm_in_large_nets}. 

\subsection{Louvain algorithm for community detection}

The Louvain method introduced in Ref.\cite{Louvain} is a heuristic community detection algorithm that aims to find a partition of a given network that maximizes the modularity function $Q$ in Eq.(\ref{eq:Q_def}). Although finding the global optimum of the modularity is an \textbf{NP}-hard problem, the Louvain method has become widely popular mostly due to its efficiency, both in terms of quality and running times. The method performs the optimisation task by using a greedy strategy in a hierarchical manner, which basically consists of two major phases~\cite{Louvain}; first at a given hierarchical level each node is moved to the community of its neighbors yielding the largest gain in $Q$ (local moving phase), and then based on this, a weighted super-graph is built up whose super-nodes correspond to the communities identified earlier (coarsening phase). Subsequently, the super-nodes are treated as nodes again, and the previous greedy strategy is re-exploited to obtain the clusters at a higher level. By using this procedure iteratively, smaller communities are merged into larger ones based on local modularity optimisation, and after multiple iterations, a complete hierarchical community structure unfolds. The starting point for the algorithm at the lower level of this hierarchy may be any legitimate partition of the network nodes, whereas the top level usually corresponds to the partition having the highest $Q$ value that the algorithm could achieve. The algorithm stops when moving nodes to other communities can not increase the modularity any further (more precisely, when the gain in the modularity falls below a predefined tolerance threshold).

\subsection{Leiden algorithm as an improvement of Louvain}
Despite its remarkable efficiency, the Louvain method has a peculiar imperfection; as demonstrated by Traag \emph{et al.}~\cite{leiden}, it may find arbitrarily badly connected communities or, in some cases even disconnected communities. The Leiden algorithm proposed in Ref.~\cite{leiden} circumvene these flaws by introducing an additional refinement phase in-between the local moving and coarsening phase, which significantly reduces the number of badly connected communities and guarantees that the obtained communities never become disconnected. The underlying concept of the refinement phase is that the communities identified right after local moving phase may break into several smaller communities. Consequently, the super-graph in the coarsening phase may encompass multiple super-nodes corresponding to the same community, rather than each community being represented by a single super-node~\cite{leiden}. Including the refinement phase between the local moving and coarsening phase provides greater opportunities for improving the modularity function $Q$ in Eq.(\ref{eq:Q_def}).

\section{The pre-weighting scheme of CLOVE}
In this section we demonstrate that the pre-weighting scheme that defines the proximity of the (sub-)communities obeys the triangle inequality. To prove this, we begin by assigning weights to the super-edges between the communities (super-nodes) of the network in the following manner: 
\begin{equation}
        W_{IJ}=\exp\left(-\frac{2EC_{IJ}}{K_IK_J}\right),
        \label{eq:naive_weigths}
\end{equation}
where $E$ is the total number of edges in the network, and $C_{IJ}$ denotes the count of inter-community edges between communities $I$ and $J$, each community having $K_I$ and $K_J$ intra-community edges, respectively. Even though the weights defined by Eq.(\ref{eq:naive_weigths}) have the advantageous property of ranging between $0$ and $1$, they do not inherently satisfy the triangle inequality. However, one can address this concern with a minor adjustment. Specifically, let us redefine the weights in Eq.(\ref{eq:naive_weigths}) as $w_{IJ}=c+W_{IJ}$, where $c$ is a constant satisfying $c\geq \max\limits_{IJ}W_{IJ}$. With this slight modification of the weights, the triangle inequality now holds, which can be verified through the following algebraic manipulations
\begin{align}
    w_{IK} + w_{KJ} - w_{IJ} & = (c + W_{IK}) + (c 
    + W_{KJ}) - (c + W_{IJ})
    \label{eq:TI_1}
    \\
    &= c + W_{IK} + W_{KJ} - W_{IJ} 
    \\
    &\geq c- W_{IJ}
    \\
    &\geq \max_{IJ}W_{IJ} - W_{IJ} \geq 0, 
    \label{eq:TI_4}
\end{align}
where we exploited that $c\geq \max\limits_{IJ}W_{IJ}$, as well as the fact that the weights defined by Eq.(\ref{eq:naive_weigths}) are strictly non-negative. Ensuring that the triangle inequality holds is essential in this setup, since it allows the edge-weights to be interpreted as a measure of proximity or distance between the corresponding communities. Finally, this interpretation opens up the possibility for leveraging approximating algorithms explicitly tailored for the metric version of TSP, which then find the optimal configuration of the communities.

\section{Christofides algorithm for solving the Traveling Salesman Problem}
The TSP is considered \textbf{NP}-hard, meaning that no known algorithm can find the optimal solution in a polynomial time~\cite{ancient_tsp_karp}. However, there exist many heuristic and approximation algorithms, such as the greedy method (also known as nearest neighbour algorithm)~\cite{reinelt1994traveling}, ant colony optimization~\cite{ant_colony_tsp}, or the threshold accepting method~\cite{tsp_solution_ta}, which can find near-optimal solutions in a reasonable amount of time. Besides the previous examples, another notable method is given by the Christofides algorithm, sometimes referred to as Christofides-Serdyukov algorithm~\cite{tsp_solution_christofides, tsp_solution_serdyukov}, which has gained widespread popularity, mainly owing to its simplicity, guaranteed performance and efficiency. A detailed description of this method consists of the following steps:

\begin{enumerate}
    \item Let $G$ be a complete weighted graph of size $N$, in which every node $i=1,..,N$ corresponds to a city and each edge connecting the nodes is assigned a weight that corresponds to the distance between the cities they represent. Furthermore, let us assume the we study the so-called metric TSP, a version of TSP in which the distances satisfy the triangle-inequality, i.e.  
    \begin{equation}
       d_{ij}\leq d_{ik}+d_{kj} \ \ \ \ \forall i,j,k=1,...,N. 
    \end{equation}
    \item Construct the minimum spanning tree of $G$ and denote it as $G_{T}$.
    \item Create a subgraph of $G_{T}$ induced by the set of odd-degree nodes and denote it as $G_{O}$. 
    \item Find a minimum-weight perfect matching $G_{M}$ in $G_{O}$.
    \item Combine the edges of the minimum spanning tree $G_T$ with that of the minimum-weight perfect matching $G_M$ in order to obtain an Eulerian graph $G_E$.
    \item Find an Eulerian circuit in $G_E$.
    \item Convert the Eulerian circuit found in  step 6 into a Hamiltonian circuit by simply neglecting repeated vertices.
\end{enumerate}
Given $N$ number of nodes in $G$, the time complexity of the Christofides algorithm is $\mathcal{O}(N^3)$, which is significantly better than exact algorithms characterized by exponential worst-case time complexity~\cite{ancient_tsp_karp}. Moreover, it is proven to have a worst-case approximation ratio of $3/2$, meaning that the solution found by the algorithm is guaranteed to be at most $3/2$ times the value of the global optimum~\cite{tsp_solution_christofides}. Owing to the previous advantageous properties, Christofides algorithm has been successfully used in many real-world applications~\cite{tsp_aerospace,tsp_crystal,TSP_in_astronomy,WJCook_book_on_TSP,carlo_tsp} and often regarded as a good compromise between runtime and solution quality. Although the CLOVE method involves solving multiple TSPs at progressively smaller scales, with their approximate solutions characterized by superlinear time complexity, the method remains scalable. This scalability is achieved because the TSP is never applied to all nodes simultaneously; instead, it is only applied to supergraphs consisting of modules that co-occur at a specific level within a particular branch of the module hierarchy.

\section{Evaluation of the embedding quality metrics}

As discussed in the main text, embedding quality metrics (explained in the Methods section) are used to assess how well a network's embedding fits into the two-dimensional hyperbolic space.  Evaluating these metric scores is typically computationally very intensive, often scaling super-linearly with the system size. While this evaluation is still manageable for embeddings of small networks ($N_{\text{small}} \sim 10^3 - 10^4$), it becomes practically infeasible in the large network regime ($N_{\text{large}} \sim 10^5 - 10^6$). Consequently, for the latter case, we adopt an efficient sampling strategy to obtain approximate values of the aforementioned quality scores. The detailed methodology of this sampling procedure is outlined below for both greedy-routing and non-greedy-routing based metrics.

\subsection{Sampling strategy for the evaluation of greedy-routing based metrics}

The greedy-routing-based metric scores consist of the greedy routing score (GR), the greedy success rate (GS), and the greedy routing efficiency (GE), as defined by Eqs.(8)-(10) in the main text. To approximate these scores in large network embeddings, we only consider greedy paths that originate from any arbitrary source vertex, but terminate at one of a randomly chosen fraction $f$ of all possible target vertices. This sampling procedure reduces the computational complexity to $\mathcal{O}(fN^2\log N)$, and carefully selecting the value of $f$ provides reasonable approximations while keeping the evaluation times relatively low. 

\subsection{Sampling strategy for the evaluation of non-greedy-routing based metrics}

Non-greedy routing-based metric scores include mapping accuracy (MA), edge prediction precision (EPP), and the area under the receiver operating characteristic curve (AUC). These scores are similar to those based on greedy-routing, but focus on potential edges between vertex pairs instead of the paths connecting them (see the Methods section for more details). Apart from this slight difference, the sampling procedure described above can be easily adapted for non-greedy routing-based scores as well. Specifically, at sampling parameter $f$, only a subset of vertex pairs is considered: the first member of the pair can be any node in the network, while the second member is chosen from a randomly selected, but fixed subset of nodes comprising $f$ fraction of the network's nodes. Metric scores are then evaluated accordingly.

\subsection{The choice for the value of $f$}

Opting for a smaller value of $f$ can significantly reduce the time required for evaluating metric scores. Nevertheless, setting 
the $f$ parameter too low may result in considerable inaccuracies. To find the right balance between the accuracy of evaluations and computational efficiency, we adjust the $f$ parameter based on the size $N$ of the embedded network.

In scenarios involving small networks ($N < 2 \cdot 10^{4}$), where computations remain manageable, metric scores can be evaluated exactly by setting $f = 1$. However, this approach quickly becomes infeasible as network size increases. Consequently, for larger networks exceeding the threshold of $N = 2 \cdot 10^{4}$, the value of $f$ is chosen such that the number of sampled vertex pairs during the evaluation of the quality scores does not decrease with the system size.
When considering only a subset of all possible vertex pairs with the second members taken from a subset comprising an $f$ fraction of the total nodes, the number of samples grows roughly as $fN^2$. This matches the previously mentioned criterion if $fN^2 = (2 \cdot 10^{4})^2 = 4 \cdot 10^8$, therefore, the size-dependent sampling parameter $f$ for a network embedding of size $N$ is generally defined as follows:
\begin{equation}
    f(N)=\left\{\begin{matrix}
1, \ \ \ \text{if}\ \ \ N\leq  2 \cdot 10^{4}\\ 
\frac{4\cdot 10^8}{N^2} ,\ \ \ \text{if} \ \ \ N> 2 \cdot 10^{4}. \\
\end{matrix}\right. 
\end{equation}

\begin{figure}
    \centering
    \includegraphics[width=\linewidth]{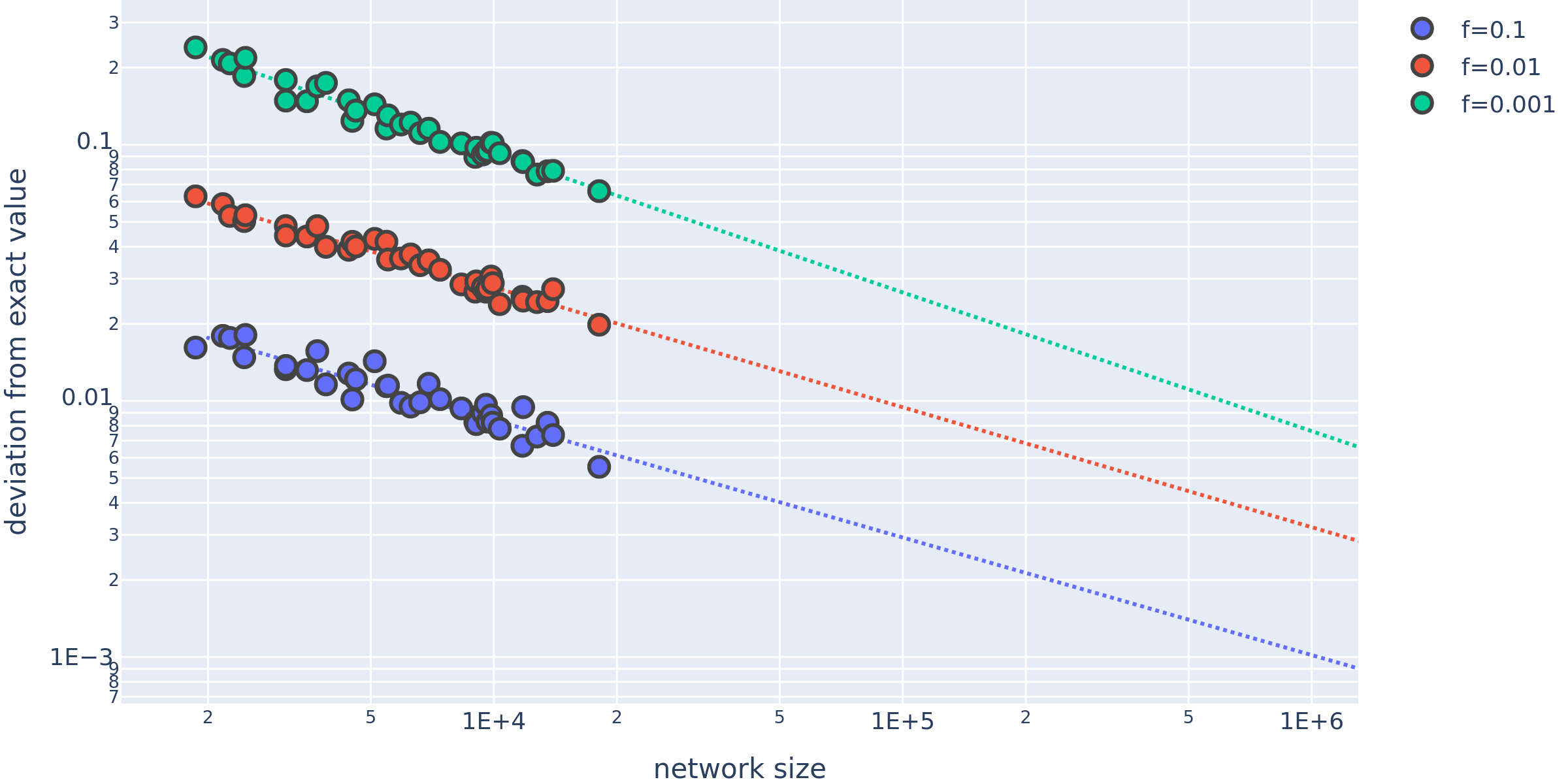}
    \caption{Precision of the embedding quality metric \textit{Greedy Routing Success Rate} at various values of sampling ratio \textit{f} (indicated by different colours) with fitted curves on exp-exp scales.}
    \label{fig:metric_precision_greety_routing_success_rate}
\end{figure}

\begin{figure}
    \centering
    \includegraphics[width=\linewidth]{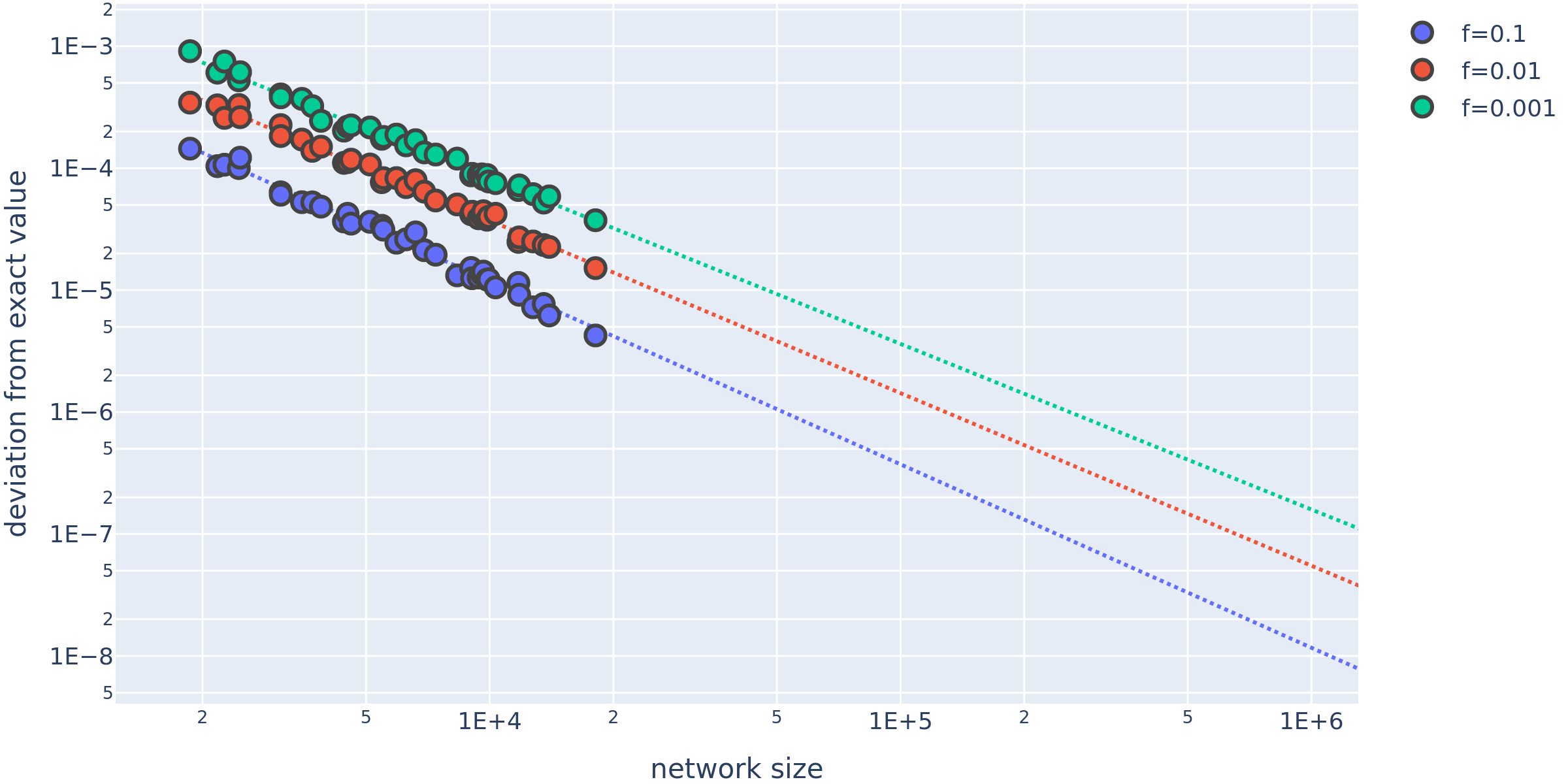}
    \caption{Precision of the embedding quality metric \textit{Edge Prediction ROC AUC} at various values of sampling ratio \textit{f} (indicated by different colours) with fitted curves on exp-exp scales.}
    \label{fig:metric_precision_edge_prediction_roc_auc}
\end{figure}

In order to validate the usage of smaller values of $f$ in large networks, or equivalently, to demonstrate that the accuracy of the above sampling procedure at a given $f<1$ increases as network size tends to infinity, we conducted a series of numerical validation experiments for each metric score individually. The results of these experiments are shown in Fig.\ref{fig:metric_precision_greety_routing_success_rate}.-\ref{fig:metric_precision_edge_prediction_roc_auc}. for GR, GS, GE, MA, EPP, and AUC, respectively. In these figures, for each metric score, we display the deviations of the corresponding score evaluated at different $f\leq1$ values from the exact metric score (evaluated at $f=1$) for PSO networks with parameters $m=4, \beta =2/3, T = 0.1$ as a function of network size. A closer inspection of Fig.\ref{fig:metric_precision_greety_routing_success_rate}.-\ref{fig:metric_precision_edge_prediction_roc_auc}. reveals that the deviation from the exact value appears to decay as a power-law with the network size $N$, becoming very low -- mostly below 0.05 -- by $N=2\cdot 10^4$ for each studied metric score. This observation strongly supports the use of smaller $f$ values for very large network embeddings.

In the upcoming section, we showcase the quality scores obtained using the above sampling strategy for a set of real-world networks embedded by different variants of the CLOVE and other state-of-the-art embedding methods. The applied quality metrics and embedding techniques are further detailed in the Results section of the main article.

\begin{figure}
    \centering
    \includegraphics[width=\linewidth]{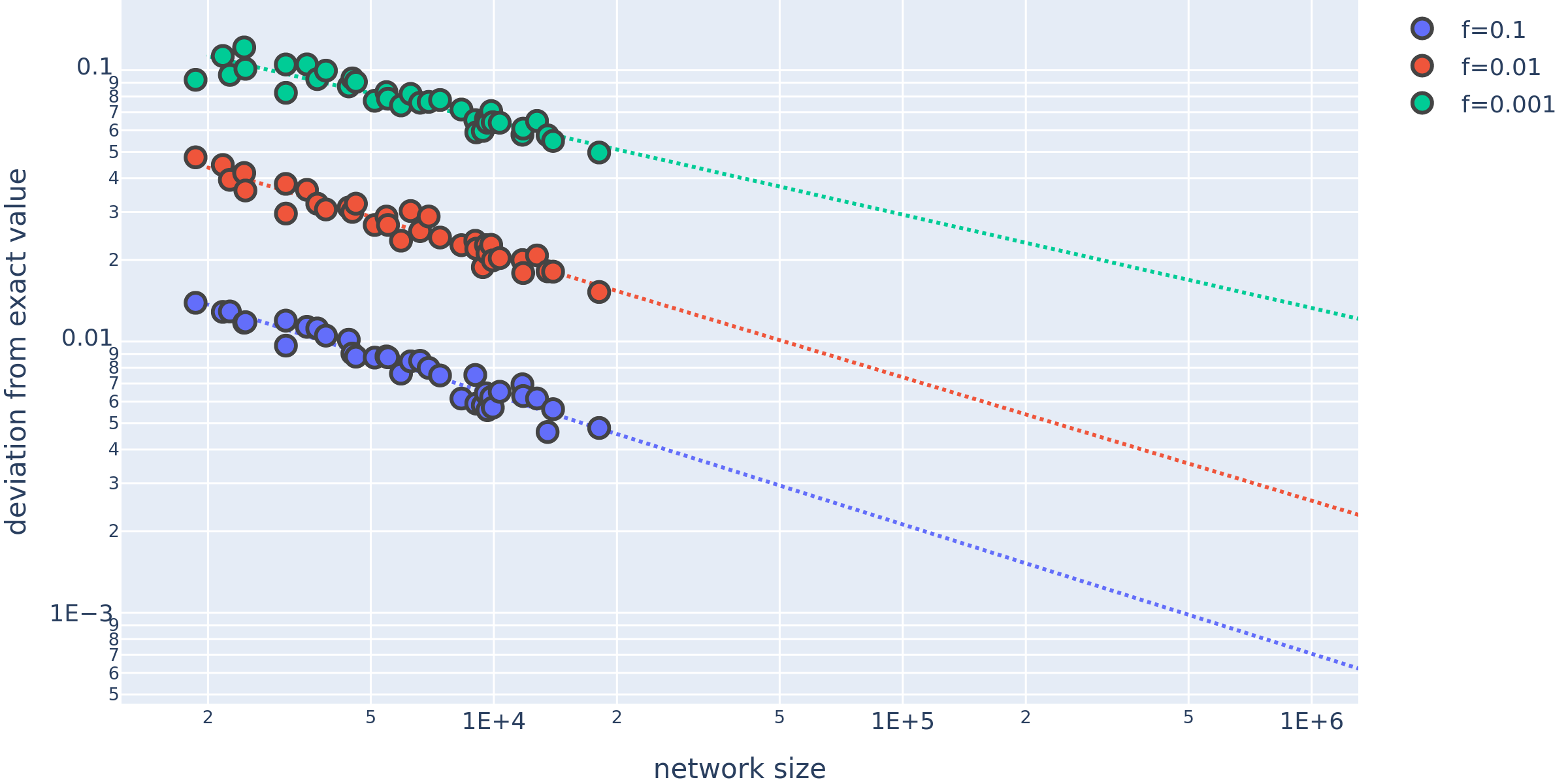}
    \caption{Precision of the embedding quality metric \textit{Mapping Accuracy} (MA) at various values of sampling ratio \textit{f} (indicated by different colours) with fitted curves on exp-exp scales.}
    \label{fig:metric_precision_mapping_accuracy}
\end{figure}

\begin{figure}
    \centering
    \includegraphics[width=\linewidth]{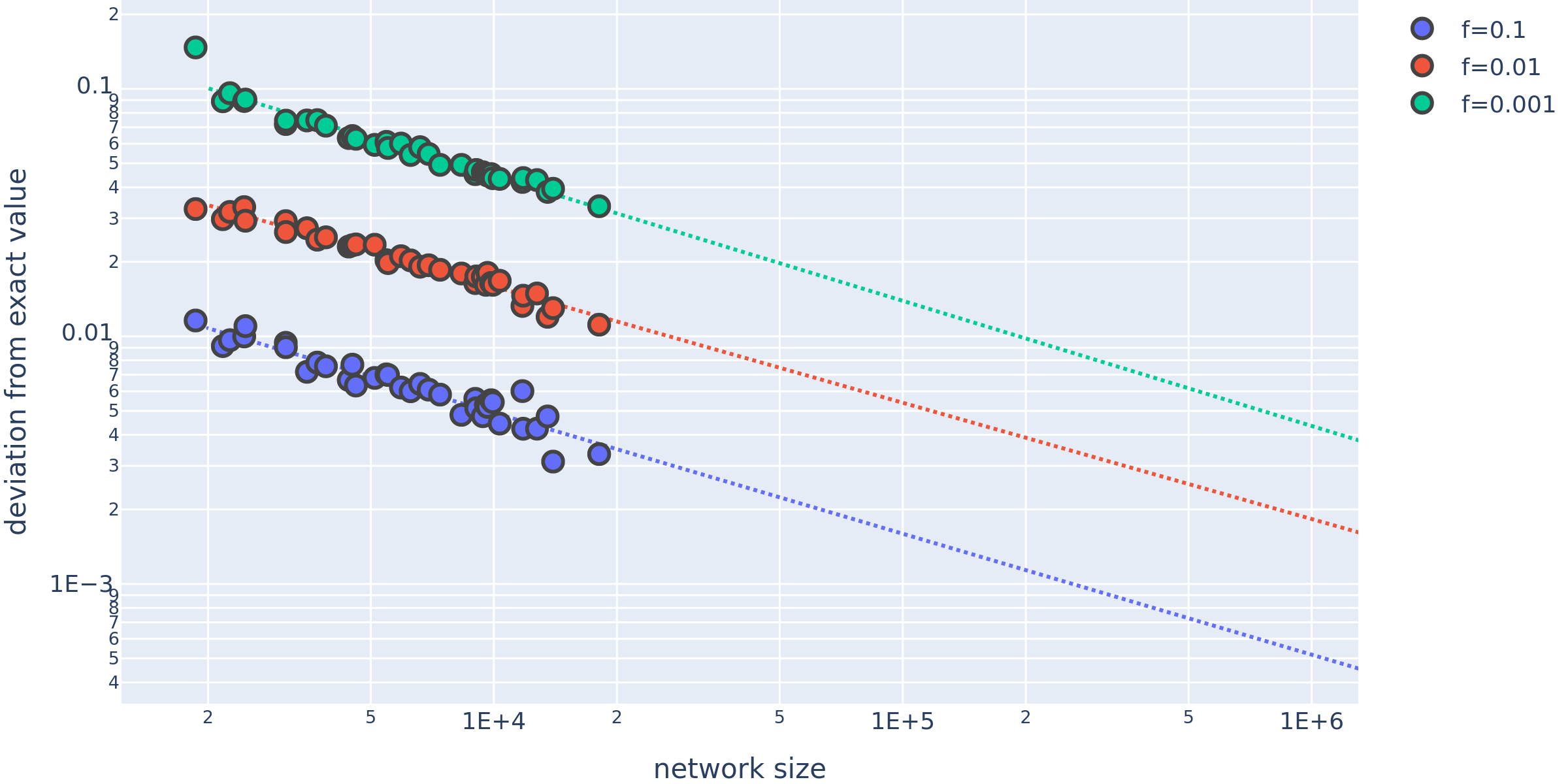}
    \caption{Precision of the embedding quality metric \textit{Edge Prediction Precision}  (EPP) at various values of sampling ratio \textit{f} (indicated by different colours) with fitted curves on exp-exp scales.}
    \label{fig:metric_precision_edge_prediction_precision}
\end{figure}

\begin{figure}
    \centering
    \includegraphics[width=\linewidth]{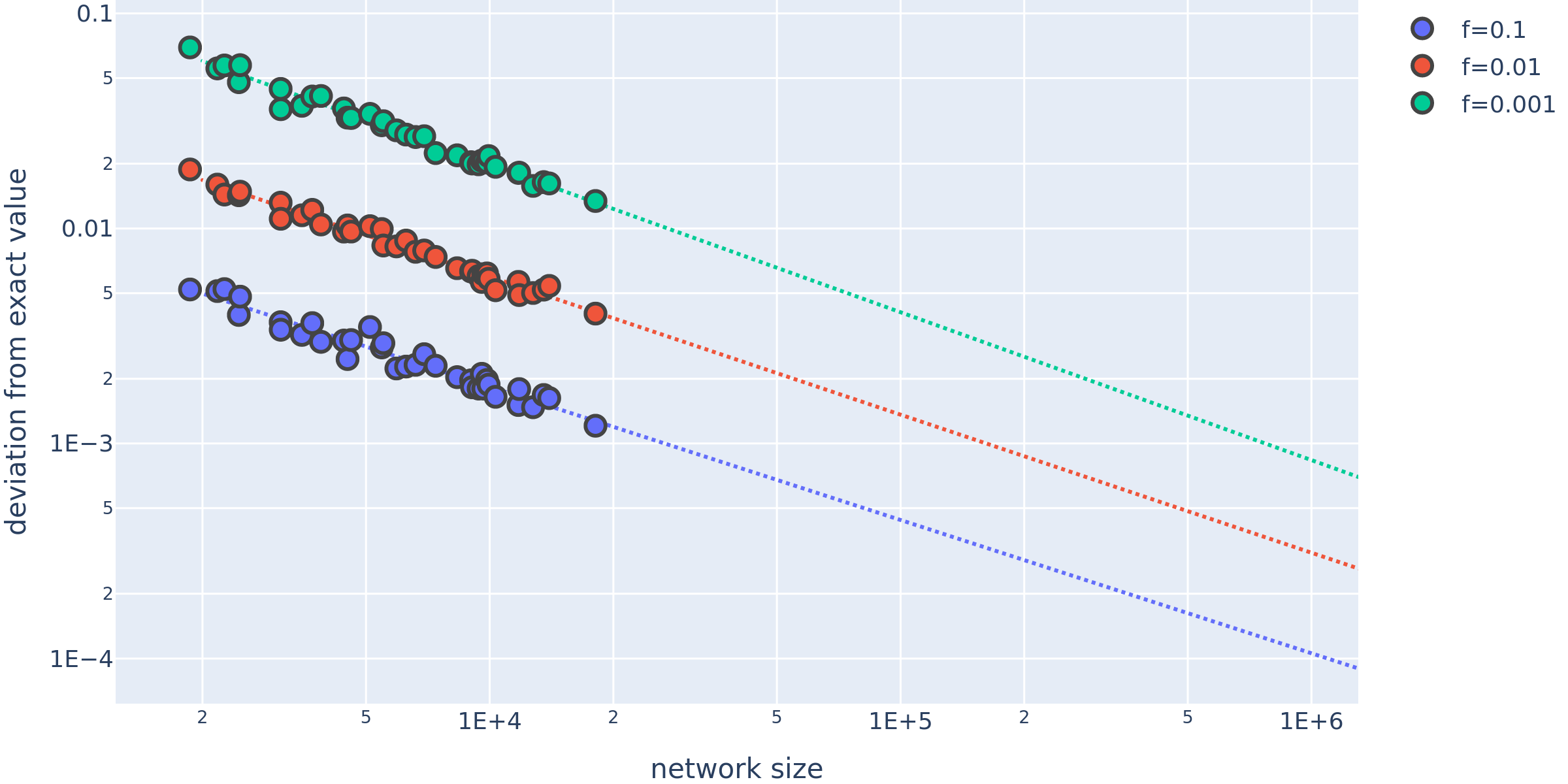}
    \caption{Precision of the embedding quality metric \textit{Greedy Routing Efficiency} (GRE) at various values of sampling ratio \textit{f} (indicated by different colours) with fitted curves on exp-exp scales.}
    \label{fig:metric_precision_greety_routing_efficiency}
\end{figure}

\begin{figure}
    \centering
    \includegraphics[width=\linewidth]{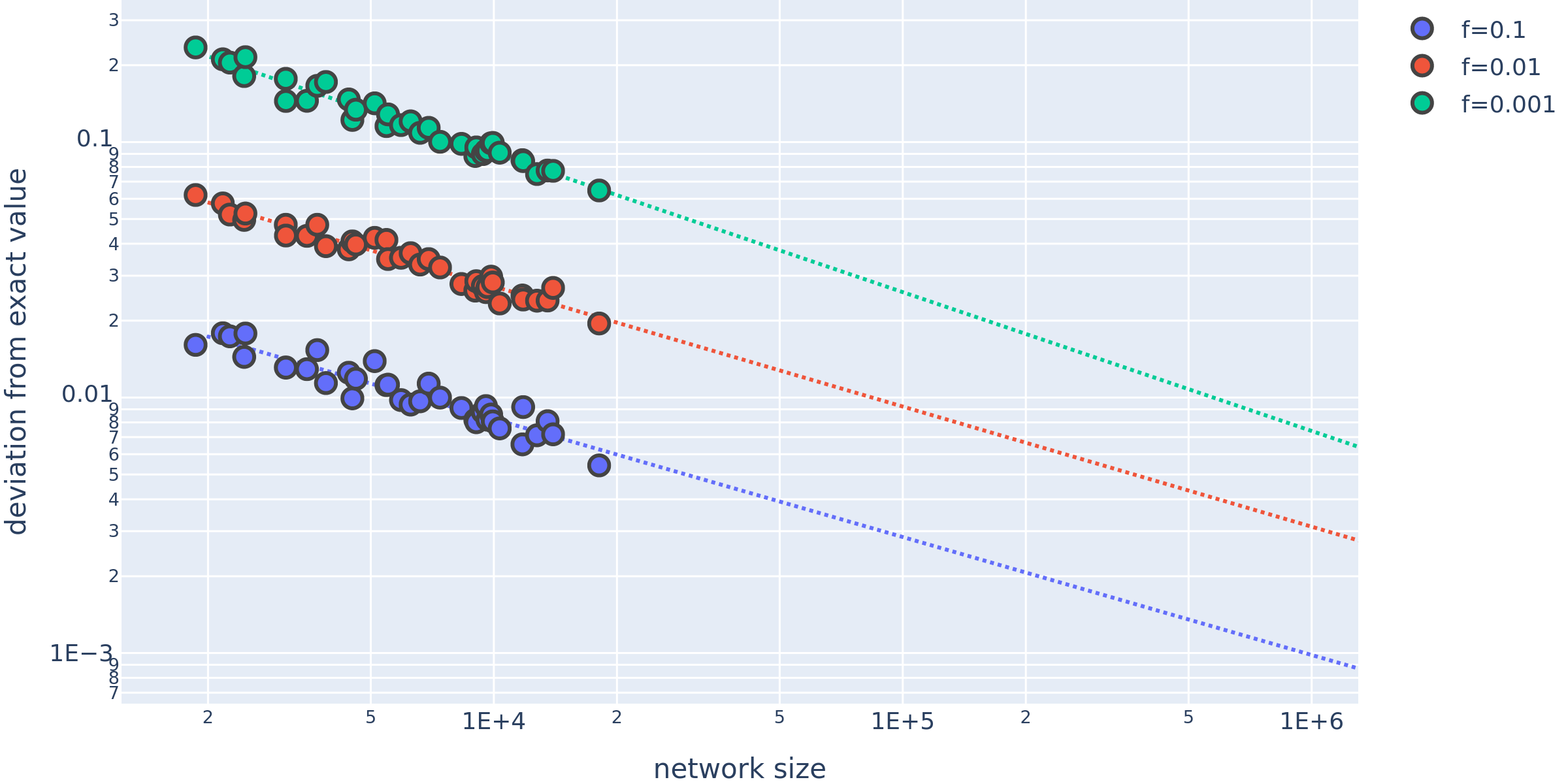}
    \caption{Precision of the embedding quality metric \textit{Greedy Routing Score} (GR) at various values of sampling ratio \textit{f} (indicated by different colours) with fitted curves on exp-exp scales.}
    \label{fig:metric_precision_greety_routing_score}
\end{figure}

\section{Embedding results for individual real-world networks}

In this section, we systematically compare the performance of CLOVE with other well-known hyperbolic embedding methods by evaluating various scores for a set of real-world networks. Table \ref{table:network_list}. provides a list of the networks studied, along with the links to their data sources. According to the 3$^{\rm rd}$ and 4$^{th}$ columns, the size of the considered systems is ranging between $N=1,133$ and $N=1,353,703$ in terms of the number of nodes and between $L=5,451$ and $L=13,126,172$ in terms of the number of links.
\begin{table}
\begin{tabular}{ |c|c|c|c|}
    \hline
    Network & Domain & \# of vertices & \# of edges \\
    \hline
    \href{http://konect.cc/networks/arenas-email}{arenas-email} & communication & 1,133 & 5,451 \\
    \hline
    \href{http://konect.cc/networks/moreno_propro}{moreno\_propro} & metabolic & 1,458 & 1,948 \\
    \hline
    \href{http://konect.cc/networks/maayan-vidal}{maayan-vidal} & metabolic & 2,783 & 6,007 \\
    \hline
    \href{http://konect.cc/networks/reactome}{reactome} & metabolic & 5,973 & 145,778 \\
    \hline
    \href{http://konect.cc/networks/as20000102}{as20000102} & computer & 6,474 & 12,572 \\
    \hline
    \href{http://konect.cc/networks/pajek-erdos}{pajek-erdos} & co-authorship & 6,927 & 11,850 \\
    \hline
    \href{http://konect.cc/networks/twin}{twin} & miscellaneous & 10,320 & 17,988 \\
    \hline
    \href{http://konect.cc/networks/arenas-pgp}{arenas-pgp} & online contact & 10,680 & 24,316 \\
    \hline
    \href{http://konect.cc/networks/dimacs10-astro-ph}{dimacs10-astro-ph} & co-authorship & 14,845 & 119,652 \\
    \hline
    \href{http://konect.cc/networks/ca-AstroPh}{ca-AstroPh} & co-authorship & 17,903 & 196,972 \\
    \hline
    \href{http://konect.cc/networks/dimacs10-cond-mat-2003}{dimacs10-cond-mat-2003} & co-authorship & 27,519 & 116,181 \\
    \hline
    \href{http://konect.cc/networks/dimacs10-cond-mat-2005}{dimacs10-cond-mat-2005} & co-authorship & 36,458 & 171,734 \\
    \hline
    \href{http://konect.cc/networks/loc-brightkite_edges}{loc-brightkite\_edges} & online social & 56,739 & 212,945 \\
    \hline
    \href{http://konect.cc/networks/facebook-wosn-links}{facebook-wosn-links} & online social & 63,392 & 816,831 \\
    \hline
    \href{http://konect.cc/networks/livemocha}{livemocha} & online social & 104,103 & 2,193,083 \\
    \hline
    \href{http://konect.cc/networks/flickrEdges}{flickrEdges} & miscellaneous & 105,722 & 2,316,668 \\
    \hline
    \href{http://konect.cc/networks/wordnet-words}{wordnet-words} & lexical & 145,145 & 656,230 \\
    \hline
    \href{http://konect.cc/networks/douban}{douban} & online social & 154,908 & 327,162 \\
    \hline
    \href{http://konect.cc/networks/loc-gowalla_edges}{loc-gowalla\_edges} & online social & 196,591 & 950,327 \\
    \hline
    \href{http://konect.cc/networks/com-dblp}{com-dblp} & co-authorship & 317,080 & 1,049,866 \\
    \hline
    \href{http://konect.cc/networks/dimacs10-cnr-2000}{dimacs10-cnr-2000} & hyperlink & 325,557 & 2,738,969 \\
    \hline
    \href{http://konect.cc/networks/com-amazon}{com-amazon} & co-purchase & 334,863 & 925,872 \\
    \hline
    \href{http://konect.cc/networks/amazon}{amazon} & rating & 524,366 & 1,491,774 \\
    \hline
    \href{http://konect.cc/networks/dimacs10-in-2004}{dimacs10-in-2004} & hyperlink & 1,353,703 & 13,126,172 \\
    \hline
\end{tabular}
\caption{{\bf The studied networks.} We list the network name, also providing a hyperlink to the data source in the 1$^{\rm st}$ column, followed by the domain of the network in the 2$^{\rm nd}$ column. The network size in terms of the number of nodes is provided in the 3$^{\rm rd}$ column, with the number of links shown in the 4$^{th}$ column.}
\label{table:network_list}
\end{table}

The embedding quality scores obtained for the different networks are presented in Tables S2-S24. 
\begin{landscape}
\begin{table}
\begin{tabular}{|p{0.19\textwidth}|*{8}{>{\centering\arraybackslash}p{0.11\textwidth}|}}
    \thickhline
    & \makecell{MA} & \makecell{EPP} & \makecell{AUC} & \makecell{GR} & \makecell{GS} & \makecell{GE} & \makecell{Running \\ Time \\ (min)} & \makecell{Peak \\ Mem. \\ (GB)} \\
    \thickhline
    CLOVE & \cellcolor[RGB]{176.05038263276012, 219.83972205986183, 124.37677897649911}$0.428 \pm 0.004$ & \cellcolor[RGB]{189.31223628691984, 227.30966795083472, 142.54577141808846}$0.377 \pm 0.001$ & \cellcolor[RGB]{58.10645727903275, 123.1288693377765, 29.022412058743736}$0.899 \pm 0.002$ & \cellcolor[RGB]{76.008234944411, 144.79944230112912, 32.791207356718104}$0.805 \pm 0.015$ & \cellcolor[RGB]{122.32956516681492, 184.07683474012453, 62.01092170676155}$0.619 \pm 0.009$ & \cellcolor[RGB]{220.06256050704684, 240.67937413349864, 192.01368429394492}$0.243 \pm 0.003$ & \cellcolor[RGB]{247.00035140605902, 246.9986529434404, 246.99953145858797}$0.084 \pm 0.009$ & \cellcolor[RGB]{247.28703713417053, 245.89969098567963, 246.61728382110596}$0.383 \pm 0.105$ \\
    CLOVE (with SA) & \cellcolor[RGB]{175.70598429076804, 219.61616524137574, 123.9598757204034}$0.429 \pm 0.016$ & \cellcolor[RGB]{189.9873417721519, 227.60319207484866, 143.63181067694}$0.374 \pm 0.006$ & \cellcolor[RGB]{58.48816484005072, 123.59093638532457, 29.102771545273836}$0.897 \pm 0.004$ & \cellcolor[RGB]{75.07796027850345, 143.67332033713578, 32.59536005863231}$0.810 \pm 0.018$ & \cellcolor[RGB]{120.81807609163968, 182.80718391697735, 61.0435686986494}$0.625 \pm 0.011$ & \cellcolor[RGB]{219.62813756390656, 240.49049459300284, 191.31482999411054}$0.245 \pm 0.004$ & \cellcolor[RGB]{247.00035997963542, 246.99862007806422, 246.99952002715278}$0.086 \pm 0.007$ & \cellcolor[RGB]{247.29505228996277, 245.8689662218094, 246.6065969467163}$0.393 \pm 0.088$ \\
    CLOVE (dendr.) & \cellcolor[RGB]{176.66725583564022, 220.24014852488926, 125.12352022209078}$0.426 \pm 0.007$ & \cellcolor[RGB]{215.05063291139237, 238.50027517886625, 183.95101816180514}$0.265 \pm 0.011$ & \cellcolor[RGB]{59.58008063133855, 124.91272918530456, 29.33264855396601}$0.892 \pm 0.002$ & \cellcolor[RGB]{186.23765701991263, 225.97289435648375, 137.5997091189899}$0.390 \pm 0.030$ & \cellcolor[RGB]{204.5096330694523, 233.9172317693271, 166.9937575465102}$0.311 \pm 0.021$ & \cellcolor[RGB]{236.2751821576015, 245.7382567244237, 222.39600612626225}$0.126 \pm 0.008$ & \cellcolor[RGB]{247.00006856907987, 246.99973715186053, 246.99990857456018}$0.016 \pm 0.002$ & \cellcolor[RGB]{247.30670022964478, 245.8243157863617, 246.59106636047363}$0.409 \pm 0.062$ \\
    CLOVE (Louvain) & \cellcolor[RGB]{174.37510911069356, 218.75226380869583, 122.3488162918922}$0.434 \pm 0.010$ & \cellcolor[RGB]{190.0928270042194, 227.64905521922583, 143.80150431113555}$0.374 \pm 0.005$ & \cellcolor[RGB]{58.65881782008357, 123.79751630852222, 29.138698488438646}$0.897 \pm 0.001$ & \cellcolor[RGB]{78.15670221616887, 146.97162986158185, 33.74028941834808}$0.795 \pm 0.018$ & \cellcolor[RGB]{123.82092348817008, 185.32957573006288, 62.965391032428855}$0.613 \pm 0.009$ & \cellcolor[RGB]{220.64455134157123, 240.93241362677009, 192.94993041904937}$0.241 \pm 0.003$ & \cellcolor[RGB]{247.0003791884375, 246.99854644432293, 246.99949441541668}$0.091 \pm 0.018$ & \cellcolor[RGB]{247.27950620651245, 245.92855954170227, 246.6273250579834}$0.373 \pm 0.091$ \\
    CLOVE (k1 decomp.) & \cellcolor[RGB]{172.84904609799887, 217.7616615022098, 120.5014768554723}$0.439 \pm 0.006$ & \cellcolor[RGB]{190.96835443037972, 228.02971931755638, 145.20996147495867}$0.370 \pm 0.002$ & \cellcolor[RGB]{57.94973809227348, 122.93915663801526, 28.989418545741785}$0.900 \pm 0.001$ & \cellcolor[RGB]{73.17049566941954, 141.3642842314026, 32.19378856198306}$0.820 \pm 0.018$ & \cellcolor[RGB]{118.81525397543624, 181.12481333936643, 59.76176254427919}$0.633 \pm 0.011$ & \cellcolor[RGB]{216.1776842981375, 238.99029752092935, 185.7641008274386}$0.260 \pm 0.004$ & \cellcolor[RGB]{247.0003179486111, 246.99878119699073, 246.9995760685185}$0.076 \pm 0.010$ & \cellcolor[RGB]{247.25711154937744, 246.01440572738647, 246.65718460083008}$0.343 \pm 0.023$ \\
    HMCS & \cellcolor[RGB]{184.79175500240362, 225.34424130539287, 135.27369282995363}$0.397 \pm 0.000$ & \cellcolor[RGB]{222.5928270042194, 241.77949000183452, 196.08411300678773}$0.232 \pm 0.006$ & \cellcolor[RGB]{60.941093650220935, 126.56027126079375, 29.619177610572827}$0.885 \pm 0.000$ & \cellcolor[RGB]{171.33413019579223, 216.77829503937392, 118.66763128964323}$0.444 \pm 0.013$ & \cellcolor[RGB]{193.74651631075562, 229.23761578728505, 149.67917841295468}$0.358 \pm 0.010$ & \cellcolor[RGB]{234.7688746227595, 245.56104407326583, 218.94035942868356}$0.144 \pm 0.004$ & \cellcolor[RGB]{247.00006511399306, 246.99975039635996, 246.9999131813426}$0.016 \pm 0.003$ & \cellcolor[RGB]{247.22588992118835, 246.13408863544464, 246.69881343841553}$0.301 \pm 0.001$ \\
    Mercator & \cellcolor[RGB]{106.773777356294, 171.00997297928697, 52.055217508028164}$0.681 \pm 0.001$ & \cellcolor[RGB]{192.7194092827004, 228.79104751421758, 148.02687580260502}$0.362 \pm 0.004$ & \cellcolor[RGB]{49.59056631107277, 112.82015921866704, 27.229592907594267}$0.944 \pm 0.000$ & \cellcolor[RGB]{137.17788466831388, 194.60669706539673, 77.32059723006415}$0.564 \pm 0.019$ & \cellcolor[RGB]{166.32785141962398, 213.5286053074752, 112.60739908691323}$0.462 \pm 0.013$ & \cellcolor[RGB]{230.9140603146744, 245.10753650760876, 210.0969618983707}$0.189 \pm 0.006$ & \cellcolor[RGB]{247.00568742838541, 246.97819819118922, 246.99241676215277}$1.365 \pm 0.028$ & \cellcolor[RGB]{247.71066451072693, 244.27578604221344, 246.05244731903076}$0.948 \pm 0.431$ \\
    ncMCE (hypyperbolic) & \cellcolor[RGB]{194.8986962845805, 229.73856360199153, 151.53268532736863}$0.353 \pm 0.000$ & \cellcolor[RGB]{220.26160337552741, 240.76591451109888, 192.33388369106586}$0.242 \pm 0.000$ & \cellcolor[RGB]{64.53860306344941, 130.9151510768072, 30.376548013357773}$0.866 \pm 0.000$ & \cellcolor[RGB]{195.64630316107994, 230.06361007003474, 152.7353572591286}$0.349 \pm 0.000$ & \cellcolor[RGB]{211.54945356816353, 236.9780232905059, 178.31868617487177}$0.280 \pm 0.000$ & \cellcolor[RGB]{237.39365540564563, 245.8698418124289, 224.9619153423635}$0.113 \pm 0.000$ & \cellcolor[RGB]{247.00059072100694, 246.99773556947338, 246.99921237199075}$0.142 \pm 0.051$ & \cellcolor[RGB]{247.59164452552795, 244.7320293188095, 246.2111406326294}$0.789 \pm 0.450$ \\
    \thickhline
\end{tabular}
\caption{Average quality scores of the considered hyperbolic embedding algorithms for the 'arenas-email' network. We show the results for the mapping accuracy, MA ($1^{\rm st}$ column), the edge prediction precision, EPP ($2^{\rm nd}$ column), the area under the receiver operating characteristic curve, AUC ($3^{\rm rd}$ column), the greedy routing score, GR ($4^{\rm th}$ column), the greedy success rate, GS ($5^{\rm th}$ column) and the greedy routing efficiency, GE ($6^{\rm th}$ column averaged over 4 different realizations. Beside the quality scores, we also display the real time in minutes ($7^{\rm th}$ column), the user time in minutes ($8^{\rm th}$ column) and the peak memory usage in GB ($9^{\rm th}$ column). In the top part of the table, we list the scores obtained for CLOVE with default settings ($1^{\rm st}$ row), for CLOVE with simulated annealing optimisation during the solution of the TSP problem ($2^{\rm nd}$ row) and for CLOVE with Louvain communities ($3^{\rm rd}$ row). For comparison, in the  $4^{\rm th}$ row we give the results for HMCS, followed by the scores for Mercator in the  $5^{\rm th}$ row and by hyperbolic ncMCE in the bottom row.}
\end{table}
\begin{table}
\begin{tabular}{|p{0.19\textwidth}|*{8}{>{\centering\arraybackslash}p{0.11\textwidth}|}}
    \thickhline
    & \makecell{MA} & \makecell{EPP} & \makecell{AUC} & \makecell{GR} & \makecell{GS} & \makecell{GE} & \makecell{Running \\ Time \\ (min)} & \makecell{Peak \\ Mem. \\ (GB)} \\
    \thickhline
    CLOVE & \cellcolor[RGB]{206.1914687167733, 234.64846465946664, 169.69931924002663}$0.304 \pm 0.014$ & \cellcolor[RGB]{133.93525179195126, 192.50183011056484, 73.3953048007831}$0.576 \pm 0.003$ & \cellcolor[RGB]{40.76424627976685, 102.13566654919146, 25.3714202694246}$0.991 \pm 0.001$ & \cellcolor[RGB]{230.62181116971146, 245.07315425526016, 209.42650797757335}$0.193 \pm 0.019$ & \cellcolor[RGB]{232.66270126246383, 245.31325897205457, 214.10854995506403}$0.169 \pm 0.017$ & \cellcolor[RGB]{243.0319943683418, 246.5331758080402, 237.8969282567841}$0.047 \pm 0.004$ & \cellcolor[RGB]{247.00281151826388, 246.98922251332175, 246.99625130898147}$0.675 \pm 0.020$ & \cellcolor[RGB]{247.30300164222717, 245.83849370479584, 246.59599781036377}$0.404 \pm 0.055$ \\
    CLOVE (with SA) & \cellcolor[RGB]{209.43531914990132, 236.05883441300057, 174.91768732810215}$0.289 \pm 0.006$ & \cellcolor[RGB]{133.47327826785704, 192.20195255983703, 72.83607369266906}$0.577 \pm 0.002$ & \cellcolor[RGB]{40.778950484238415, 102.15346637565703, 25.374515891418614}$0.991 \pm 0.000$ & \cellcolor[RGB]{231.3176930983659, 245.1550227174548, 211.02294299036882}$0.184 \pm 0.013$ & \cellcolor[RGB]{233.26159859837023, 245.38371748216122, 215.48249090214347}$0.162 \pm 0.009$ & \cellcolor[RGB]{243.18009250515004, 246.55059911825293, 238.23668280593242}$0.045 \pm 0.002$ & \cellcolor[RGB]{247.00302326388888, 246.98841082175926, 246.9959689814815}$0.726 \pm 0.044$ & \cellcolor[RGB]{247.28838324546814, 245.89453089237213, 246.61548900604248}$0.385 \pm 0.057$ \\
    CLOVE (dendr.) & \cellcolor[RGB]{209.1737216109776, 235.94509635259894, 174.49685650461612}$0.291 \pm 0.003$ & \cellcolor[RGB]{153.42102246206227, 205.15048826484744, 96.98334298039117}$0.507 \pm 0.005$ & \cellcolor[RGB]{40.86251366718547, 102.25462180764558, 25.3921081404601}$0.990 \pm 0.000$ & \cellcolor[RGB]{236.9537590157207, 245.81808929596716, 223.95274127135926}$0.118 \pm 0.013$ & \cellcolor[RGB]{238.24533205496084, 245.9700390652895, 226.91576177314545}$0.103 \pm 0.011$ & \cellcolor[RGB]{244.54833432026214, 246.71156874356026, 241.37559049942487}$0.029 \pm 0.003$ & \cellcolor[RGB]{247.0014606130382, 246.9944009833536, 246.99805251594907}$0.351 \pm 0.016$ & \cellcolor[RGB]{247.3488438129425, 245.6627653837204, 246.53487491607666}$0.465 \pm 0.110$ \\
    CLOVE (Louvain) & \cellcolor[RGB]{210.30455724056264, 236.43676401763594, 176.31602686525298}$0.286 \pm 0.005$ & \cellcolor[RGB]{133.5688659474049, 192.26400070270142, 72.951785094227}$0.577 \pm 0.002$ & \cellcolor[RGB]{40.762278865343106, 102.13328494225743, 25.371006076914338}$0.991 \pm 0.000$ & \cellcolor[RGB]{231.52912487116137, 245.17989704366605, 211.50799235148787}$0.182 \pm 0.015$ & \cellcolor[RGB]{233.48696119058374, 245.41023072830396, 215.99949920192742}$0.159 \pm 0.012$ & \cellcolor[RGB]{243.23375435541553, 246.5569122771077, 238.35978940360033}$0.044 \pm 0.003$ & \cellcolor[RGB]{247.002939141875, 246.98873328947917, 246.99608114416668}$0.705 \pm 0.048$ & \cellcolor[RGB]{247.3417263031006, 245.69004917144775, 246.54436492919922}$0.456 \pm 0.115$ \\
    CLOVE (k1 decomp.) & \cellcolor[RGB]{206.95684399908006, 234.98123652133916, 170.93057512895487}$0.300 \pm 0.018$ & \cellcolor[RGB]{164.4560234504268, 212.313559081856, 110.34150207156931}$0.469 \pm 0.001$ & \cellcolor[RGB]{40.65233575797436, 102.00019591754791, 25.34786015957355}$0.991 \pm 0.000$ & \cellcolor[RGB]{229.12390954708172, 244.61909110742684, 206.5906370974793}$0.204 \pm 0.006$ & \cellcolor[RGB]{231.95080516449394, 245.22950648994046, 212.47537655383908}$0.177 \pm 0.006$ & \cellcolor[RGB]{242.37449778546235, 246.45582326887794, 236.38855374311953}$0.054 \pm 0.001$ & \cellcolor[RGB]{247.00181464461807, 246.99304386229744, 246.9975804738426}$0.436 \pm 0.012$ & \cellcolor[RGB]{247.26073789596558, 246.00050473213196, 246.6523494720459}$0.348 \pm 0.033$ \\
    HMCS & \cellcolor[RGB]{210.133125107534, 236.36222830762347, 176.04024473820687}$0.286 \pm 0.010$ & \cellcolor[RGB]{227.95598299589682, 244.11129695473775, 204.71179873252967}$0.209 \pm 0.003$ & \cellcolor[RGB]{40.86360138020918, 102.2559385128848, 25.392337132675618}$0.990 \pm 0.000$ & \cellcolor[RGB]{239.09090999578888, 246.06951882303397, 228.85561704916273}$0.093 \pm 0.009$ & \cellcolor[RGB]{240.02939809308552, 246.17992918742183, 231.00861915472564}$0.082 \pm 0.007$ & \cellcolor[RGB]{244.96681407867365, 246.76080165631456, 242.33563229813365}$0.024 \pm 0.002$ & \cellcolor[RGB]{247.0009501905903, 246.99635760273728, 246.99873307921297}$0.228 \pm 0.001$ & \cellcolor[RGB]{247.25409960746765, 246.02595150470734, 246.66120052337646}$0.339 \pm 0.006$ \\
    Mercator & \cellcolor[RGB]{187.0801874448725, 226.33921193255327, 138.95508415044708}$0.387 \pm 0.002$ & \cellcolor[RGB]{163.89366783606565, 211.9485212269198, 109.66075580155315}$0.471 \pm 0.003$ & \cellcolor[RGB]{39.81004877067994, 100.98058535398098, 25.17053658330104}$0.996 \pm 0.000$ & \cellcolor[RGB]{237.30897949229234, 245.85987994026968, 224.7676588352589}$0.114 \pm 0.002$ & \cellcolor[RGB]{238.27623470799801, 245.97367467152918, 226.98665609481895}$0.103 \pm 0.002$ & \cellcolor[RGB]{244.5390430320958, 246.7104756508348, 241.3542751912786}$0.029 \pm 0.000$ & \cellcolor[RGB]{247.66015733944445, 244.46939686546295, 246.11979021407407}$158.438 \pm 0.082$ & \cellcolor[RGB]{252.77318239212036, 224.86946749687195, 239.30242347717285}$7.698 \pm 6.216$ \\
    ncMCE (hypyperbolic) & \cellcolor[RGB]{212.45247803766503, 237.37064262507175, 179.77137771276549}$0.276 \pm 0.000$ & \cellcolor[RGB]{236.23964479308486, 245.73407585801, 222.3144792311947}$0.127 \pm 0.000$ & \cellcolor[RGB]{41.36853256025286, 102.8671709939903, 25.498638433737444}$0.988 \pm 0.000$ & \cellcolor[RGB]{242.62541641415888, 246.4853431075481, 236.964190597188}$0.051 \pm 0.000$ & \cellcolor[RGB]{243.04638231657427, 246.53486850783227, 237.92993590272917}$0.047 \pm 0.000$ & \cellcolor[RGB]{245.8351264292759, 246.86295605050304, 244.32764298480944}$0.014 \pm 0.000$ & \cellcolor[RGB]{247.02890273609376, 246.8892061783073, 246.96146301854168}$6.937 \pm 0.185$ & \cellcolor[RGB]{246.76172971725464, 202.16605401039124, 228.08302700519562}$12.159 \pm 5.335$ \\
    \thickhline
\end{tabular}
\caption{Average quality scores of the considered hyperbolic embedding algorithms for the 'arenas-pgp' network. We show the results for the mapping accuracy, MA ($1^{\rm st}$ column), the edge prediction precision, EPP ($2^{\rm nd}$ column), the area under the receiver operating characteristic curve, AUC ($3^{\rm rd}$ column), the greedy routing score, GR ($4^{\rm th}$ column), the greedy success rate, GS ($5^{\rm th}$ column) and the greedy routing efficiency, GE ($6^{\rm th}$ column averaged over 4 different realizations. Beside the quality scores, we also display the real time in minutes ($7^{\rm th}$ column), the user time in minutes ($8^{\rm th}$ column) and the peak memory usage in GB ($9^{\rm th}$ column). In the top part of the table, we list the scores obtained for CLOVE with default settings ($1^{\rm st}$ row), for CLOVE with simulated annealing optimisation during the solution of the TSP problem ($2^{\rm nd}$ row) and for CLOVE with Louvain communities ($3^{\rm rd}$ row). For comparison, in the  $4^{\rm th}$ row we give the results for HMCS, followed by the scores for Mercator in the  $5^{\rm th}$ row and by hyperbolic ncMCE in the bottom row.}
\end{table}
\begin{table}
\begin{tabular}{|p{0.19\textwidth}|*{8}{>{\centering\arraybackslash}p{0.11\textwidth}|}}
    \thickhline
    & \makecell{MA} & \makecell{EPP} & \makecell{AUC} & \makecell{GR} & \makecell{GS} & \makecell{GE} & \makecell{Running \\ Time \\ (min)} & \makecell{Peak \\ Mem. \\ (GB)} \\
    \thickhline
    CLOVE & \cellcolor[RGB]{191.73092625606915, 228.36127228524745, 146.43670745541556}$0.366 \pm 0.007$ & \cellcolor[RGB]{158.2713370983137, 208.29893811644922, 102.85477648743236}$0.490 \pm 0.002$ & \cellcolor[RGB]{39.78526378261445, 100.95058247369117, 25.165318691076727}$0.996 \pm 0.000$ & \cellcolor[RGB]{84.29640354357814, 152.12897897660562, 37.669698267890006}$0.771 \pm 0.012$ & \cellcolor[RGB]{98.66913418900963, 164.20207271876808, 46.86824588096616}$0.713 \pm 0.009$ & \cellcolor[RGB]{165.11742924890365, 212.74289267034095, 111.14215119604124}$0.466 \pm 0.005$ & \cellcolor[RGB]{247.0006910519965, 246.99735096734665, 246.99907859733796}$0.166 \pm 0.002$ & \cellcolor[RGB]{247.32156538963318, 245.76733267307281, 246.57124614715576}$0.429 \pm 0.067$ \\
    CLOVE (with SA) & \cellcolor[RGB]{189.7212237891733, 227.4874886039884, 143.20370783475707}$0.375 \pm 0.005$ & \cellcolor[RGB]{156.68447741011772, 207.2688713013045, 100.93384107540567}$0.496 \pm 0.001$ & \cellcolor[RGB]{39.773286397791885, 100.93608353416913, 25.16279713637724}$0.996 \pm 0.000$ & \cellcolor[RGB]{78.18664265835883, 146.9967798330214, 33.75945130134965}$0.795 \pm 0.008$ & \cellcolor[RGB]{93.28770914067917, 159.6816756781705, 43.42413385003468}$0.735 \pm 0.008$ & \cellcolor[RGB]{161.30098035538944, 210.265548651744, 106.52223937757668}$0.480 \pm 0.005$ & \cellcolor[RGB]{247.0007096288889, 246.99727975592594, 246.99905382814816}$0.170 \pm 0.002$ & \cellcolor[RGB]{247.3554563522339, 245.63741731643677, 246.52605819702148}$0.474 \pm 0.126$ \\
    CLOVE (dendr.) & \cellcolor[RGB]{190.44326546384548, 227.80141976688935, 144.36525313749058}$0.372 \pm 0.002$ & \cellcolor[RGB]{162.49351734012092, 211.03965160674514, 107.96583678014636}$0.475 \pm 0.003$ & \cellcolor[RGB]{39.572974283386955, 100.69360044831052, 25.12062616492357}$0.997 \pm 0.000$ & \cellcolor[RGB]{87.45704778995793, 154.78392014356467, 39.692510585573075}$0.758 \pm 0.004$ & \cellcolor[RGB]{102.15570381408179, 167.1307912038287, 49.09965044101234}$0.699 \pm 0.005$ & \cellcolor[RGB]{168.43437042438805, 214.89599483688346, 115.1573957768908}$0.455 \pm 0.003$ & \cellcolor[RGB]{247.00061109953126, 246.99765745179687, 246.999185200625}$0.147 \pm 0.009$ & \cellcolor[RGB]{247.29165983200073, 245.8819706439972, 246.61112022399902}$0.389 \pm 0.071$ \\
    CLOVE (Louvain) & \cellcolor[RGB]{190.6338056482809, 227.88426332533953, 144.67177430375625}$0.371 \pm 0.008$ & \cellcolor[RGB]{158.4583598472797, 208.4203388482342, 103.081172446707}$0.490 \pm 0.004$ & \cellcolor[RGB]{39.53435978727463, 100.64685658459561, 25.112496797320976}$0.997 \pm 0.000$ & \cellcolor[RGB]{82.23219172256822, 150.39504104695732, 36.348602702443664}$0.779 \pm 0.011$ & \cellcolor[RGB]{97.58441942859108, 163.2909123200165, 46.17402843429829}$0.718 \pm 0.010$ & \cellcolor[RGB]{165.24315202956595, 212.82450219463053, 111.29434193052721}$0.466 \pm 0.006$ & \cellcolor[RGB]{247.0007528075, 246.99711423791666, 246.99899625666666}$0.181 \pm 0.009$ & \cellcolor[RGB]{247.32634902000427, 245.74899542331696, 246.56486797332764}$0.435 \pm 0.111$ \\
    CLOVE (k1 decomp.) & \cellcolor[RGB]{192.42515942994595, 228.66311279562868, 147.55351734382612}$0.363 \pm 0.003$ & \cellcolor[RGB]{246.99492920776328, 246.99940343620744, 246.9883670060452}$0.000 \pm 0.000$ & \cellcolor[RGB]{40.27476577249068, 101.54313751406767, 25.268371741576985}$0.993 \pm 0.000$ & \cellcolor[RGB]{84.30806155852132, 152.1387717091579, 37.67715939745364}$0.771 \pm 0.010$ & \cellcolor[RGB]{98.73993026505961, 164.26154142265008, 46.913555369638146}$0.713 \pm 0.008$ & \cellcolor[RGB]{153.53386291015414, 205.22373557325795, 97.11993931229186}$0.507 \pm 0.005$ & \cellcolor[RGB]{247.00058679850696, 246.99775060572338, 246.99921760199075}$0.141 \pm 0.004$ & \cellcolor[RGB]{247.26391005516052, 245.98834478855133, 246.64811992645264}$0.352 \pm 0.021$ \\
    HMCS & \cellcolor[RGB]{193.74744399684502, 229.23801912906305, 149.6806707775333}$0.358 \pm 0.011$ & \cellcolor[RGB]{197.67614540248172, 230.94615017499206, 156.00075564747058}$0.341 \pm 0.012$ & \cellcolor[RGB]{39.80479488034625, 100.97422538147178, 25.169430501125525}$0.996 \pm 0.000$ & \cellcolor[RGB]{126.09545503840901, 187.24018223226358, 64.42109122458177}$0.604 \pm 0.016$ & \cellcolor[RGB]{137.53980765303479, 194.84162952916293, 77.75871452735788}$0.563 \pm 0.014$ & \cellcolor[RGB]{190.11132316805814, 227.6570970295905, 143.83125900948482}$0.373 \pm 0.009$ & \cellcolor[RGB]{247.00022741118056, 246.9991282571412, 246.9996967850926}$0.055 \pm 0.004$ & \cellcolor[RGB]{247.23911714553833, 246.0833842754364, 246.68117713928223}$0.319 \pm 0.001$ \\
    Mercator & \cellcolor[RGB]{184.58620588434962, 225.25487212363026, 134.943026857432}$0.397 \pm 0.002$ & \cellcolor[RGB]{158.82107063315303, 208.65578269169583, 103.52024339802736}$0.488 \pm 0.001$ & \cellcolor[RGB]{39.754433502153134, 100.91326160786957, 25.15882810571645}$0.996 \pm 0.000$ & \cellcolor[RGB]{130.11462017029274, 190.02177098773387, 68.77032967982804}$0.589 \pm 0.001$ & \cellcolor[RGB]{138.28889559774342, 195.3278795985352, 78.66550519726833}$0.560 \pm 0.001$ & \cellcolor[RGB]{214.090044866679, 238.08262820290392, 182.4057243507445}$0.269 \pm 0.003$ & \cellcolor[RGB]{247.20642315208335, 246.20871125034722, 246.72476913055556}$49.542 \pm 1.276$ & \cellcolor[RGB]{251.50812673568726, 229.71884751319885, 240.989164352417}$6.011 \pm 7.139$ \\
    ncMCE (hypyperbolic) & \cellcolor[RGB]{197.53225592200295, 230.88358953130563, 155.76928126583084}$0.341 \pm 0.000$ & \cellcolor[RGB]{233.6705774737512, 245.43183264397072, 216.42073655742922}$0.157 \pm 0.000$ & \cellcolor[RGB]{39.95075391364625, 101.15091263230862, 25.20015871866237}$0.995 \pm 0.000$ & \cellcolor[RGB]{196.34242126834522, 230.36627011667184, 153.8551994316858}$0.346 \pm 0.000$ & \cellcolor[RGB]{202.0063307114703, 232.8288394397697, 162.96670592714784}$0.322 \pm 0.000$ & \cellcolor[RGB]{226.9163540236322, 243.65928435810096, 203.03935212497356}$0.213 \pm 0.000$ & \cellcolor[RGB]{247.01126109706598, 246.9568324612471, 246.98498520391203}$2.703 \pm 0.072$ & \cellcolor[RGB]{250.5022246837616, 233.57480537891388, 242.33036708831787}$4.670 \pm 0.697$ \\
    \thickhline
\end{tabular}
\caption{Average quality scores of the considered hyperbolic embedding algorithms for the 'as20000102' network. We show the results for the mapping accuracy, MA ($1^{\rm st}$ column), the edge prediction precision, EPP ($2^{\rm nd}$ column), the area under the receiver operating characteristic curve, AUC ($3^{\rm rd}$ column), the greedy routing score, GR ($4^{\rm th}$ column), the greedy success rate, GS ($5^{\rm th}$ column) and the greedy routing efficiency, GE ($6^{\rm th}$ column averaged over 4 different realizations. Beside the quality scores, we also display the real time in minutes ($7^{\rm th}$ column), the user time in minutes ($8^{\rm th}$ column) and the peak memory usage in GB ($9^{\rm th}$ column). In the top part of the table, we list the scores obtained for CLOVE with default settings ($1^{\rm st}$ row), for CLOVE with simulated annealing optimisation during the solution of the TSP problem ($2^{\rm nd}$ row) and for CLOVE with Louvain communities ($3^{\rm rd}$ row). For comparison, in the  $4^{\rm th}$ row we give the results for HMCS, followed by the scores for Mercator in the  $5^{\rm th}$ row and by hyperbolic ncMCE in the bottom row.}
\end{table}
\begin{table}
\begin{tabular}{|p{0.19\textwidth}|*{8}{>{\centering\arraybackslash}p{0.11\textwidth}|}}
    \thickhline
    & \makecell{MA} & \makecell{EPP} & \makecell{AUC} & \makecell{GR} & \makecell{GS} & \makecell{GE} & \makecell{Running \\ Time \\ (min)} & \makecell{Peak \\ Mem. \\ (GB)} \\
    \thickhline
    CLOVE & \cellcolor[RGB]{192.86797390943525, 228.85564083018923, 148.26587107170016}$0.361 \pm 0.005$ & \cellcolor[RGB]{179.32445039057126, 221.96499411317782, 128.3401241570073}$0.416 \pm 0.003$ & \cellcolor[RGB]{52.902515221246034, 116.82936053098204, 27.926845309736006}$0.927 \pm 0.002$ & \cellcolor[RGB]{140.20711141532743, 196.5730372345108, 80.98755592381742}$0.554 \pm 0.021$ & \cellcolor[RGB]{180.94024290504973, 223.01384188573402, 130.29608351663913}$0.411 \pm 0.014$ & \cellcolor[RGB]{234.3478679317604, 245.51151387432475, 217.97452054933265}$0.149 \pm 0.004$ & \cellcolor[RGB]{247.00594700076388, 246.9772031637384, 246.99207066564816}$1.427 \pm 0.020$ & \cellcolor[RGB]{247.37015056610107, 245.58108949661255, 246.50646591186523}$0.494 \pm 0.023$ \\
    CLOVE (with SA) & \cellcolor[RGB]{193.5376699773843, 229.14681303364534, 149.3432082244878}$0.359 \pm 0.005$ & \cellcolor[RGB]{179.10308545743572, 221.82130108640564, 128.0721560800538}$0.417 \pm 0.001$ & \cellcolor[RGB]{53.044579062154774, 117.00133254892421, 27.956753486769426}$0.926 \pm 0.001$ & \cellcolor[RGB]{138.66486774681533, 195.57193169530117, 79.12062937772382}$0.559 \pm 0.016$ & \cellcolor[RGB]{180.13347226475543, 222.49014866308684, 129.31946642575656}$0.414 \pm 0.009$ & \cellcolor[RGB]{234.29684763123873, 245.5055114860281, 217.85747397754767}$0.149 \pm 0.003$ & \cellcolor[RGB]{247.00582403274305, 246.9776745411516, 246.99223462300927}$1.398 \pm 0.063$ & \cellcolor[RGB]{247.3919608592987, 245.4974833726883, 246.47738552093506}$0.523 \pm 0.037$ \\
    CLOVE (dendr.) & \cellcolor[RGB]{193.3140015181939, 229.0495658774756, 148.98339374665971}$0.360 \pm 0.003$ & \cellcolor[RGB]{197.12220161635622, 230.70530505058966, 155.10962868718175}$0.343 \pm 0.006$ & \cellcolor[RGB]{53.58587753894929, 117.65658859978072, 28.070711060831428}$0.923 \pm 0.001$ & \cellcolor[RGB]{225.04332433155284, 242.84492362241429, 200.02621740293284}$0.222 \pm 0.011$ & \cellcolor[RGB]{232.61390593422988, 245.3075183452035, 213.99660773146854}$0.169 \pm 0.008$ & \cellcolor[RGB]{241.56986959258532, 246.36116112853944, 234.5426420065193}$0.064 \pm 0.003$ & \cellcolor[RGB]{247.0024436387847, 246.9906327179919, 246.9967418149537}$0.586 \pm 0.027$ & \cellcolor[RGB]{247.39334845542908, 245.49216425418854, 246.47553539276123}$0.524 \pm 0.012$ \\
    CLOVE (Louvain) & \cellcolor[RGB]{193.8038876199406, 229.2625598347568, 149.7714713886001}$0.357 \pm 0.004$ & \cellcolor[RGB]{180.0613350417862, 222.4433227464226, 129.23214241900433}$0.414 \pm 0.003$ & \cellcolor[RGB]{53.32902849287909, 117.34566607032733, 28.01663757744823}$0.925 \pm 0.001$ & \cellcolor[RGB]{142.38709493979366, 197.98811425916432, 83.62648334817129}$0.546 \pm 0.015$ & \cellcolor[RGB]{183.4789906258955, 224.66180093259885, 133.3693044418735}$0.402 \pm 0.010$ & \cellcolor[RGB]{234.6392047594924, 245.54578879523442, 218.64288150707083}$0.145 \pm 0.003$ & \cellcolor[RGB]{247.00685092409722, 246.973738124294, 246.99086543453703}$1.644 \pm 0.055$ & \cellcolor[RGB]{247.41159176826477, 245.42223155498505, 246.45121097564697}$0.549 \pm 0.032$ \\
    CLOVE (k1 decomp.) & \cellcolor[RGB]{193.09885401063744, 228.95602348288583, 148.6372868866776}$0.360 \pm 0.002$ & \cellcolor[RGB]{180.62890105692426, 222.8117427913368, 129.91919601627674}$0.412 \pm 0.001$ & \cellcolor[RGB]{52.901787234654606, 116.82847928405558, 27.92669204940097}$0.927 \pm 0.002$ & \cellcolor[RGB]{137.930629333301, 195.09532079530067, 78.23181445610123}$0.562 \pm 0.014$ & \cellcolor[RGB]{179.5518759857589, 222.11262125391366, 128.61542882486603}$0.416 \pm 0.009$ & \cellcolor[RGB]{234.06514891873093, 245.47825281396834, 217.3259298723827}$0.152 \pm 0.003$ & \cellcolor[RGB]{247.00537891288195, 246.97938083395255, 246.9928281161574}$1.291 \pm 0.014$ & \cellcolor[RGB]{247.3855757713318, 245.52195954322815, 246.48589897155762}$0.514 \pm 0.039$ \\
    HMCS & \cellcolor[RGB]{204.56596907916102, 233.94172568659175, 167.0843850403895}$0.311 \pm 0.015$ & \cellcolor[RGB]{233.4573829386618, 245.4067509339602, 215.93164321222412}$0.159 \pm 0.002$ & \cellcolor[RGB]{54.064128553332175, 118.23552403824422, 28.171395484912036}$0.921 \pm 0.001$ & \cellcolor[RGB]{224.07227811461766, 242.42272961505117, 198.4640995756893}$0.226 \pm 0.004$ & \cellcolor[RGB]{231.86078642680167, 245.21891605021196, 212.26886297913325}$0.178 \pm 0.003$ & \cellcolor[RGB]{241.3115626496877, 246.33077207643385, 233.95005549046002}$0.067 \pm 0.001$ & \cellcolor[RGB]{247.00305202375, 246.988300575625, 246.995930635}$0.732 \pm 0.029$ & \cellcolor[RGB]{247.36614656448364, 245.59643816947937, 246.51180458068848}$0.488 \pm 0.003$ \\
    Mercator & \cellcolor[RGB]{118.07599046895878, 180.50383199392536, 59.28863390013362}$0.636 \pm 0.002$ & \cellcolor[RGB]{207.81206523596694, 235.35307184172476, 172.30636581438162}$0.296 \pm 0.005$ & \cellcolor[RGB]{48.56154115093973, 111.57449718271653, 27.012956031776785}$0.950 \pm 0.000$ & \cellcolor[RGB]{232.81971996084843, 245.3317317600998, 214.46876932194635}$0.167 \pm 0.003$ & \cellcolor[RGB]{235.52296517051087, 245.6497606082954, 220.67033186176025}$0.135 \pm 0.002$ & \cellcolor[RGB]{243.06654670890373, 246.53724078928278, 237.97619539101441}$0.046 \pm 0.001$ & \cellcolor[RGB]{248.94480133140624, 239.54492822960938, 244.406931558125}$466.752 \pm 0.699$ & \cellcolor[RGB]{250.4874095916748, 233.63159656524658, 242.3501205444336}$4.650 \pm 8.110$ \\
    ncMCE (hypyperbolic) & \cellcolor[RGB]{200.2015217038115, 232.0441398712224, 160.06331752352284}$0.330 \pm 0.000$ & \cellcolor[RGB]{236.9352772758382, 245.81591497362803, 223.9103419857464}$0.118 \pm 0.000$ & \cellcolor[RGB]{60.25514370432715, 125.72991079997497, 29.47476709564782}$0.888 \pm 0.000$ & \cellcolor[RGB]{240.94483433327534, 246.28762756862062, 233.10873758810226}$0.071 \pm 0.000$ & \cellcolor[RGB]{242.0493109561339, 246.4175659948393, 235.642536899366}$0.058 \pm 0.000$ & \cellcolor[RGB]{245.10500445481597, 246.7770593476254, 242.6526572786955}$0.022 \pm 0.000$ & \cellcolor[RGB]{247.0988908021875, 246.62091859161458, 246.86814559708333}$23.734 \pm 1.114$ & \cellcolor[RGB]{210.75904977321625, 77.63330316543576, 151.96773755550385}$27.597 \pm 3.667$ \\
    \thickhline
\end{tabular}
\caption{Average quality scores of the considered hyperbolic embedding algorithms for the 'ca-AstroPh' network. We show the results for the mapping accuracy, MA ($1^{\rm st}$ column), the edge prediction precision, EPP ($2^{\rm nd}$ column), the area under the receiver operating characteristic curve, AUC ($3^{\rm rd}$ column), the greedy routing score, GR ($4^{\rm th}$ column), the greedy success rate, GS ($5^{\rm th}$ column) and the greedy routing efficiency, GE ($6^{\rm th}$ column averaged over 4 different realizations. Beside the quality scores, we also display the real time in minutes ($7^{\rm th}$ column), the user time in minutes ($8^{\rm th}$ column) and the peak memory usage in GB ($9^{\rm th}$ column). In the top part of the table, we list the scores obtained for CLOVE with default settings ($1^{\rm st}$ row), for CLOVE with simulated annealing optimisation during the solution of the TSP problem ($2^{\rm nd}$ row) and for CLOVE with Louvain communities ($3^{\rm rd}$ row). For comparison, in the  $4^{\rm th}$ row we give the results for HMCS, followed by the scores for Mercator in the  $5^{\rm th}$ row and by hyperbolic ncMCE in the bottom row.}
\end{table}
\begin{table}
\begin{tabular}{|p{0.19\textwidth}|*{8}{>{\centering\arraybackslash}p{0.11\textwidth}|}}
    \thickhline
    & \makecell{MA} & \makecell{EPP} & \makecell{AUC} & \makecell{GR} & \makecell{GS} & \makecell{GE} & \makecell{Running \\ Time \\ (min)} & \makecell{Peak \\ Mem. \\ (GB)} \\
    \thickhline
    CLOVE & \cellcolor[RGB]{199.022759450216, 231.53163454357218, 158.16704781121706}$0.335 \pm 0.003$ & \cellcolor[RGB]{153.37703918577597, 205.12193771708263, 96.93010006699194}$0.507 \pm 0.002$ & \cellcolor[RGB]{48.78333359087342, 111.8429827678994, 27.05964917702598}$0.949 \pm 0.000$ & \cellcolor[RGB]{179.15547684571652, 221.85530953143004, 128.13557723428843}$0.417 \pm 0.022$ & \cellcolor[RGB]{203.59495915060245, 233.51954745678367, 165.5223255900996}$0.315 \pm 0.014$ & \cellcolor[RGB]{238.01684695679404, 245.94315846550518, 226.39159007735105}$0.106 \pm 0.005$ & \cellcolor[RGB]{247.00478510305555, 246.98165710495368, 246.9936198625926}$1.148 \pm 0.061$ & \cellcolor[RGB]{247.3188259601593, 245.77783381938934, 246.5748987197876}$0.425 \pm 0.013$ \\
    CLOVE (with SA) & \cellcolor[RGB]{198.40416173066006, 231.26267901333046, 157.17191234932272}$0.337 \pm 0.009$ & \cellcolor[RGB]{153.48611373211614, 205.19274049277715, 97.06213767571955}$0.507 \pm 0.001$ & \cellcolor[RGB]{48.53462806419002, 111.54191818296687, 27.007290118776847}$0.950 \pm 0.002$ & \cellcolor[RGB]{175.78920991782675, 219.6701888940279, 124.06062253210607}$0.429 \pm 0.004$ & \cellcolor[RGB]{201.6133024231068, 232.65795757526382, 162.33444302847613}$0.323 \pm 0.002$ & \cellcolor[RGB]{237.782878710066, 245.91563278941953, 225.85483939368083}$0.108 \pm 0.001$ & \cellcolor[RGB]{247.00453411269098, 246.98261923468462, 246.99395451641203}$1.088 \pm 0.047$ & \cellcolor[RGB]{247.3570957183838, 245.6311330795288, 246.52387237548828}$0.476 \pm 0.074$ \\
    CLOVE (dendr.) & \cellcolor[RGB]{200.11578621462283, 232.00686357157514, 159.92539521482803}$0.330 \pm 0.003$ & \cellcolor[RGB]{172.84669100750227, 217.76013275925587, 120.49862595645013}$0.439 \pm 0.007$ & \cellcolor[RGB]{49.36848158027598, 112.5513198077025, 27.182838227426522}$0.945 \pm 0.001$ & \cellcolor[RGB]{227.1766323750791, 243.77244885873003, 203.45806077730117}$0.212 \pm 0.004$ & \cellcolor[RGB]{233.16410286061435, 245.3722473953664, 215.25882420964467}$0.163 \pm 0.002$ & \cellcolor[RGB]{242.17541109975085, 246.43240130585303, 235.93182546413428}$0.057 \pm 0.001$ & \cellcolor[RGB]{247.00158857381945, 246.99391046702544, 246.99788190157406}$0.381 \pm 0.007$ & \cellcolor[RGB]{247.36160254478455, 245.61385691165924, 246.5178632736206}$0.482 \pm 0.055$ \\
    CLOVE (Louvain) & \cellcolor[RGB]{199.39530768944033, 231.6936120388871, 158.76636454388228}$0.333 \pm 0.004$ & \cellcolor[RGB]{153.99417805322406, 205.52253663104017, 97.67716290653436}$0.505 \pm 0.002$ & \cellcolor[RGB]{49.18291224610534, 112.32668324528541, 27.14377099918007}$0.946 \pm 0.001$ & \cellcolor[RGB]{180.41873630257723, 222.6753200560589, 129.66478605048823}$0.413 \pm 0.017$ & \cellcolor[RGB]{204.59657574538255, 233.95503293277503, 167.13362185126758}$0.310 \pm 0.011$ & \cellcolor[RGB]{238.1364344545809, 245.95722758289187, 226.66593786639143}$0.104 \pm 0.003$ & \cellcolor[RGB]{247.00514036543402, 246.98029526583622, 246.9931461794213}$1.234 \pm 0.056$ & \cellcolor[RGB]{247.356294631958, 245.63420391082764, 246.52494049072266}$0.475 \pm 0.057$ \\
    CLOVE (k1 decomp.) & \cellcolor[RGB]{198.6301302962227, 231.360926215749, 157.5354269982713}$0.336 \pm 0.007$ & \cellcolor[RGB]{155.988578706739, 206.8171475815674, 100.09143738184193}$0.498 \pm 0.003$ & \cellcolor[RGB]{48.6255074867659, 111.65193011555871, 27.02642262879282}$0.949 \pm 0.002$ & \cellcolor[RGB]{168.63232367984796, 215.02449080972588, 115.39702340192122}$0.454 \pm 0.015$ & \cellcolor[RGB]{197.82040675604432, 231.00887250262795, 156.23282825972348}$0.340 \pm 0.012$ & \cellcolor[RGB]{237.20592931779194, 245.84775639032847, 224.53124961140506}$0.115 \pm 0.003$ & \cellcolor[RGB]{247.0041113088021, 246.98423998292535, 246.99451825493054}$0.987 \pm 0.049$ & \cellcolor[RGB]{247.36059403419495, 245.61772286891937, 246.51920795440674}$0.481 \pm 0.084$ \\
    HMCS & \cellcolor[RGB]{205.13951971097214, 234.19109552650963, 168.00705344808563}$0.308 \pm 0.006$ & \cellcolor[RGB]{226.33120926269916, 243.40487359247788, 202.0980322921682}$0.216 \pm 0.003$ & \cellcolor[RGB]{49.56084486166503, 112.78418062201557, 27.223335760350533}$0.944 \pm 0.001$ & \cellcolor[RGB]{231.0500802503632, 245.1235388529839, 210.40900763318618}$0.188 \pm 0.004$ & \cellcolor[RGB]{234.35995210779978, 245.51293554209408, 218.00224307083477}$0.149 \pm 0.003$ & \cellcolor[RGB]{242.5859471286403, 246.48069966219296, 236.87364341276302}$0.052 \pm 0.001$ & \cellcolor[RGB]{247.00216585453126, 246.9916975576302, 246.99711219395834}$0.520 \pm 0.011$ & \cellcolor[RGB]{247.30916571617126, 245.81486475467682, 246.58777904510498}$0.412 \pm 0.007$ \\
    Mercator & \cellcolor[RGB]{134.76219089056292, 193.03861513948823, 74.39633634120776}$0.573 \pm 0.002$ & \cellcolor[RGB]{192.68144080667238, 228.77453948116192, 147.96579608029907}$0.362 \pm 0.001$ & \cellcolor[RGB]{46.754434715458046, 109.38694728713342, 26.632512571675377}$0.959 \pm 0.000$ & \cellcolor[RGB]{235.65651298000006, 245.66547211529414, 220.97670624823542}$0.133 \pm 0.002$ & \cellcolor[RGB]{237.6291422154908, 245.89754614299892, 225.50214978847885}$0.110 \pm 0.001$ & \cellcolor[RGB]{243.94685577669574, 246.6408065619642, 239.99572795830196}$0.036 \pm 0.000$ & \cellcolor[RGB]{248.25856927291667, 242.17548445381945, 245.32190763611112}$302.057 \pm 9.427$ & \cellcolor[RGB]{248.9017834663391, 239.70983004570007, 244.46428871154785}$2.536 \pm 3.135$ \\
    ncMCE (hypyperbolic) & \cellcolor[RGB]{206.45233237770637, 234.76188364248102, 170.1189694771798}$0.302 \pm 0.000$ & \cellcolor[RGB]{233.89774885324312, 245.45855868861685, 216.94189442802832}$0.154 \pm 0.000$ & \cellcolor[RGB]{54.20780009853594, 118.40944222454351, 28.201642126007567}$0.920 \pm 0.000$ & \cellcolor[RGB]{241.84897885255032, 246.39399751206474, 235.18295148526252}$0.061 \pm 0.000$ & \cellcolor[RGB]{242.71457665921395, 246.49583254814283, 237.16873468878492}$0.050 \pm 0.000$ & \cellcolor[RGB]{245.46198874611432, 246.81905749954285, 243.47162124108578}$0.018 \pm 0.000$ & \cellcolor[RGB]{247.0638909075521, 246.75508485438368, 246.91481212326389}$15.334 \pm 0.862$ & \cellcolor[RGB]{227.8974529504776, 138.55471241474152, 187.65725946426392}$21.517 \pm 3.470$ \\
    \thickhline
\end{tabular}
\caption{Average quality scores of the considered hyperbolic embedding algorithms for the 'dimacs10-astro-ph' network. We show the results for the mapping accuracy, MA ($1^{\rm st}$ column), the edge prediction precision, EPP ($2^{\rm nd}$ column), the area under the receiver operating characteristic curve, AUC ($3^{\rm rd}$ column), the greedy routing score, GR ($4^{\rm th}$ column), the greedy success rate, GS ($5^{\rm th}$ column) and the greedy routing efficiency, GE ($6^{\rm th}$ column averaged over 4 different realizations. Beside the quality scores, we also display the real time in minutes ($7^{\rm th}$ column), the user time in minutes ($8^{\rm th}$ column) and the peak memory usage in GB ($9^{\rm th}$ column). In the top part of the table, we list the scores obtained for CLOVE with default settings ($1^{\rm st}$ row), for CLOVE with simulated annealing optimisation during the solution of the TSP problem ($2^{\rm nd}$ row) and for CLOVE with Louvain communities ($3^{\rm rd}$ row). For comparison, in the  $4^{\rm th}$ row we give the results for HMCS, followed by the scores for Mercator in the  $5^{\rm th}$ row and by hyperbolic ncMCE in the bottom row.}
\end{table}
\begin{table}
\begin{tabular}{|p{0.19\textwidth}|*{8}{>{\centering\arraybackslash}p{0.11\textwidth}|}}
    \thickhline
    & \makecell{MA} & \makecell{EPP} & \makecell{AUC} & \makecell{GR} & \makecell{GS} & \makecell{GE} & \makecell{Running \\ Time \\ (min)} & \makecell{Peak \\ Mem. \\ (GB)} \\
    \thickhline
    CLOVE & \cellcolor[RGB]{184.56333566915825, 225.24492855180793, 134.90623564168936}$0.398 \pm 0.014$ & \cellcolor[RGB]{161.08660729149327, 210.1263942067588, 106.26273514233397}$0.480 \pm 0.002$ & \cellcolor[RGB]{46.381562258327136, 108.93557536534337, 26.55401310701624}$0.961 \pm 0.001$ & \cellcolor[RGB]{150.57719808643714, 203.30449700347674, 93.5408187362134}$0.517 \pm 0.018$ & \cellcolor[RGB]{178.39412629850267, 221.36109952709822, 127.21394236134533}$0.420 \pm 0.016$ & \cellcolor[RGB]{234.07896127266108, 245.47987779678365, 217.3576170372813}$0.152 \pm 0.005$ & \cellcolor[RGB]{247.00059918347222, 246.99770313002315, 246.9992010887037}$0.144 \pm 0.003$ & \cellcolor[RGB]{247.32794404029846, 245.7428811788559, 246.56274127960205}$0.437 \pm 0.059$ \\
    CLOVE (with SA) & \cellcolor[RGB]{183.70106585755042, 224.80595503033976, 133.63813235387684}$0.401 \pm 0.016$ & \cellcolor[RGB]{160.09026968536705, 209.479648743133, 105.05664225070748}$0.484 \pm 0.003$ & \cellcolor[RGB]{46.378941125448364, 108.93240241501644, 26.553461289568077}$0.961 \pm 0.002$ & \cellcolor[RGB]{147.5920556752578, 201.3667729821849, 89.92722529110156}$0.528 \pm 0.019$ & \cellcolor[RGB]{176.3941319993332, 220.06285761360226, 124.79289663077178}$0.427 \pm 0.014$ & \cellcolor[RGB]{233.87537487335908, 245.4559264556893, 216.89056588594144}$0.154 \pm 0.004$ & \cellcolor[RGB]{247.00066311847223, 246.9974580458565, 246.99911584203704}$0.159 \pm 0.010$ & \cellcolor[RGB]{247.34688544273376, 245.67027246952057, 246.53748607635498}$0.463 \pm 0.158$ \\
    CLOVE (dendr.) & \cellcolor[RGB]{184.29007055773906, 225.1261176337996, 134.4666352450585}$0.399 \pm 0.006$ & \cellcolor[RGB]{180.04074413184614, 222.4299567171633, 129.20721658065585}$0.414 \pm 0.014$ & \cellcolor[RGB]{46.46848549972899, 109.04079823651404, 26.57231273678505}$0.961 \pm 0.000$ & \cellcolor[RGB]{177.78375250896084, 220.9648919795009, 126.4750688266368}$0.422 \pm 0.027$ & \cellcolor[RGB]{197.5261243905645, 230.8809236480715, 155.7594174978646}$0.341 \pm 0.022$ & \cellcolor[RGB]{236.42280415199585, 245.75562401788187, 222.73466834869635}$0.124 \pm 0.009$ & \cellcolor[RGB]{247.00033588953124, 246.99871242346353, 246.99955214729167}$0.081 \pm 0.008$ & \cellcolor[RGB]{247.29785680770874, 245.85821557044983, 246.60285758972168}$0.397 \pm 0.080$ \\
    CLOVE (Louvain) & \cellcolor[RGB]{185.46080472766724, 225.63513249029012, 136.3499902140734}$0.394 \pm 0.012$ & \cellcolor[RGB]{161.88130514399865, 210.64225070750788, 107.22473780589308}$0.478 \pm 0.002$ & \cellcolor[RGB]{46.26879080090281, 108.79906254846131, 26.530271747558487}$0.962 \pm 0.002$ & \cellcolor[RGB]{152.6646981446496, 204.6595409009129, 96.06779249089163}$0.510 \pm 0.013$ & \cellcolor[RGB]{180.82334708833386, 222.9379621450588, 130.15457805429887}$0.411 \pm 0.010$ & \cellcolor[RGB]{234.3479566311022, 245.51152430954144, 217.974724036058}$0.149 \pm 0.003$ & \cellcolor[RGB]{247.0007019419618, 246.99730922247974, 246.99906407738425}$0.168 \pm 0.011$ & \cellcolor[RGB]{247.26614665985107, 245.97977113723755, 246.64513778686523}$0.355 \pm 0.034$ \\
    CLOVE (k1 decomp.) & \cellcolor[RGB]{178.9014294500843, 221.69040157286173, 127.82804617641783}$0.418 \pm 0.008$ & \cellcolor[RGB]{184.6049608789745, 225.26302646911935, 134.9731979357416}$0.397 \pm 0.001$ & \cellcolor[RGB]{46.11097428524464, 108.60802150319088, 26.49704721794624}$0.963 \pm 0.001$ & \cellcolor[RGB]{139.07744005659816, 195.83974179112514, 79.62005901588199}$0.558 \pm 0.018$ & \cellcolor[RGB]{169.08763107519445, 215.32004122424905, 115.94818498576171}$0.452 \pm 0.013$ & \cellcolor[RGB]{231.66800717422936, 245.19623613814463, 211.82660469382031}$0.180 \pm 0.004$ & \cellcolor[RGB]{247.00054524369793, 246.99790989915797, 246.99927300840278}$0.131 \pm 0.023$ & \cellcolor[RGB]{247.35842537879944, 245.6260360479355, 246.52209949493408}$0.478 \pm 0.134$ \\
    HMCS & \cellcolor[RGB]{189.76848934697142, 227.50803884650932, 143.27974373208448}$0.375 \pm 0.004$ & \cellcolor[RGB]{222.90319627101715, 241.91443316131182, 196.58340269685368}$0.231 \pm 0.002$ & \cellcolor[RGB]{46.77648969324303, 109.4136454181363, 26.63715572489327}$0.959 \pm 0.001$ & \cellcolor[RGB]{207.84113743698165, 235.36571192912245, 172.3531341377531}$0.296 \pm 0.019$ & \cellcolor[RGB]{219.08752414144593, 240.25544527888954, 190.44514753189128}$0.247 \pm 0.014$ & \cellcolor[RGB]{239.20096224468594, 246.08246614643363, 229.10808985545594}$0.092 \pm 0.005$ & \cellcolor[RGB]{247.00014159328126, 246.9994572257552, 246.99981120895833}$0.034 \pm 0.006$ & \cellcolor[RGB]{247.22920727729797, 246.1213721036911, 246.69439029693604}$0.306 \pm 0.001$ \\
    Mercator & \cellcolor[RGB]{160.04346309930983, 209.44926552060463, 104.99998164653294}$0.484 \pm 0.002$ & \cellcolor[RGB]{195.67038455135673, 230.0740802397203, 152.77409688696517}$0.349 \pm 0.002$ & \cellcolor[RGB]{45.05750058507515, 107.33276386614361, 26.275263281068455}$0.968 \pm 0.000$ & \cellcolor[RGB]{220.77636355780606, 240.98972328600263, 193.16197615820977}$0.240 \pm 0.006$ & \cellcolor[RGB]{228.19001021747405, 244.2130479206409, 205.0882773063713}$0.208 \pm 0.005$ & \cellcolor[RGB]{240.7527715698171, 246.26503194939025, 232.66812301310978}$0.073 \pm 0.002$ & \cellcolor[RGB]{247.03409564651042, 246.86930002171007, 246.9545391379861}$8.183 \pm 0.014$ & \cellcolor[RGB]{247.54687237739563, 244.90365588665009, 246.27083683013916}$0.729 \pm 0.492$ \\
    ncMCE (hypyperbolic) & \cellcolor[RGB]{195.9383185654432, 230.19057328932314, 153.2051211704956}$0.348 \pm 0.000$ & \cellcolor[RGB]{228.08173797236554, 244.1659730314633, 204.91410021641414}$0.208 \pm 0.000$ & \cellcolor[RGB]{47.144468499749294, 109.85909344706494, 26.71462494731564}$0.957 \pm 0.000$ & \cellcolor[RGB]{232.25185303478833, 245.2649238864457, 213.1660157856909}$0.174 \pm 0.000$ & \cellcolor[RGB]{234.1667536415308, 245.49020631076831, 217.5590230599824}$0.151 \pm 0.000$ & \cellcolor[RGB]{242.1339646859525, 246.42752525717088, 235.83674251483217}$0.057 \pm 0.000$ & \cellcolor[RGB]{247.00204782932292, 246.9921499875955, 246.9972695609028}$0.491 \pm 0.017$ & \cellcolor[RGB]{247.9199390411377, 243.47356700897217, 245.7734146118164}$1.227 \pm 0.177$ \\
    \thickhline
\end{tabular}
\caption{Average quality scores of the considered hyperbolic embedding algorithms for the 'maayan-vidal' network. We show the results for the mapping accuracy, MA ($1^{\rm st}$ column), the edge prediction precision, EPP ($2^{\rm nd}$ column), the area under the receiver operating characteristic curve, AUC ($3^{\rm rd}$ column), the greedy routing score, GR ($4^{\rm th}$ column), the greedy success rate, GS ($5^{\rm th}$ column) and the greedy routing efficiency, GE ($6^{\rm th}$ column averaged over 4 different realizations. Beside the quality scores, we also display the real time in minutes ($7^{\rm th}$ column), the user time in minutes ($8^{\rm th}$ column) and the peak memory usage in GB ($9^{\rm th}$ column). In the top part of the table, we list the scores obtained for CLOVE with default settings ($1^{\rm st}$ row), for CLOVE with simulated annealing optimisation during the solution of the TSP problem ($2^{\rm nd}$ row) and for CLOVE with Louvain communities ($3^{\rm rd}$ row). For comparison, in the  $4^{\rm th}$ row we give the results for HMCS, followed by the scores for Mercator in the  $5^{\rm th}$ row and by hyperbolic ncMCE in the bottom row.}
\end{table}
\begin{table}
\begin{tabular}{|p{0.19\textwidth}|*{8}{>{\centering\arraybackslash}p{0.11\textwidth}|}}
    \thickhline
    & \makecell{MA} & \makecell{EPP} & \makecell{AUC} & \makecell{GR} & \makecell{GS} & \makecell{GE} & \makecell{Running \\ Time \\ (min)} & \makecell{Peak \\ Mem. \\ (GB)} \\
    \thickhline
    CLOVE & \cellcolor[RGB]{204.78631288861536, 234.03752734287625, 167.4388511686421}$0.310 \pm 0.019$ & \cellcolor[RGB]{121.52002053388088, 183.39681724845994, 61.49281314168377}$0.622 \pm 0.002$ & \cellcolor[RGB]{42.39112545908679, 104.10504660836821, 25.71392114928143}$0.982 \pm 0.001$ & \cellcolor[RGB]{227.89897118953408, 244.0865092128409, 204.62008408751134}$0.209 \pm 0.009$ & \cellcolor[RGB]{231.0139260714666, 245.11928542017253, 210.3260656933646}$0.188 \pm 0.008$ & \cellcolor[RGB]{241.97810568636018, 246.40918890427767, 235.47918363341455}$0.059 \pm 0.002$ & \cellcolor[RGB]{247.00030326744792, 246.998837474783, 246.99959564340278}$0.073 \pm 0.009$ & \cellcolor[RGB]{247.25960564613342, 246.0048450231552, 246.65385913848877}$0.346 \pm 0.079$ \\
    CLOVE (with SA) & \cellcolor[RGB]{205.59819552978422, 234.39051979555836, 168.74492324356595}$0.306 \pm 0.005$ & \cellcolor[RGB]{121.3275154004107, 183.235112936345, 61.36960985626285}$0.623 \pm 0.002$ & \cellcolor[RGB]{42.43736290413868, 104.1610182523784, 25.723655348239724}$0.982 \pm 0.001$ & \cellcolor[RGB]{227.35486061657886, 243.84993939851253, 203.74477577449642}$0.212 \pm 0.009$ & \cellcolor[RGB]{230.83221207296958, 245.0979073027023, 209.90919240269494}$0.190 \pm 0.009$ & \cellcolor[RGB]{241.94520231889368, 246.40531791986984, 235.40369943746197}$0.059 \pm 0.002$ & \cellcolor[RGB]{247.00035049875, 246.99865642145832, 246.99953266833333}$0.084 \pm 0.017$ & \cellcolor[RGB]{247.28807854652405, 245.89569890499115, 246.61589527130127}$0.384 \pm 0.106$ \\
    CLOVE (dendr.) & \cellcolor[RGB]{206.40726942262117, 234.74229105331355, 170.04647689726013}$0.303 \pm 0.017$ & \cellcolor[RGB]{125.851386036961, 187.03516427104725, 64.26488706365504}$0.605 \pm 0.007$ & \cellcolor[RGB]{42.194703760993406, 103.86727297383412, 25.672569212840717}$0.983 \pm 0.000$ & \cellcolor[RGB]{228.25182276069248, 244.23992293943152, 205.1877148758966}$0.208 \pm 0.006$ & \cellcolor[RGB]{231.21544505164007, 245.14299353548708, 210.78837394199783}$0.186 \pm 0.005$ & \cellcolor[RGB]{242.03903884393515, 246.4163575110512, 235.61897146549825}$0.058 \pm 0.001$ & \cellcolor[RGB]{247.00022773513888, 246.99912701530093, 246.99969635314815}$0.055 \pm 0.002$ & \cellcolor[RGB]{247.35682678222656, 245.63216400146484, 246.52423095703125}$0.476 \pm 0.099$ \\
    CLOVE (Louvain) & \cellcolor[RGB]{207.784318724019, 235.34100814087782, 172.26173012124792}$0.297 \pm 0.006$ & \cellcolor[RGB]{120.71791581108832, 182.7230492813142, 60.97946611909652}$0.625 \pm 0.003$ & \cellcolor[RGB]{42.14762932504812, 103.81028813032141, 25.66265880527329}$0.983 \pm 0.000$ & \cellcolor[RGB]{226.6052001735513, 243.5240000754571, 202.53880027919126}$0.215 \pm 0.019$ & \cellcolor[RGB]{230.73072265694742, 245.0859673714056, 209.6763637424088}$0.191 \pm 0.016$ & \cellcolor[RGB]{241.8894037938924, 246.39875338751676, 235.27569105657665}$0.060 \pm 0.004$ & \cellcolor[RGB]{247.00034219112848, 246.99868826734087, 246.99954374516204}$0.082 \pm 0.016$ & \cellcolor[RGB]{247.2484748363495, 246.04751312732697, 246.66870021820068}$0.331 \pm 0.060$ \\
    CLOVE (k1 decomp.) & \cellcolor[RGB]{197.2665959861559, 230.76808521137212, 155.3419152820769}$0.342 \pm 0.010$ & \cellcolor[RGB]{245.99640657084188, 246.88193018480493, 244.6976386036961}$0.012 \pm 0.002$ & \cellcolor[RGB]{42.53409239062059, 104.27811184127756, 25.744019450656968}$0.981 \pm 0.000$ & \cellcolor[RGB]{220.82914035493135, 241.01266971953538, 193.24687796228088}$0.240 \pm 0.023$ & \cellcolor[RGB]{226.87810850942057, 243.6426558736611, 202.97782673254616}$0.214 \pm 0.020$ & \cellcolor[RGB]{240.5523757078306, 246.24145596562713, 232.20839132972904}$0.076 \pm 0.006$ & \cellcolor[RGB]{247.000200279184, 246.9992322631279, 246.99973296108797}$0.048 \pm 0.005$ & \cellcolor[RGB]{247.29123497009277, 245.88359928131104, 246.61168670654297}$0.388 \pm 0.039$ \\
    HMCS & \cellcolor[RGB]{204.40987187064738, 233.8738573350641, 166.83327213973706}$0.311 \pm 0.000$ & \cellcolor[RGB]{190.39938398357287, 227.78234086242298, 144.2946611909651}$0.372 \pm 0.006$ & \cellcolor[RGB]{42.30033409548369, 103.99514127348026, 25.694807177996566}$0.983 \pm 0.000$ & \cellcolor[RGB]{232.71357131403832, 245.3192436840045, 214.22525183808793}$0.168 \pm 0.010$ & \cellcolor[RGB]{234.2237214360948, 245.49690840424645, 217.68971388280568}$0.150 \pm 0.008$ & \cellcolor[RGB]{242.95495980594092, 246.524112918346, 237.7202019077468}$0.048 \pm 0.002$ & \cellcolor[RGB]{247.0000732296354, 246.99971928639758, 246.99990236048612}$0.018 \pm 0.002$ & \cellcolor[RGB]{247.2253270149231, 246.1362464427948, 246.69956398010254}$0.300 \pm 0.000$ \\
    Mercator & \cellcolor[RGB]{192.0899754111578, 228.51738061354686, 147.01430827012342}$0.365 \pm 0.002$ & \cellcolor[RGB]{135.38321355236138, 193.44173511293633, 75.14810061601642}$0.571 \pm 0.004$ & \cellcolor[RGB]{41.47669009268256, 102.99809853324732, 25.52140844056475}$0.987 \pm 0.000$ & \cellcolor[RGB]{236.89795959272027, 245.81152465796708, 223.82473083035828}$0.119 \pm 0.007$ & \cellcolor[RGB]{237.42722299433785, 245.87379094051033, 225.03892333995154}$0.113 \pm 0.006$ & \cellcolor[RGB]{243.99322544236318, 246.64626181674862, 240.1021054265979}$0.035 \pm 0.002$ & \cellcolor[RGB]{247.0086601954514, 246.966802584103, 246.9884530727315}$2.078 \pm 0.015$ & \cellcolor[RGB]{247.43863916397095, 245.3185498714447, 246.41514778137207}$0.585 \pm 0.397$ \\
    ncMCE (hypyperbolic) & \cellcolor[RGB]{210.89255192598475, 236.69241388086294, 177.26193135919283}$0.283 \pm 0.000$ & \cellcolor[RGB]{214.75128336755645, 238.3701232032854, 183.46945585215605}$0.266 \pm 0.000$ & \cellcolor[RGB]{42.335210525322914, 104.03736010960142, 25.70214958427851}$0.982 \pm 0.000$ & \cellcolor[RGB]{237.6320853042572, 245.89789238873615, 225.50890158035475}$0.110 \pm 0.000$ & \cellcolor[RGB]{238.28122133220958, 245.97426133320113, 226.99809599742198}$0.103 \pm 0.000$ & \cellcolor[RGB]{244.13566879156974, 246.66301985783173, 240.42888722771886}$0.034 \pm 0.000$ & \cellcolor[RGB]{247.00078725246527, 246.9969821988831, 246.99895033004628}$0.189 \pm 0.041$ & \cellcolor[RGB]{247.66335248947144, 244.4571487903595, 246.1155300140381}$0.884 \pm 0.391$ \\
    \thickhline
\end{tabular}
\caption{Average quality scores of the considered hyperbolic embedding algorithms for the 'moreno\_propro' network. We show the results for the mapping accuracy, MA ($1^{\rm st}$ column), the edge prediction precision, EPP ($2^{\rm nd}$ column), the area under the receiver operating characteristic curve, AUC ($3^{\rm rd}$ column), the greedy routing score, GR ($4^{\rm th}$ column), the greedy success rate, GS ($5^{\rm th}$ column) and the greedy routing efficiency, GE ($6^{\rm th}$ column averaged over 4 different realizations. Beside the quality scores, we also display the real time in minutes ($7^{\rm th}$ column), the user time in minutes ($8^{\rm th}$ column) and the peak memory usage in GB ($9^{\rm th}$ column). In the top part of the table, we list the scores obtained for CLOVE with default settings ($1^{\rm st}$ row), for CLOVE with simulated annealing optimisation during the solution of the TSP problem ($2^{\rm nd}$ row) and for CLOVE with Louvain communities ($3^{\rm rd}$ row). For comparison, in the  $4^{\rm th}$ row we give the results for HMCS, followed by the scores for Mercator in the  $5^{\rm th}$ row and by hyperbolic ncMCE in the bottom row.}
\end{table}
\begin{table}
\begin{tabular}{|p{0.19\textwidth}|*{8}{>{\centering\arraybackslash}p{0.11\textwidth}|}}
    \thickhline
    & \makecell{MA} & \makecell{EPP} & \makecell{AUC} & \makecell{GR} & \makecell{GS} & \makecell{GE} & \makecell{Running \\ Time \\ (min)} & \makecell{Peak \\ Mem. \\ (GB)} \\
    \thickhline
    CLOVE & \cellcolor[RGB]{161.58893530203466, 210.45246677500495, 106.8708164182525}$0.479 \pm 0.008$ & \cellcolor[RGB]{178.76297468354431, 221.60052742616034, 127.66044303797469}$0.418 \pm 0.001$ & \cellcolor[RGB]{40.80548791465549, 102.18559063353032, 25.380102718874838}$0.990 \pm 0.000$ & \cellcolor[RGB]{62.969996310114084, 129.0163113227697, 30.046315012655597}$0.874 \pm 0.009$ & \cellcolor[RGB]{70.55081547438483, 138.1930924163606, 31.642276941975755}$0.834 \pm 0.008$ & \cellcolor[RGB]{192.12081769034057, 228.53079030014806, 147.06392411054784}$0.365 \pm 0.004$ & \cellcolor[RGB]{247.00040614578126, 246.99844310783854, 246.99945847229168}$0.097 \pm 0.004$ & \cellcolor[RGB]{247.32792687416077, 245.74294698238373, 246.56276416778564}$0.437 \pm 0.061$ \\
    CLOVE (with SA) & \cellcolor[RGB]{162.1066408867, 210.7885212773316, 107.49751265232108}$0.477 \pm 0.007$ & \cellcolor[RGB]{178.6066455696202, 221.49905063291135, 127.4712025316455}$0.419 \pm 0.001$ & \cellcolor[RGB]{40.82427445641945, 102.20833223671829, 25.384057780298832}$0.990 \pm 0.000$ & \cellcolor[RGB]{65.75581784734467, 132.3886216046804, 30.632803757335722}$0.859 \pm 0.011$ & \cellcolor[RGB]{72.94036151730239, 141.0857007841029, 32.145339266800505}$0.821 \pm 0.010$ & \cellcolor[RGB]{193.4211372125455, 229.0961466141502, 149.15574247235577}$0.359 \pm 0.004$ & \cellcolor[RGB]{247.0004724676389, 246.99818887405092, 246.99937004314816}$0.113 \pm 0.033$ & \cellcolor[RGB]{247.34947395324707, 245.66034984588623, 246.5340347290039}$0.466 \pm 0.113$ \\
    CLOVE (dendr.) & \cellcolor[RGB]{162.21276982691202, 210.8574119929078, 107.62598452731456}$0.476 \pm 0.004$ & \cellcolor[RGB]{181.41455696202527, 223.3217299578059, 130.8702531645569}$0.409 \pm 0.004$ & \cellcolor[RGB]{40.677359735029626, 102.03048810029902, 25.353128365269395}$0.991 \pm 0.000$ & \cellcolor[RGB]{66.55134601174298, 133.35162938263625, 30.800283370893258}$0.855 \pm 0.006$ & \cellcolor[RGB]{74.43753374983947, 142.89806717085833, 32.46053342101884}$0.813 \pm 0.006$ & \cellcolor[RGB]{194.41067062625507, 229.5263785331544, 150.74760057267122}$0.355 \pm 0.002$ & \cellcolor[RGB]{247.00029459479165, 246.99887071996528, 246.99960720694443}$0.071 \pm 0.003$ & \cellcolor[RGB]{247.2964005470276, 245.8637979030609, 246.60479927062988}$0.395 \pm 0.064$ \\
    CLOVE (Louvain) & \cellcolor[RGB]{162.6596148521096, 211.14746928996587, 108.16690218939583}$0.475 \pm 0.004$ & \cellcolor[RGB]{178.89525316455695, 221.6863924050633, 127.82056962025315}$0.418 \pm 0.001$ & \cellcolor[RGB]{40.68238906694152, 102.0365762389292, 25.35418717198769}$0.991 \pm 0.000$ & \cellcolor[RGB]{64.02848233661578, 130.2976365127454, 30.269154176129636}$0.868 \pm 0.006$ & \cellcolor[RGB]{72.14130700525502, 140.11842426951924, 31.977117264264212}$0.826 \pm 0.006$ & \cellcolor[RGB]{193.11853844243254, 228.9645819314924, 148.66895314652191}$0.360 \pm 0.003$ & \cellcolor[RGB]{247.00044812435763, 246.9982821899624, 246.9994025008565}$0.108 \pm 0.003$ & \cellcolor[RGB]{247.30631113052368, 245.82580733299255, 246.59158515930176}$0.408 \pm 0.133$ \\
    CLOVE (k1 decomp.) & \cellcolor[RGB]{163.4702352296423, 211.6736614648555, 109.14817948851434}$0.472 \pm 0.009$ & \cellcolor[RGB]{247.0, 247.0, 247.0}$0.000 \pm 0.000$ & \cellcolor[RGB]{41.14345240584286, 102.59470554391504, 25.451253138072182}$0.989 \pm 0.001$ & \cellcolor[RGB]{76.45109973156465, 145.3355417803151, 32.88444204875045}$0.803 \pm 0.013$ & \cellcolor[RGB]{84.68481471974391, 152.45524436458487, 37.918281420636106}$0.769 \pm 0.013$ & \cellcolor[RGB]{166.33490860279002, 213.5331862860216, 112.61594199285108}$0.462 \pm 0.008$ & \cellcolor[RGB]{247.00040030397568, 246.9984655014265, 246.99946626136574}$0.096 \pm 0.015$ & \cellcolor[RGB]{247.31739473342896, 245.78332018852234, 246.57680702209473}$0.423 \pm 0.115$ \\
    HMCS & \cellcolor[RGB]{162.87460291494187, 211.28702294478683, 108.42715089703488}$0.474 \pm 0.005$ & \cellcolor[RGB]{219.47046413502107, 240.42194092827003, 191.06118143459912}$0.246 \pm 0.006$ & \cellcolor[RGB]{40.78268037068678, 102.15798150135768, 25.375301130670902}$0.991 \pm 0.000$ & \cellcolor[RGB]{130.08169390538205, 190.00039779823044, 68.73047156967301}$0.589 \pm 0.015$ & \cellcolor[RGB]{139.23588026495898, 195.94258894392073, 79.81185505758191}$0.557 \pm 0.014$ & \cellcolor[RGB]{220.06103080717608, 240.6787090465983, 192.0112234724137}$0.243 \pm 0.006$ & \cellcolor[RGB]{247.00022399076389, 246.99914136873844, 246.99970134564813}$0.054 \pm 0.007$ & \cellcolor[RGB]{247.24258399009705, 246.070094704628, 246.6765546798706}$0.323 \pm 0.007$ \\
    Mercator & \cellcolor[RGB]{153.04954144917235, 204.90935146700662, 96.5336554384718}$0.509 \pm 0.001$ & \cellcolor[RGB]{140.4443037974684, 196.7270042194093, 81.27468354430383}$0.553 \pm 0.002$ & \cellcolor[RGB]{39.667365456880866, 100.80786344780316, 25.140497990922288}$0.996 \pm 0.000$ & \cellcolor[RGB]{63.534307107449315, 129.69942439322813, 30.165117285778802}$0.871 \pm 0.005$ & \cellcolor[RGB]{66.06174239937421, 132.75895132555826, 30.697208926184043}$0.858 \pm 0.005$ & \cellcolor[RGB]{193.4530299340943, 229.1100130148236, 149.20704815484737}$0.359 \pm 0.003$ & \cellcolor[RGB]{247.22911128659723, 246.12174006804398, 246.69451828453703}$54.987 \pm 0.066$ & \cellcolor[RGB]{252.29051065444946, 226.71970915794373, 239.94598579406738}$7.054 \pm 6.917$ \\
    ncMCE (hypyperbolic) & \cellcolor[RGB]{169.2173689435893, 215.40425703355794, 116.10523608960807}$0.452 \pm 0.000$ & \cellcolor[RGB]{239.78037974683545, 246.1506329113924, 230.43734177215188}$0.085 \pm 0.000$ & \cellcolor[RGB]{41.461918745163146, 102.98021742835539, 25.518298683192242}$0.987 \pm 0.000$ & \cellcolor[RGB]{222.6673543755678, 241.81189320676862, 196.20400486504388}$0.232 \pm 0.000$ & \cellcolor[RGB]{226.66955549872938, 243.55198065162148, 202.64232841099945}$0.214 \pm 0.000$ & \cellcolor[RGB]{239.0569857333973, 246.06552773334087, 228.7777908001468}$0.093 \pm 0.000$ & \cellcolor[RGB]{247.01164979772568, 246.9553424420515, 246.98446693636575}$2.796 \pm 0.026$ & \cellcolor[RGB]{252.69610357284546, 222.93636250495908, 238.46818125247955}$8.203 \pm 6.624$ \\
    \thickhline
\end{tabular}
\caption{Average quality scores of the considered hyperbolic embedding algorithms for the 'pajek-erdos' network. We show the results for the mapping accuracy, MA ($1^{\rm st}$ column), the edge prediction precision, EPP ($2^{\rm nd}$ column), the area under the receiver operating characteristic curve, AUC ($3^{\rm rd}$ column), the greedy routing score, GR ($4^{\rm th}$ column), the greedy success rate, GS ($5^{\rm th}$ column) and the greedy routing efficiency, GE ($6^{\rm th}$ column averaged over 4 different realizations. Beside the quality scores, we also display the real time in minutes ($7^{\rm th}$ column), the user time in minutes ($8^{\rm th}$ column) and the peak memory usage in GB ($9^{\rm th}$ column). In the top part of the table, we list the scores obtained for CLOVE with default settings ($1^{\rm st}$ row), for CLOVE with simulated annealing optimisation during the solution of the TSP problem ($2^{\rm nd}$ row) and for CLOVE with Louvain communities ($3^{\rm rd}$ row). For comparison, in the  $4^{\rm th}$ row we give the results for HMCS, followed by the scores for Mercator in the  $5^{\rm th}$ row and by hyperbolic ncMCE in the bottom row.}
\end{table}
\begin{table}
\begin{tabular}{|p{0.19\textwidth}|*{8}{>{\centering\arraybackslash}p{0.11\textwidth}|}}
    \thickhline
    & \makecell{MA} & \makecell{EPP} & \makecell{AUC} & \makecell{GR} & \makecell{GS} & \makecell{GE} & \makecell{Running \\ Time \\ (min)} & \makecell{Peak \\ Mem. \\ (GB)} \\
    \thickhline
    CLOVE & \cellcolor[RGB]{188.90714079375886, 227.13353947554734, 141.8940960595251}$0.379 \pm 0.008$ & \cellcolor[RGB]{163.91346087887058, 211.96136934242477, 109.68471580073808}$0.470 \pm 0.006$ & \cellcolor[RGB]{44.67945340339984, 106.8751278041156, 26.195674400715756}$0.970 \pm 0.001$ & \cellcolor[RGB]{134.5308985576271, 192.88847801109128, 74.1163508855486}$0.574 \pm 0.015$ & \cellcolor[RGB]{159.24705836206275, 208.93230104204073, 104.03591275407597}$0.487 \pm 0.010$ & \cellcolor[RGB]{231.43724753928967, 245.1690879457988, 211.2972149430763}$0.183 \pm 0.003$ & \cellcolor[RGB]{247.00181975935763, 246.99302425579572, 246.9975736541898}$0.437 \pm 0.025$ & \cellcolor[RGB]{247.38660788536072, 245.51800310611725, 246.48452281951904}$0.515 \pm 0.103$ \\
    CLOVE (with SA) & \cellcolor[RGB]{190.04087164517352, 227.62646593268414, 143.71792395093132}$0.374 \pm 0.024$ & \cellcolor[RGB]{162.99606422093868, 211.36586624867948, 108.57418300429418}$0.474 \pm 0.006$ & \cellcolor[RGB]{44.8034047078631, 107.02517412004481, 26.221769412181708}$0.969 \pm 0.002$ & \cellcolor[RGB]{139.16350189518135, 195.89560649336335, 79.72423913627216}$0.557 \pm 0.015$ & \cellcolor[RGB]{162.6957479490903, 211.17092410730424, 108.21064225416194}$0.475 \pm 0.016$ & \cellcolor[RGB]{231.76112869482674, 245.20719161115608, 212.04023641754367}$0.179 \pm 0.005$ & \cellcolor[RGB]{247.00182970399305, 246.9929861346933, 246.99756039467593}$0.439 \pm 0.028$ & \cellcolor[RGB]{247.3913450241089, 245.49984407424927, 246.47820663452148}$0.522 \pm 0.123$ \\
    CLOVE (dendr.) & \cellcolor[RGB]{192.5491848275591, 228.71703688154741, 147.7530364617255}$0.363 \pm 0.014$ & \cellcolor[RGB]{169.51314841745665, 215.5962542358929, 116.46328492639488}$0.451 \pm 0.001$ & \cellcolor[RGB]{44.8815275461289, 107.11974387162971, 26.238216325500822}$0.969 \pm 0.001$ & \cellcolor[RGB]{192.365203989915, 228.6370452130065, 147.4570672881241}$0.364 \pm 0.014$ & \cellcolor[RGB]{201.98020983253286, 232.81748253588387, 162.9246853827703}$0.322 \pm 0.012$ & \cellcolor[RGB]{236.19925071052086, 245.72932361300246, 222.22181045354785}$0.127 \pm 0.004$ & \cellcolor[RGB]{247.00075886440973, 246.99709101976273, 246.99898818078705}$0.182 \pm 0.008$ & \cellcolor[RGB]{247.38290429115295, 245.532200217247, 246.4894609451294}$0.511 \pm 0.088$ \\
    CLOVE (Louvain) & \cellcolor[RGB]{188.4524988950198, 226.93586908479122, 141.16271561372753}$0.381 \pm 0.021$ & \cellcolor[RGB]{163.4867761253413, 211.68439853750226, 109.16820267804474}$0.472 \pm 0.002$ & \cellcolor[RGB]{44.708472309316164, 106.91025595338272, 26.20178364406656}$0.970 \pm 0.003$ & \cellcolor[RGB]{137.02803592930314, 194.50942683130202, 77.1392013881038}$0.565 \pm 0.024$ & \cellcolor[RGB]{160.6851336427417, 209.8657885049376, 105.77674072542416}$0.482 \pm 0.016$ & \cellcolor[RGB]{231.5480146562876, 245.18211937132796, 211.5513277408951}$0.182 \pm 0.005$ & \cellcolor[RGB]{247.00200662133682, 246.99230795154224, 246.99732450488426}$0.482 \pm 0.011$ & \cellcolor[RGB]{247.42177414894104, 245.383199095726, 246.4376344680786}$0.562 \pm 0.087$ \\
    CLOVE (k1 decomp.) & \cellcolor[RGB]{192.97852301009934, 228.90370565656494, 148.44371092929023}$0.361 \pm 0.011$ & \cellcolor[RGB]{169.23651202513406, 215.41668324438527, 116.12840929358336}$0.452 \pm 0.001$ & \cellcolor[RGB]{45.08746868875565, 107.36904104428315, 26.281572355527505}$0.968 \pm 0.001$ & \cellcolor[RGB]{139.66718714668502, 196.22256007767274, 80.3339633880924}$0.556 \pm 0.019$ & \cellcolor[RGB]{164.66266839762494, 212.44769703003726, 110.59165121817757}$0.468 \pm 0.016$ & \cellcolor[RGB]{231.74232308745403, 245.2049791867593, 211.99709414180631}$0.180 \pm 0.006$ & \cellcolor[RGB]{247.00168411319444, 246.99354423275463, 246.99775451574075}$0.404 \pm 0.011$ & \cellcolor[RGB]{247.38398790359497, 245.5280463695526, 246.48801612854004}$0.512 \pm 0.113$ \\
    HMCS & \cellcolor[RGB]{214.7658069073898, 238.37643778582165, 183.49281980754012}$0.266 \pm 0.026$ & \cellcolor[RGB]{224.34868773065895, 242.54290770898214, 198.90875852323396}$0.225 \pm 0.008$ & \cellcolor[RGB]{48.608984822077986, 111.63192899514704, 27.02294417306905}$0.949 \pm 0.003$ & \cellcolor[RGB]{198.1964501741136, 231.17236964091896, 156.83776767140017}$0.338 \pm 0.015$ & \cellcolor[RGB]{207.47712098545037, 235.20744390671754, 171.76754245485495}$0.298 \pm 0.015$ & \cellcolor[RGB]{236.88705019331726, 245.8102411992138, 223.79970338466904}$0.119 \pm 0.005$ & \cellcolor[RGB]{247.00136870159722, 246.99475331054398, 246.99817506453704}$0.328 \pm 0.007$ & \cellcolor[RGB]{247.3207242488861, 245.77055704593658, 246.57236766815186}$0.428 \pm 0.001$ \\
    Mercator & \cellcolor[RGB]{127.43952379330335, 188.28530491846007, 65.53205511820933}$0.598 \pm 0.003$ & \cellcolor[RGB]{98.66762131460166, 164.2008019042654, 46.867277641345055}$0.713 \pm 0.002$ & \cellcolor[RGB]{41.843055408031404, 103.44159338866959, 25.59853798063819}$0.985 \pm 0.000$ & \cellcolor[RGB]{171.37747903429133, 216.8064337591014, 118.7201061994053}$0.444 \pm 0.011$ & \cellcolor[RGB]{184.53684535470268, 225.2334110237838, 134.863620788}$0.398 \pm 0.008$ & \cellcolor[RGB]{234.67967283753921, 245.55054974559286, 218.73572003906054}$0.145 \pm 0.003$ & \cellcolor[RGB]{247.18743721543402, 246.28149067416956, 246.75008371275464}$44.985 \pm 1.202$ & \cellcolor[RGB]{251.18567872047424, 230.95489823818207, 241.41909503936768}$5.581 \pm 7.492$ \\
    ncMCE (hypyperbolic) & \cellcolor[RGB]{206.99330322883662, 234.99708836036376, 170.9892269333459}$0.300 \pm 0.000$ & \cellcolor[RGB]{201.3333836381347, 232.53625375571073, 161.88413889612974}$0.325 \pm 0.000$ & \cellcolor[RGB]{53.60184519023376, 117.67591786186192, 28.07407267162816}$0.923 \pm 0.000$ & \cellcolor[RGB]{215.69019197720692, 238.77834433791605, 184.9798740502894}$0.262 \pm 0.000$ & \cellcolor[RGB]{225.2757851036404, 242.9459935233219, 200.4001760362911}$0.221 \pm 0.000$ & \cellcolor[RGB]{239.52351559594402, 246.12041359952283, 229.84806519069508}$0.088 \pm 0.000$ & \cellcolor[RGB]{247.01113504902779, 246.95731564539352, 246.98515326796297}$2.672 \pm 0.083$ & \cellcolor[RGB]{249.80485606193542, 236.24805176258087, 243.26019191741943}$3.740 \pm 0.050$ \\
    \thickhline
\end{tabular}
\caption{Average quality scores of the considered hyperbolic embedding algorithms for the 'reactome' network. We show the results for the mapping accuracy, MA ($1^{\rm st}$ column), the edge prediction precision, EPP ($2^{\rm nd}$ column), the area under the receiver operating characteristic curve, AUC ($3^{\rm rd}$ column), the greedy routing score, GR ($4^{\rm th}$ column), the greedy success rate, GS ($5^{\rm th}$ column) and the greedy routing efficiency, GE ($6^{\rm th}$ column averaged over 4 different realizations. Beside the quality scores, we also display the real time in minutes ($7^{\rm th}$ column), the user time in minutes ($8^{\rm th}$ column) and the peak memory usage in GB ($9^{\rm th}$ column). In the top part of the table, we list the scores obtained for CLOVE with default settings ($1^{\rm st}$ row), for CLOVE with simulated annealing optimisation during the solution of the TSP problem ($2^{\rm nd}$ row) and for CLOVE with Louvain communities ($3^{\rm rd}$ row). For comparison, in the  $4^{\rm th}$ row we give the results for HMCS, followed by the scores for Mercator in the  $5^{\rm th}$ row and by hyperbolic ncMCE in the bottom row.}
\end{table}
\begin{table}
\begin{tabular}{|p{0.19\textwidth}|*{8}{>{\centering\arraybackslash}p{0.11\textwidth}|}}
    \thickhline
    & \makecell{MA} & \makecell{EPP} & \makecell{AUC} & \makecell{GR} & \makecell{GS} & \makecell{GE} & \makecell{Running \\ Time \\ (min)} & \makecell{Peak \\ Mem. \\ (GB)} \\
    \thickhline
    CLOVE & \cellcolor[RGB]{208.29700171320775, 235.56391378835121, 173.08648101689943}$0.294 \pm 0.001$ & \cellcolor[RGB]{157.11892256448078, 207.55087955939982, 101.45974836752936}$0.494 \pm 0.002$ & \cellcolor[RGB]{46.906043128125056, 109.57047326036191, 26.664430132236852}$0.958 \pm 0.001$ & \cellcolor[RGB]{235.15668242432375, 245.6066685205087, 219.8300361499192}$0.139 \pm 0.005$ & \cellcolor[RGB]{237.82019377543892, 245.92002279711048, 225.94044454365402}$0.108 \pm 0.004$ & \cellcolor[RGB]{244.4624984770324, 246.70147040906264, 241.17867297672134}$0.030 \pm 0.001$ & \cellcolor[RGB]{247.00280923902778, 246.98923125039352, 246.99625434796297}$0.674 \pm 0.029$ & \cellcolor[RGB]{247.2906563282013, 245.8858174085617, 246.61245822906494}$0.388 \pm 0.067$ \\
    CLOVE (with SA) & \cellcolor[RGB]{208.6631458723569, 235.72310690102475, 173.67549553379155}$0.293 \pm 0.001$ & \cellcolor[RGB]{156.68065405278023, 207.26638947285733, 100.92921280073395}$0.496 \pm 0.001$ & \cellcolor[RGB]{46.954963987989956, 109.62969324861942, 26.674729260629462}$0.958 \pm 0.000$ & \cellcolor[RGB]{234.99958543241448, 245.58818652146053, 219.46963716848023}$0.141 \pm 0.005$ & \cellcolor[RGB]{237.68992626410534, 245.9046972075418, 225.6415955470652}$0.110 \pm 0.004$ & \cellcolor[RGB]{244.43286504626056, 246.69798412308947, 241.11069040024483}$0.030 \pm 0.001$ & \cellcolor[RGB]{247.0029110384375, 246.98884101932293, 246.99611861541666}$0.699 \pm 0.038$ & \cellcolor[RGB]{247.33171486854553, 245.72842633724213, 246.55771350860596}$0.442 \pm 0.112$ \\
    CLOVE (dendr.) & \cellcolor[RGB]{209.3493361605008, 236.02145050456556, 174.77936686689256}$0.290 \pm 0.002$ & \cellcolor[RGB]{181.18862803072622, 223.17507433573456, 130.5967602477212}$0.410 \pm 0.008$ & \cellcolor[RGB]{47.23291769849723, 109.96616352975981, 26.733245831262575}$0.957 \pm 0.001$ & \cellcolor[RGB]{240.6655737349734, 246.2547733805851, 232.4680809214096}$0.075 \pm 0.015$ & \cellcolor[RGB]{241.9986846080388, 246.4116099538869, 235.52639410079487}$0.059 \pm 0.010$ & \cellcolor[RGB]{245.66379061093912, 246.8427988954046, 243.93457846038976}$0.016 \pm 0.003$ & \cellcolor[RGB]{247.00237476081597, 246.99089675020542, 246.99683365224536}$0.570 \pm 0.056$ & \cellcolor[RGB]{247.32971787452698, 245.73608148097992, 246.56037616729736}$0.440 \pm 0.081$ \\
    CLOVE (Louvain) & \cellcolor[RGB]{209.1226569704493, 235.92289433497794, 174.41470903941843}$0.291 \pm 0.001$ & \cellcolor[RGB]{155.98945735233434, 206.81771793046263, 100.09250100545736}$0.498 \pm 0.002$ & \cellcolor[RGB]{47.069165104028855, 109.76793670487703, 26.69877160084818}$0.958 \pm 0.001$ & \cellcolor[RGB]{235.9005349683847, 245.69418058451586, 221.53652139805905}$0.131 \pm 0.004$ & \cellcolor[RGB]{238.33728956297205, 245.98085759564378, 227.1267231150535}$0.102 \pm 0.003$ & \cellcolor[RGB]{244.5889361152301, 246.71634542532118, 241.46873579376313}$0.028 \pm 0.001$ & \cellcolor[RGB]{247.00304071524306, 246.9883439249016, 246.99594571300926}$0.730 \pm 0.051$ & \cellcolor[RGB]{247.32703709602356, 245.74635779857635, 246.56395053863525}$0.436 \pm 0.108$ \\
    CLOVE (k1 decomp.) & \cellcolor[RGB]{212.78885291726544, 237.51689257272412, 180.3125025190792}$0.275 \pm 0.001$ & \cellcolor[RGB]{159.6952997090765, 209.22326472343562, 104.57852070046103}$0.485 \pm 0.002$ & \cellcolor[RGB]{46.87306457704304, 109.53055185642052, 26.65748727937748}$0.959 \pm 0.001$ & \cellcolor[RGB]{233.43752731642329, 245.40441497840274, 215.88609207885344}$0.160 \pm 0.005$ & \cellcolor[RGB]{236.55742823449765, 245.771462145235, 223.0435118320828}$0.123 \pm 0.003$ & \cellcolor[RGB]{243.86925383126714, 246.63167692132555, 239.81769996584816}$0.037 \pm 0.001$ & \cellcolor[RGB]{247.00181485479166, 246.99304305663196, 246.9975801936111}$0.436 \pm 0.011$ & \cellcolor[RGB]{247.35558152198792, 245.63693749904633, 246.5258913040161}$0.474 \pm 0.088$ \\
    HMCS & \cellcolor[RGB]{223.46959685483574, 242.16069428471118, 197.49456885343142}$0.228 \pm 0.011$ & \cellcolor[RGB]{234.8824148021283, 245.57440174142687, 219.20083395782373}$0.143 \pm 0.002$ & \cellcolor[RGB]{48.36111700046338, 111.33187847424514, 26.970761473781764}$0.951 \pm 0.001$ & \cellcolor[RGB]{242.8393632130291, 246.5105133191799, 237.45500972400794}$0.049 \pm 0.005$ & \cellcolor[RGB]{243.5750411086088, 246.59706365983632, 239.14274136680842}$0.040 \pm 0.004$ & \cellcolor[RGB]{246.02416930885906, 246.88519638927752, 244.76132959091197}$0.011 \pm 0.001$ & \cellcolor[RGB]{247.0007620285243, 246.99707889065684, 246.99898396196758}$0.183 \pm 0.002$ & \cellcolor[RGB]{247.25423622131348, 246.02542781829834, 246.66101837158203}$0.339 \pm 0.007$ \\
    Mercator & \cellcolor[RGB]{176.05041353056964, 219.8397421163347, 124.37681637911062}$0.428 \pm 0.005$ & \cellcolor[RGB]{201.91558346989248, 232.78938411734455, 162.82072123417487}$0.322 \pm 0.002$ & \cellcolor[RGB]{43.77170234691111, 105.7762712620503, 26.00456891513918}$0.975 \pm 0.000$ & \cellcolor[RGB]{242.7339313300438, 246.49810956824044, 237.21313658068868}$0.050 \pm 0.001$ & \cellcolor[RGB]{243.2938923452213, 246.5639873347319, 238.4977530272724}$0.044 \pm 0.001$ & \cellcolor[RGB]{245.95608179913748, 246.87718609401617, 244.60512883331538}$0.012 \pm 0.000$ & \cellcolor[RGB]{247.61489962661457, 244.6428847646441, 246.18013383118057}$147.576 \pm 0.060$ & \cellcolor[RGB]{251.47567343711853, 229.84325182437897, 241.0324354171753}$5.968 \pm 7.313$ \\
    ncMCE (hypyperbolic) & \cellcolor[RGB]{209.27958413920805, 235.9911235387861, 174.6671570935086}$0.290 \pm 0.000$ & \cellcolor[RGB]{243.02011789127062, 246.5317785754436, 237.8696822211502}$0.047 \pm 0.000$ & \cellcolor[RGB]{48.65922585969216, 111.69274709331155, 27.0335212336194}$0.949 \pm 0.000$ & \cellcolor[RGB]{244.75899311043733, 246.7363521306397, 241.85886654747387}$0.026 \pm 0.000$ & \cellcolor[RGB]{245.0240645360683, 246.76753700424334, 242.4669715827449}$0.023 \pm 0.000$ & \cellcolor[RGB]{246.45392249334415, 246.93575558745226, 245.74723395531893}$0.006 \pm 0.000$ & \cellcolor[RGB]{247.02714197909722, 246.895955746794, 246.96381069453705}$6.514 \pm 0.124$ & \cellcolor[RGB]{244.04728269577026, 192.66548943519592, 223.33274471759796}$13.968 \pm 6.750$ \\
    \thickhline
\end{tabular}
\caption{Average quality scores of the considered hyperbolic embedding algorithms for the 'twin' network. We show the results for the mapping accuracy, MA ($1^{\rm st}$ column), the edge prediction precision, EPP ($2^{\rm nd}$ column), the area under the receiver operating characteristic curve, AUC ($3^{\rm rd}$ column), the greedy routing score, GR ($4^{\rm th}$ column), the greedy success rate, GS ($5^{\rm th}$ column) and the greedy routing efficiency, GE ($6^{\rm th}$ column averaged over 4 different realizations. Beside the quality scores, we also display the real time in minutes ($7^{\rm th}$ column), the user time in minutes ($8^{\rm th}$ column) and the peak memory usage in GB ($9^{\rm th}$ column). In the top part of the table, we list the scores obtained for CLOVE with default settings ($1^{\rm st}$ row), for CLOVE with simulated annealing optimisation during the solution of the TSP problem ($2^{\rm nd}$ row) and for CLOVE with Louvain communities ($3^{\rm rd}$ row). For comparison, in the  $4^{\rm th}$ row we give the results for HMCS, followed by the scores for Mercator in the  $5^{\rm th}$ row and by hyperbolic ncMCE in the bottom row.}
\end{table}

\begin{table}
\begin{tabular}{|p{0.19\textwidth}|*{8}{>{\centering\arraybackslash}p{0.11\textwidth}|}}
    \thickhline
    & \makecell{MA} & \makecell{EPP} & \makecell{AUC} & \makecell{GR} & \makecell{GS} & \makecell{GE} & \makecell{Running \\ Time \\ (min)} & \makecell{Peak \\ Mem. \\ (GB)} \\
    \thickhline
    CLOVE & \cellcolor[RGB]{235.71922224079728, 245.6728496753879, 221.1205686700644}$0.133 \pm 0.012$ & \cellcolor[RGB]{141.33082261938364, 197.3024638055648, 82.34783790767491}$0.550 \pm 0.005$ & \cellcolor[RGB]{42.42361278252044, 104.14437336831423, 25.720760585793776}$0.982 \pm 0.001$ & \cellcolor[RGB]{246.33500751157422, 246.92176558959696, 245.47442899714085}$0.008 \pm 0.000$ & \cellcolor[RGB]{246.48569749602328, 246.93949382306155, 245.82012954970048}$0.006 \pm 0.000$ & \cellcolor[RGB]{246.88460459880264, 246.98642407044736, 246.7352693737237}$0.001 \pm 0.000$ & \cellcolor[RGB]{252.6565056425347, 225.31672837028356, 239.45799247662038}$1357.561 \pm 250.064$ & \cellcolor[RGB]{249.72193002700806, 236.56593489646912, 243.37075996398926}$3.629 \pm 1.353$ \\
    CLOVE (with SA) & \cellcolor[RGB]{235.98547127903015, 245.7041730916506, 221.73137528718684}$0.130 \pm 0.005$ & \cellcolor[RGB]{143.85283349294886, 198.93955858314223, 85.40079843883282}$0.541 \pm 0.001$ & \cellcolor[RGB]{42.1985046790781, 103.8718740851998, 25.673369406121704}$0.983 \pm 0.000$ & \cellcolor[RGB]{246.3486634803871, 246.92337217416318, 245.5057573961822}$0.008 \pm 0.000$ & \cellcolor[RGB]{246.49522025732327, 246.94061414792037, 245.8419758844475}$0.006 \pm 0.000$ & \cellcolor[RGB]{246.8865605236694, 246.98665417925523, 246.73975649547685}$0.001 \pm 0.000$ & \cellcolor[RGB]{252.03021516083334, 220.60575306291665, 237.30287653145834}$1556.374 \pm 76.087$ & \cellcolor[RGB]{252.1007661819458, 220.8526816368103, 237.42634081840515}$8.599 \pm 6.017$ \\
    CLOVE (dendr.) & \cellcolor[RGB]{233.19733093038218, 245.37615658004498, 215.3350533108768}$0.162 \pm 0.015$ & \cellcolor[RGB]{172.2701015848533, 217.38585541472935, 119.8006492869277}$0.441 \pm 0.008$ & \cellcolor[RGB]{42.25918428699488, 103.94532834741486, 25.686144060419977}$0.983 \pm 0.000$ & \cellcolor[RGB]{246.5568620643546, 246.94786612521818, 245.98338944175467}$0.005 \pm 0.000$ & \cellcolor[RGB]{246.65258273132466, 246.95912738015585, 246.20298391303893}$0.004 \pm 0.000$ & \cellcolor[RGB]{246.9217857730443, 246.9907983262405, 246.82056736168988}$0.001 \pm 0.000$ & \cellcolor[RGB]{247.56470893447917, 244.83528241782986, 246.24705475402777}$135.530 \pm 21.045$ & \cellcolor[RGB]{248.94679737091064, 239.53727674484253, 244.40427017211914}$2.596 \pm 0.000$ \\
    CLOVE (Louvain) & \cellcolor[RGB]{233.50426603854433, 245.41226659276992, 216.03919855901347}$0.159 \pm 0.027$ & \cellcolor[RGB]{143.4915388109375, 198.70503396499453, 84.96344171850328}$0.542 \pm 0.008$ & \cellcolor[RGB]{42.37059141588605, 104.08018960870416, 25.709598192818113}$0.982 \pm 0.002$ & \cellcolor[RGB]{246.28213033716176, 246.91554474554843, 245.3531225381946}$0.008 \pm 0.000$ & \cellcolor[RGB]{246.44986607094933, 246.93527836128817, 245.73792804511905}$0.006 \pm 0.000$ & \cellcolor[RGB]{246.87805385620496, 246.98565339484765, 246.72024119952903}$0.001 \pm 0.000$ & \cellcolor[RGB]{251.34658946645834, 218.21306313260416, 236.1065315663021}$1638.409 \pm 0.132$ & \cellcolor[RGB]{251.07576751708984, 231.37622451782227, 241.56564331054688}$5.434 \pm 4.161$ \\
    CLOVE (k1 decomp.) & \cellcolor[RGB]{234.4376958011933, 245.52208185896393, 218.18059624979642}$0.148 \pm 0.008$ & \cellcolor[RGB]{152.99155140947437, 204.8717088096588, 96.4634569693637}$0.509 \pm 0.004$ & \cellcolor[RGB]{42.03033129725469, 103.66829578088726, 25.637964483632565}$0.984 \pm 0.001$ & \cellcolor[RGB]{246.2904379294889, 246.91652210935163, 245.3721811323569}$0.008 \pm 0.000$ & \cellcolor[RGB]{246.45274100873075, 246.93561658926245, 245.7445234906176}$0.006 \pm 0.000$ & \cellcolor[RGB]{246.87446412035027, 246.98523107298237, 246.71200592315648}$0.001 \pm 0.000$ & \cellcolor[RGB]{251.20185057461805, 230.8929061306308, 241.39753256717592}$1008.444 \pm 26.436$ & \cellcolor[RGB]{249.4542899131775, 237.59188866615295, 243.72761344909668}$3.272 \pm 1.674$ \\
    HMCS & \cellcolor[RGB]{230.01372615095738, 245.0016148412891, 208.0314894051375}$0.200 \pm 0.039$ & \cellcolor[RGB]{240.943246537851, 246.28744076915893, 233.10509499859936}$0.071 \pm 0.006$ & \cellcolor[RGB]{41.6010688763627, 103.148662324018, 25.547593447655306}$0.986 \pm 0.002$ & \cellcolor[RGB]{246.83305496861334, 246.98035940807216, 246.6170084574071}$0.002 \pm 0.000$ & \cellcolor[RGB]{246.8619522080283, 246.98375908329746, 246.68330212430024}$0.002 \pm 0.000$ & \cellcolor[RGB]{246.968633617231, 246.9963098373213, 246.92804182776527}$0.000 \pm 0.000$ & \cellcolor[RGB]{247.22875510451388, 246.12310543269678, 246.69499319398147}$54.901 \pm 7.182$ & \cellcolor[RGB]{252.51041316986084, 222.28644609451294, 238.14322304725647}$8.326 \pm 0.543$ \\
    \thickhline
\end{tabular}
\caption{Average quality scores of the considered hyperbolic embedding algorithms for the 'amazon' network. We show the results for the mapping accuracy, MA ($1^{\rm st}$ column), the edge prediction precision, EPP ($2^{\rm nd}$ column), the area under the receiver operating characteristic curve, AUC ($3^{\rm rd}$ column), the greedy routing score, GR ($4^{\rm th}$ column), the greedy success rate, GS ($5^{\rm th}$ column) and the greedy routing efficiency, GE ($6^{\rm th}$ column averaged over 2 different realizations. Beside the quality scores, we also display the real time in minutes ($7^{\rm th}$ column), the user time in minutes ($8^{\rm th}$ column) and the peak memory usage in GB ($9^{\rm th}$ column). In the top part of the table, we list the scores obtained for CLOVE with default settings ($1^{\rm st}$ row), for CLOVE with simulated annealing optimisation during the solution of the TSP problem ($2^{\rm nd}$ row) and for CLOVE with Louvain communities ($3^{\rm rd}$ row). For comparison, in the  $4^{\rm th}$ row we give the results for HMCS.}
\end{table}
\begin{table}
\begin{tabular}{|p{0.19\textwidth}|*{8}{>{\centering\arraybackslash}p{0.11\textwidth}|}}
    \thickhline
    & \makecell{MA} & \makecell{EPP} & \makecell{AUC} & \makecell{GR} & \makecell{GS} & \makecell{GE} & \makecell{Running \\ Time \\ (min)} & \makecell{Peak \\ Mem. \\ (GB)} \\
    \thickhline
    CLOVE & \cellcolor[RGB]{236.64854390861183, 245.78218163630729, 223.25254190799188}$0.122 \pm 0.012$ & \cellcolor[RGB]{137.0747633977244, 194.53975869676847, 77.19576621829796}$0.565 \pm 0.004$ & \cellcolor[RGB]{41.99391251973664, 103.62420989231276, 25.630297372576134}$0.984 \pm 0.001$ & \cellcolor[RGB]{246.47275865204685, 246.93797160612317, 245.79044631940155}$0.006 \pm 0.000$ & \cellcolor[RGB]{246.57536354127652, 246.95004276956195, 246.02583400645787}$0.005 \pm 0.000$ & \cellcolor[RGB]{246.8990227560363, 246.98812032423956, 246.7683463226715}$0.001 \pm 0.000$ & \cellcolor[RGB]{251.33441700407985, 230.38473481769387, 241.22077732789353}$1040.260 \pm 65.676$ & \cellcolor[RGB]{248.3384072780609, 241.86943876743317, 245.21545696258545}$1.785 \pm 0.631$ \\
    CLOVE (with SA) & \cellcolor[RGB]{237.24733101385064, 245.85262717810008, 224.62622997295145}$0.115 \pm 0.006$ & \cellcolor[RGB]{136.73862274641652, 194.3215621336388, 76.78885911408318}$0.566 \pm 0.005$ & \cellcolor[RGB]{42.11381402327182, 103.76935381764483, 25.655539794373013}$0.984 \pm 0.000$ & \cellcolor[RGB]{246.45992611766107, 246.93646189619542, 245.76100697581066}$0.006 \pm 0.000$ & \cellcolor[RGB]{246.56399737901756, 246.94870557400208, 245.99975869304032}$0.005 \pm 0.000$ & \cellcolor[RGB]{246.89675022071378, 246.9878529671428, 246.76313285928458}$0.001 \pm 0.000$ & \cellcolor[RGB]{251.33828262505207, 230.369916603967, 241.21562316659723}$1041.188 \pm 147.363$ & \cellcolor[RGB]{248.4898397922516, 241.28894746303558, 245.01354694366455}$1.986 \pm 0.587$ \\
    CLOVE (dendr.) & \cellcolor[RGB]{235.6125078250818, 245.66029503824493, 220.8757532457759}$0.134 \pm 0.007$ & \cellcolor[RGB]{163.3586270381462, 211.60121404230543, 109.01307483565068}$0.472 \pm 0.006$ & \cellcolor[RGB]{41.97911859192321, 103.60630145338072, 25.62718286145752}$0.984 \pm 0.001$ & \cellcolor[RGB]{246.55129848167817, 246.9472115860798, 245.97062592855582}$0.005 \pm 0.000$ & \cellcolor[RGB]{246.63839637310224, 246.95745839683556, 246.17043873829337}$0.004 \pm 0.000$ & \cellcolor[RGB]{246.914861682558, 246.98998372735977, 246.80468268351544}$0.001 \pm 0.000$ & \cellcolor[RGB]{247.35058406784722, 245.65609440658565, 246.5325545762037}$84.140 \pm 23.193$ & \cellcolor[RGB]{248.69096112251282, 240.51798236370087, 244.7453851699829}$2.255 \pm 0.651$ \\
    CLOVE (Louvain) & \cellcolor[RGB]{236.8372543859707, 245.80438286893772, 223.68546594428577}$0.120 \pm 0.004$ & \cellcolor[RGB]{137.29964595032047, 194.6857350905589, 77.4679924661774}$0.564 \pm 0.003$ & \cellcolor[RGB]{42.21120380259579, 103.88724670840543, 25.676042905809638}$0.983 \pm 0.001$ & \cellcolor[RGB]{246.36634657537095, 246.92545253827893, 245.54632449643924}$0.007 \pm 0.000$ & \cellcolor[RGB]{246.4977097852666, 246.9409070335608, 245.84768715443516}$0.006 \pm 0.000$ & \cellcolor[RGB]{246.88381384516236, 246.98633104060733, 246.73345529184306}$0.001 \pm 0.000$ & \cellcolor[RGB]{251.52829941895834, 229.64151889399304, 240.9622674413889}$1086.792 \pm 85.631$ & \cellcolor[RGB]{248.7532570362091, 240.27918136119843, 244.6623239517212}$2.338 \pm 0.590$ \\
    CLOVE (k1 decomp.) & \cellcolor[RGB]{237.8500258842763, 245.92353245697367, 226.00888291098678}$0.108 \pm 0.005$ & \cellcolor[RGB]{142.44605531785845, 198.02638678527654, 83.69785643740758}$0.546 \pm 0.003$ & \cellcolor[RGB]{42.06635034699568, 103.71189778846845, 25.64554744147277}$0.984 \pm 0.001$ & \cellcolor[RGB]{246.44940524283427, 246.9352241462158, 245.736870851208}$0.006 \pm 0.000$ & \cellcolor[RGB]{246.55748552093786, 246.94793947305152, 245.9848197245045}$0.005 \pm 0.000$ & \cellcolor[RGB]{246.89321627346496, 246.98743720864294, 246.7550255685373}$0.001 \pm 0.000$ & \cellcolor[RGB]{250.17349519425346, 234.83493508869503, 242.7686730743287}$761.639 \pm 206.400$ & \cellcolor[RGB]{248.41422080993652, 241.57882022857666, 245.11437225341797}$1.886 \pm 0.519$ \\
    HMCS & \cellcolor[RGB]{231.94876887056756, 245.22926692594913, 212.47070505600794}$0.177 \pm 0.023$ & \cellcolor[RGB]{237.5377533508713, 245.8867945118672, 225.29249298141065}$0.111 \pm 0.004$ & \cellcolor[RGB]{41.69383744227972, 103.26096111433861, 25.56712367205889}$0.986 \pm 0.002$ & \cellcolor[RGB]{246.8075580834517, 246.97735977452373, 246.55851560321275}$0.002 \pm 0.000$ & \cellcolor[RGB]{246.838576339011, 246.9810089810601, 246.6296751306723}$0.002 \pm 0.000$ & \cellcolor[RGB]{246.96120933950644, 246.9954363928831, 246.91100966122065}$0.000 \pm 0.000$ & \cellcolor[RGB]{247.15043829059027, 246.4233198860706, 246.79941561254628}$36.105 \pm 4.623$ & \cellcolor[RGB]{248.68232083320618, 240.55110347270966, 244.7569055557251}$2.243 \pm 0.697$ \\
    \thickhline
\end{tabular}
\caption{Average quality scores of the considered hyperbolic embedding algorithms for the 'com-amazon' network. We show the results for the mapping accuracy, MA ($1^{\rm st}$ column), the edge prediction precision, EPP ($2^{\rm nd}$ column), the area under the receiver operating characteristic curve, AUC ($3^{\rm rd}$ column), the greedy routing score, GR ($4^{\rm th}$ column), the greedy success rate, GS ($5^{\rm th}$ column) and the greedy routing efficiency, GE ($6^{\rm th}$ column averaged over 2 different realizations. Beside the quality scores, we also display the real time in minutes ($7^{\rm th}$ column), the user time in minutes ($8^{\rm th}$ column) and the peak memory usage in GB ($9^{\rm th}$ column). In the top part of the table, we list the scores obtained for CLOVE with default settings ($1^{\rm st}$ row), for CLOVE with simulated annealing optimisation during the solution of the TSP problem ($2^{\rm nd}$ row) and for CLOVE with Louvain communities ($3^{\rm rd}$ row). For comparison, in the  $4^{\rm th}$ row we give the results for HMCS.}
\end{table}
\begin{table}
\begin{tabular}{|p{0.19\textwidth}|*{8}{>{\centering\arraybackslash}p{0.11\textwidth}|}}
    \thickhline
    & \makecell{MA} & \makecell{EPP} & \makecell{AUC} & \makecell{GR} & \makecell{GS} & \makecell{GE} & \makecell{Running \\ Time \\ (min)} & \makecell{Peak \\ Mem. \\ (GB)} \\
    \thickhline
    CLOVE & \cellcolor[RGB]{231.33488120029753, 245.15704484709383, 211.06237451832965}$0.184 \pm 0.002$ & \cellcolor[RGB]{151.48266847154505, 203.89225848152924, 94.63691446555453}$0.514 \pm 0.005$ & \cellcolor[RGB]{45.689764411691925, 108.09813586678496, 26.408371455093036}$0.965 \pm 0.001$ & \cellcolor[RGB]{235.6184547094594, 245.6609946717011, 220.8893960981716}$0.134 \pm 0.002$ & \cellcolor[RGB]{239.26860290459194, 246.09042387112845, 229.26326548700504}$0.091 \pm 0.002$ & \cellcolor[RGB]{245.18789917343526, 246.78681166746298, 242.84282751552794}$0.021 \pm 0.000$ & \cellcolor[RGB]{250.63426222019098, 233.06866148926792, 242.15431703974537}$872.223 \pm 264.483$ & \cellcolor[RGB]{248.43882131576538, 241.48451828956604, 245.0815715789795}$1.918 \pm 0.376$ \\
    CLOVE (with SA) & \cellcolor[RGB]{230.80672108196728, 245.0949083625844, 209.85071307039556}$0.191 \pm 0.002$ & \cellcolor[RGB]{150.57136695494634, 203.30071188303535, 93.53375999809293}$0.517 \pm 0.002$ & \cellcolor[RGB]{45.686176079713405, 108.09379209649518, 26.40761601678177}$0.965 \pm 0.001$ & \cellcolor[RGB]{234.98489955866035, 245.5864587716071, 219.43594604633844}$0.141 \pm 0.007$ & \cellcolor[RGB]{238.89698415155664, 246.0467040178302, 228.41072834768877}$0.095 \pm 0.005$ & \cellcolor[RGB]{245.10569799823156, 246.77714094096842, 242.6542483488842}$0.022 \pm 0.001$ & \cellcolor[RGB]{250.51215502458334, 233.53673907243055, 242.31712663388888}$842.917 \pm 234.411$ & \cellcolor[RGB]{248.92906260490417, 239.605260014534, 244.42791652679443}$2.572 \pm 0.248$ \\
    CLOVE (dendr.) & \cellcolor[RGB]{230.72226380830622, 245.08497221274192, 209.65695814846723}$0.192 \pm 0.001$ & \cellcolor[RGB]{171.56094286845095, 216.92552431811728, 118.94219399865115}$0.444 \pm 0.005$ & \cellcolor[RGB]{45.90615773685039, 108.36008568145047, 26.453927944600082}$0.964 \pm 0.001$ & \cellcolor[RGB]{241.82935425968628, 246.39168873643368, 235.13793036045678}$0.061 \pm 0.003$ & \cellcolor[RGB]{243.39576484535016, 246.57597233474706, 238.73146052756798}$0.042 \pm 0.002$ & \cellcolor[RGB]{246.13728395899318, 246.89850399517567, 245.02082790592556}$0.010 \pm 0.001$ & \cellcolor[RGB]{252.00099051151042, 227.8295363725434, 240.33201265131945}$1200.238 \pm 388.088$ & \cellcolor[RGB]{252.25808453559875, 226.84400928020477, 239.98922061920166}$7.011 \pm 0.872$ \\
    CLOVE (Louvain) & \cellcolor[RGB]{230.9177701208459, 245.10797295539362, 210.10547263017585}$0.189 \pm 0.003$ & \cellcolor[RGB]{150.69202136155647, 203.37903141013317, 93.67981533241047}$0.517 \pm 0.005$ & \cellcolor[RGB]{46.137002550994566, 108.63952940383552, 26.502526852840962}$0.962 \pm 0.001$ & \cellcolor[RGB]{236.22947969441339, 245.73287996404864, 222.29115929894832}$0.127 \pm 0.007$ & \cellcolor[RGB]{239.69724886185492, 246.14085280727704, 230.24662974190244}$0.086 \pm 0.005$ & \cellcolor[RGB]{245.2799951709214, 246.79764649069665, 243.05410656858442}$0.020 \pm 0.001$ & \cellcolor[RGB]{251.25456843949652, 230.69082098192996, 241.32724208067128}$1021.096 \pm 207.259$ & \cellcolor[RGB]{248.69826889038086, 240.48996925354004, 244.7356414794922}$2.264 \pm 0.389$ \\
    CLOVE (k1 decomp.) & \cellcolor[RGB]{230.5819366062394, 245.0684631301458, 209.3350310378433}$0.193 \pm 0.002$ & \cellcolor[RGB]{161.700733386457, 210.52503746138436, 107.00615094150056}$0.478 \pm 0.005$ & \cellcolor[RGB]{45.675333545243284, 108.08066692318923, 26.405333377945954}$0.965 \pm 0.003$ & \cellcolor[RGB]{234.77939869526978, 245.5622821994435, 218.96450288914835}$0.144 \pm 0.005$ & \cellcolor[RGB]{238.74581920779926, 246.0289199067999, 228.0639381825983}$0.097 \pm 0.003$ & \cellcolor[RGB]{245.0160892924013, 246.7665987402825, 242.44867543550882}$0.023 \pm 0.001$ & \cellcolor[RGB]{249.21037953364583, 238.52687845435764, 244.05282728847223}$530.491 \pm 189.050$ & \cellcolor[RGB]{248.7357304096222, 240.34636676311493, 244.6856927871704}$2.314 \pm 0.578$ \\
    HMCS & \cellcolor[RGB]{231.26181672597025, 245.14844902658473, 210.89475601840235}$0.185 \pm 0.003$ & \cellcolor[RGB]{240.62517843542503, 246.25002099240294, 232.37540935185743}$0.075 \pm 0.007$ & \cellcolor[RGB]{46.07582840923183, 108.5654764953859, 26.48964808615407}$0.963 \pm 0.002$ & \cellcolor[RGB]{244.42311007609715, 246.69683647954085, 241.08831135104637}$0.030 \pm 0.001$ & \cellcolor[RGB]{245.13886455574544, 246.78104288891123, 242.73033633376897}$0.022 \pm 0.001$ & \cellcolor[RGB]{246.55087127010546, 246.94716132589477, 245.96964585494783}$0.005 \pm 0.000$ & \cellcolor[RGB]{247.19721489847223, 246.24400955585648, 246.73704680203704}$47.332 \pm 4.128$ & \cellcolor[RGB]{248.78580689430237, 240.15440690517426, 244.61892414093018}$2.381 \pm 0.429$ \\
    \thickhline
\end{tabular}
\caption{Average quality scores of the considered hyperbolic embedding algorithms for the 'com-dblp' network. We show the results for the mapping accuracy, MA ($1^{\rm st}$ column), the edge prediction precision, EPP ($2^{\rm nd}$ column), the area under the receiver operating characteristic curve, AUC ($3^{\rm rd}$ column), the greedy routing score, GR ($4^{\rm th}$ column), the greedy success rate, GS ($5^{\rm th}$ column) and the greedy routing efficiency, GE ($6^{\rm th}$ column averaged over 2 different realizations. Beside the quality scores, we also display the real time in minutes ($7^{\rm th}$ column), the user time in minutes ($8^{\rm th}$ column) and the peak memory usage in GB ($9^{\rm th}$ column). In the top part of the table, we list the scores obtained for CLOVE with default settings ($1^{\rm st}$ row), for CLOVE with simulated annealing optimisation during the solution of the TSP problem ($2^{\rm nd}$ row) and for CLOVE with Louvain communities ($3^{\rm rd}$ row). For comparison, in the  $4^{\rm th}$ row we give the results for HMCS.}
\end{table}
\begin{table}
\begin{tabular}{|p{0.19\textwidth}|*{8}{>{\centering\arraybackslash}p{0.11\textwidth}|}}
    \thickhline
    & \makecell{MA} & \makecell{EPP} & \makecell{AUC} & \makecell{GR} & \makecell{GS} & \makecell{GE} & \makecell{Running \\ Time \\ (min)} & \makecell{Peak \\ Mem. \\ (GB)} \\
    \thickhline
    CLOVE & \cellcolor[RGB]{212.8217808994052, 237.5312090866979, 180.3654736207823}$0.275 \pm 0.012$ & \cellcolor[RGB]{171.86123130400492, 217.12044839031898, 119.30570105221646}$0.443 \pm 0.016$ & \cellcolor[RGB]{39.12992242906607, 100.15727451939577, 25.0273520903297}$0.999 \pm 0.000$ & \cellcolor[RGB]{243.20376203189602, 246.55338376845836, 238.2909834849379}$0.045 \pm 0.002$ & \cellcolor[RGB]{243.38034841364166, 246.57415863689903, 238.69609341953083}$0.043 \pm 0.002$ & \cellcolor[RGB]{245.23278875604012, 246.79209279482825, 242.9458094991509}$0.021 \pm 0.001$ & \cellcolor[RGB]{248.93859271975694, 239.56872790759837, 244.41520970699074}$465.262 \pm 129.396$ & \cellcolor[RGB]{249.78722715377808, 236.3156292438507, 243.2836971282959}$3.716 \pm 0.220$ \\
    CLOVE (with SA) & \cellcolor[RGB]{216.39496554743755, 239.08476762932068, 186.1136402284865}$0.259 \pm 0.017$ & \cellcolor[RGB]{164.58384561953102, 212.396531367064, 110.49623417101125}$0.468 \pm 0.028$ & \cellcolor[RGB]{39.12933204172216, 100.15655983997945, 25.027227798257297}$0.999 \pm 0.000$ & \cellcolor[RGB]{243.1123558603889, 246.54263010122222, 238.0812869738334}$0.046 \pm 0.002$ & \cellcolor[RGB]{243.2944551957962, 246.5640535524466, 238.49904427270894}$0.044 \pm 0.002$ & \cellcolor[RGB]{245.2049731023317, 246.7888203649802, 242.88199711711394}$0.021 \pm 0.001$ & \cellcolor[RGB]{248.85033694815974, 239.90704169872106, 244.53288406912037}$444.081 \pm 78.868$ & \cellcolor[RGB]{249.97251987457275, 235.60534048080444, 243.03664016723633}$3.963 \pm 0.252$ \\
    CLOVE (dendr.) & \cellcolor[RGB]{213.24380966666885, 237.7146998550734, 181.0443894637716}$0.273 \pm 0.012$ & \cellcolor[RGB]{191.5273831684921, 228.27277529064872, 146.1092685754003}$0.367 \pm 0.039$ & \cellcolor[RGB]{39.30199496378258, 100.36557285089471, 25.063577887112125}$0.998 \pm 0.000$ & \cellcolor[RGB]{243.4009259312953, 246.57657952132885, 238.7433006659127}$0.042 \pm 0.001$ & \cellcolor[RGB]{243.55958754171255, 246.59524559314266, 239.10728906628174}$0.040 \pm 0.001$ & \cellcolor[RGB]{245.32309581878653, 246.80271715515136, 243.15298452545147}$0.020 \pm 0.000$ & \cellcolor[RGB]{247.19457881831596, 246.25411452978878, 246.7405615755787}$46.699 \pm 7.284$ & \cellcolor[RGB]{249.75996446609497, 236.4201362133026, 243.32004737854004}$3.680 \pm 0.110$ \\
    CLOVE (Louvain) & \cellcolor[RGB]{215.02381487397471, 238.4886151625977, 183.9078761016115}$0.265 \pm 0.014$ & \cellcolor[RGB]{167.08320737126888, 214.01892408310437, 113.5217773441676}$0.459 \pm 0.011$ & \cellcolor[RGB]{39.113776558222746, 100.13772951784858, 25.02395295962584}$0.999 \pm 0.000$ & \cellcolor[RGB]{243.22484058508923, 246.5558635982458, 238.33934016579292}$0.044 \pm 0.001$ & \cellcolor[RGB]{243.39684472289878, 246.57609937916456, 238.73393789370897}$0.042 \pm 0.001$ & \cellcolor[RGB]{245.24979950068627, 246.79409405890428, 242.9848341486332}$0.021 \pm 0.001$ & \cellcolor[RGB]{248.53885454059028, 241.10105759440393, 244.94819394587964}$369.325 \pm 76.797$ & \cellcolor[RGB]{250.03326797485352, 235.37247276306152, 242.9556427001953}$4.044 \pm 0.130$ \\
    CLOVE (k1 decomp.) & \cellcolor[RGB]{207.4925429314857, 235.21414910064595, 171.79235167239003}$0.298 \pm 0.023$ & \cellcolor[RGB]{246.92013432403783, 246.9906040381221, 246.81677874338095}$0.001 \pm 0.001$ & \cellcolor[RGB]{39.181813163157145, 100.22008961855866, 25.038276455401505}$0.999 \pm 0.000$ & \cellcolor[RGB]{243.1612708099001, 246.5483848011647, 238.19350362271194}$0.045 \pm 0.001$ & \cellcolor[RGB]{243.32552618968398, 246.56770896349224, 238.57032478809856}$0.043 \pm 0.001$ & \cellcolor[RGB]{245.12217958490564, 246.77907995116536, 242.69205904772468}$0.022 \pm 0.001$ & \cellcolor[RGB]{248.01179780940973, 243.12144173059608, 245.65093625412038}$242.831 \pm 45.986$ & \cellcolor[RGB]{249.43963932991028, 237.64804923534393, 243.74714756011963}$3.253 \pm 0.348$ \\
    HMCS & \cellcolor[RGB]{222.65162718934928, 241.80505529971708, 196.17870460895318}$0.232 \pm 0.033$ & \cellcolor[RGB]{206.89924216651048, 234.95619224630892, 170.83791131134296}$0.300 \pm 0.020$ & \cellcolor[RGB]{39.24532853925018, 100.29697665277654, 25.051648113526355}$0.999 \pm 0.000$ & \cellcolor[RGB]{244.26966754641845, 246.6787844172257, 240.73629613590117}$0.032 \pm 0.001$ & \cellcolor[RGB]{244.3728170867936, 246.69091965726983, 240.97293331676175}$0.031 \pm 0.001$ & \cellcolor[RGB]{245.60834503454416, 246.83627588641696, 243.80737978513073}$0.016 \pm 0.000$ & \cellcolor[RGB]{247.32505498057293, 245.75395590780383, 246.56659335923612}$78.013 \pm 7.730$ & \cellcolor[RGB]{249.35368061065674, 237.97755765914917, 243.86175918579102}$3.138 \pm 0.243$ \\
    \thickhline
\end{tabular}
\caption{Average quality scores of the considered hyperbolic embedding algorithms for the 'dimacs10-cnr-2000' network. We show the results for the mapping accuracy, MA ($1^{\rm st}$ column), the edge prediction precision, EPP ($2^{\rm nd}$ column), the area under the receiver operating characteristic curve, AUC ($3^{\rm rd}$ column), the greedy routing score, GR ($4^{\rm th}$ column), the greedy success rate, GS ($5^{\rm th}$ column) and the greedy routing efficiency, GE ($6^{\rm th}$ column averaged over 2 different realizations. Beside the quality scores, we also display the real time in minutes ($7^{\rm th}$ column), the user time in minutes ($8^{\rm th}$ column) and the peak memory usage in GB ($9^{\rm th}$ column). In the top part of the table, we list the scores obtained for CLOVE with default settings ($1^{\rm st}$ row), for CLOVE with simulated annealing optimisation during the solution of the TSP problem ($2^{\rm nd}$ row) and for CLOVE with Louvain communities ($3^{\rm rd}$ row). For comparison, in the  $4^{\rm th}$ row we give the results for HMCS.}
\end{table}
\begin{table}
\begin{tabular}{|p{0.19\textwidth}|*{8}{>{\centering\arraybackslash}p{0.11\textwidth}|}}
    \thickhline
    & \makecell{MA} & \makecell{EPP} & \makecell{AUC} & \makecell{GR} & \makecell{GS} & \makecell{GE} & \makecell{Running \\ Time \\ (min)} & \makecell{Peak \\ Mem. \\ (GB)} \\
    \thickhline
    CLOVE & \cellcolor[RGB]{204.54683314524812, 233.93340571532525, 167.05360114670347}$0.311 \pm 0.007$ & \cellcolor[RGB]{144.9605100604582, 199.65857670591146, 86.74167007318624}$0.537 \pm 0.001$ & \cellcolor[RGB]{46.971201049574674, 109.64934863895883, 26.678147589384142}$0.958 \pm 0.001$ & \cellcolor[RGB]{208.83610892808213, 235.79830822960093, 173.95374044952345}$0.292 \pm 0.005$ & \cellcolor[RGB]{226.8293662580601, 243.6214635904609, 202.89941528470538}$0.214 \pm 0.002$ & \cellcolor[RGB]{241.79768948723728, 246.38796346908674, 235.06528764719138}$0.061 \pm 0.001$ & \cellcolor[RGB]{247.01051307949652, 246.95969986192998, 246.9859825606713}$2.523 \pm 0.060$ & \cellcolor[RGB]{247.3989975452423, 245.47050940990448, 246.46800327301025}$0.532 \pm 0.090$ \\
    CLOVE (with SA) & \cellcolor[RGB]{203.25179252541477, 233.37034457626729, 164.97027493218897}$0.316 \pm 0.008$ & \cellcolor[RGB]{144.83407866091167, 199.57650720094267, 86.58862153689307}$0.537 \pm 0.001$ & \cellcolor[RGB]{46.79691704339007, 109.43837326305113, 26.641456219661066}$0.959 \pm 0.001$ & \cellcolor[RGB]{207.59178236866836, 235.25729668202973, 171.95199772351}$0.297 \pm 0.010$ & \cellcolor[RGB]{225.73386582570714, 243.1451590546553, 201.13708850222454}$0.219 \pm 0.007$ & \cellcolor[RGB]{241.6940776545353, 246.37577384171004, 234.8275899133457}$0.062 \pm 0.002$ & \cellcolor[RGB]{247.01019610333333, 246.9609149372222, 246.98640519555556}$2.447 \pm 0.191$ & \cellcolor[RGB]{247.36683177947998, 245.5938115119934, 246.51089096069336}$0.489 \pm 0.070$ \\
    CLOVE (dendr.) & \cellcolor[RGB]{203.9816527999686, 233.68767513042113, 166.14439798255816}$0.313 \pm 0.014$ & \cellcolor[RGB]{172.15295479275213, 217.30981276020754, 119.6588400122789}$0.442 \pm 0.002$ & \cellcolor[RGB]{47.297569777943934, 110.04442657330054, 26.746856795356617}$0.956 \pm 0.002$ & \cellcolor[RGB]{233.03142984961195, 245.3566388058367, 214.95445671381566}$0.164 \pm 0.010$ & \cellcolor[RGB]{236.5692763064481, 245.77285603605273, 223.07069270302802}$0.123 \pm 0.007$ & \cellcolor[RGB]{243.93977300605812, 246.63997329483036, 239.97947924919214}$0.036 \pm 0.002$ & \cellcolor[RGB]{247.00350339298612, 246.98657032688658, 246.99532880935186}$0.841 \pm 0.060$ & \cellcolor[RGB]{247.5143575668335, 245.02829599380493, 246.31418991088867}$0.686 \pm 0.213$ \\
    CLOVE (Louvain) & \cellcolor[RGB]{203.49409932145062, 233.47569535715246, 165.36007282146406}$0.315 \pm 0.010$ & \cellcolor[RGB]{145.46979964328244, 199.98916818949914, 87.35817851555245}$0.535 \pm 0.001$ & \cellcolor[RGB]{47.12072786680476, 109.83035478613209, 26.70962691932732}$0.957 \pm 0.001$ & \cellcolor[RGB]{213.3742079447957, 237.77139475860682, 181.25416060684526}$0.272 \pm 0.017$ & \cellcolor[RGB]{229.70234435190736, 244.8705845008293, 207.52116265306836}$0.201 \pm 0.011$ & \cellcolor[RGB]{242.09051321095518, 246.4224133189359, 235.7370597192501}$0.058 \pm 0.003$ & \cellcolor[RGB]{247.01175487282987, 246.9549396541522, 246.98432683622684}$2.821 \pm 0.137$ & \cellcolor[RGB]{247.4276900291443, 245.3605215549469, 246.42974662780762}$0.570 \pm 0.023$ \\
    CLOVE (k1 decomp.) & \cellcolor[RGB]{202.6082547469326, 233.0905455421446, 163.93501850593506}$0.319 \pm 0.014$ & \cellcolor[RGB]{148.28500990731908, 201.8165853784352, 90.76606462464943}$0.525 \pm 0.002$ & \cellcolor[RGB]{46.71546722199708, 109.33977611083857, 26.62430888884149}$0.959 \pm 0.002$ & \cellcolor[RGB]{206.19156711578302, 234.6485074416448, 169.6994775340857}$0.304 \pm 0.019$ & \cellcolor[RGB]{224.69470850124927, 242.69335152228228, 199.46540063244447}$0.223 \pm 0.012$ & \cellcolor[RGB]{241.5145158723745, 246.3546489261617, 234.41565406015323}$0.065 \pm 0.003$ & \cellcolor[RGB]{247.0090860563368, 246.96517011737558, 246.9878852582176}$2.181 \pm 0.158$ & \cellcolor[RGB]{247.4433205127716, 245.30060470104218, 246.4089059829712}$0.591 \pm 0.248$ \\
    HMCS & \cellcolor[RGB]{211.75316410206213, 237.0665930878531, 178.64639442505646}$0.279 \pm 0.008$ & \cellcolor[RGB]{233.3592815377612, 245.3952095926778, 215.70658705721686}$0.160 \pm 0.001$ & \cellcolor[RGB]{47.37038172043007, 110.13256734578377, 26.7621856253537}$0.956 \pm 0.001$ & \cellcolor[RGB]{236.3682275515861, 245.74920324136306, 222.60946320657987}$0.125 \pm 0.004$ & \cellcolor[RGB]{238.91694512285832, 246.0490523673951, 228.45652116420442}$0.095 \pm 0.003$ & \cellcolor[RGB]{244.61286308657148, 246.71916036312606, 241.52362708095808}$0.028 \pm 0.001$ & \cellcolor[RGB]{247.00340596930556, 246.9869437843287, 246.9954587075926}$0.817 \pm 0.098$ & \cellcolor[RGB]{247.41214108467102, 245.42012584209442, 246.45047855377197}$0.550 \pm 0.212$ \\
    \thickhline
\end{tabular}
\caption{Average quality scores of the considered hyperbolic embedding algorithms for the 'dimacs10-cond-mat-2003' network. We show the results for the mapping accuracy, MA ($1^{\rm st}$ column), the edge prediction precision, EPP ($2^{\rm nd}$ column), the area under the receiver operating characteristic curve, AUC ($3^{\rm rd}$ column), the greedy routing score, GR ($4^{\rm th}$ column), the greedy success rate, GS ($5^{\rm th}$ column) and the greedy routing efficiency, GE ($6^{\rm th}$ column averaged over 2 different realizations. Beside the quality scores, we also display the real time in minutes ($7^{\rm th}$ column), the user time in minutes ($8^{\rm th}$ column) and the peak memory usage in GB ($9^{\rm th}$ column). In the top part of the table, we list the scores obtained for CLOVE with default settings ($1^{\rm st}$ row), for CLOVE with simulated annealing optimisation during the solution of the TSP problem ($2^{\rm nd}$ row) and for CLOVE with Louvain communities ($3^{\rm rd}$ row). For comparison, in the  $4^{\rm th}$ row we give the results for HMCS.}
\end{table}
\begin{table}
\begin{tabular}{|p{0.19\textwidth}|*{8}{>{\centering\arraybackslash}p{0.11\textwidth}|}}
    \thickhline
    & \makecell{MA} & \makecell{EPP} & \makecell{AUC} & \makecell{GR} & \makecell{GS} & \makecell{GE} & \makecell{Running \\ Time \\ (min)} & \makecell{Peak \\ Mem. \\ (GB)} \\
    \thickhline
    CLOVE & \cellcolor[RGB]{203.50331692988792, 233.47970301299475, 165.37490114808057}$0.315 \pm 0.007$ & \cellcolor[RGB]{155.3900512106798, 206.42862973324827, 99.36690409713867}$0.500 \pm 0.001$ & \cellcolor[RGB]{48.03555822781276, 110.93778101261543, 26.902222784802685}$0.952 \pm 0.001$ & \cellcolor[RGB]{202.84335506001034, 233.1927630695697, 164.31322335740794}$0.318 \pm 0.005$ & \cellcolor[RGB]{222.95920464616296, 241.9387846287665, 196.67350312643606}$0.231 \pm 0.004$ & \cellcolor[RGB]{241.25830366427064, 246.3245063134436, 233.82787311215026}$0.068 \pm 0.001$ & \cellcolor[RGB]{247.01593758833334, 246.9389059113889, 246.97874988222222}$3.825 \pm 0.070$ & \cellcolor[RGB]{247.5335636138916, 244.95467281341553, 246.28858184814453}$0.711 \pm 0.284$ \\
    CLOVE (with SA) & \cellcolor[RGB]{204.4653421460314, 233.8979748461006, 166.9225069305722}$0.311 \pm 0.007$ & \cellcolor[RGB]{155.1799423582959, 206.29224328520962, 99.11256180214767}$0.501 \pm 0.001$ & \cellcolor[RGB]{48.01811099664815, 110.91666068015303, 26.898549683504875}$0.953 \pm 0.001$ & \cellcolor[RGB]{202.75408340699292, 233.15394930738822, 164.1696124373364}$0.318 \pm 0.015$ & \cellcolor[RGB]{222.9563002197678, 241.93752183468166, 196.6688307883221}$0.231 \pm 0.011$ & \cellcolor[RGB]{241.24697794630353, 246.3231738760357, 233.8018905826963}$0.068 \pm 0.003$ & \cellcolor[RGB]{247.01544174324653, 246.9408066508883, 246.97941100900462}$3.706 \pm 0.354$ & \cellcolor[RGB]{247.5337131023407, 244.95409977436066, 246.2883825302124}$0.712 \pm 0.225$ \\
    CLOVE (dendr.) & \cellcolor[RGB]{203.55669278171587, 233.50290990509387, 165.46076664884728}$0.315 \pm 0.008$ & \cellcolor[RGB]{183.9795379232041, 224.98671759927285, 133.97523011756286}$0.400 \pm 0.003$ & \cellcolor[RGB]{48.32728108537844, 111.290919208616, 26.963638123237565}$0.951 \pm 0.001$ & \cellcolor[RGB]{232.83231472357497, 245.33321349689118, 214.49766318937787}$0.167 \pm 0.012$ & \cellcolor[RGB]{236.65891293004202, 245.78340152118142, 223.2763296630376}$0.122 \pm 0.009$ & \cellcolor[RGB]{243.9177708449039, 246.6373848052828, 239.92900370301487}$0.036 \pm 0.002$ & \cellcolor[RGB]{247.00505914671874, 246.9806066042448, 246.99325447104167}$1.214 \pm 0.215$ & \cellcolor[RGB]{247.43369102478027, 245.33751773834229, 246.42174530029297}$0.578 \pm 0.044$ \\
    CLOVE (Louvain) & \cellcolor[RGB]{202.9405413534727, 233.23501797977073, 164.46956652515172}$0.318 \pm 0.004$ & \cellcolor[RGB]{155.74448583810195, 206.65870133350478, 99.79595654086026}$0.499 \pm 0.003$ & \cellcolor[RGB]{48.319513903222216, 111.28151683021636, 26.962002926994153}$0.951 \pm 0.001$ & \cellcolor[RGB]{205.15185161635813, 234.19645722450352, 168.02689173066307}$0.308 \pm 0.016$ & \cellcolor[RGB]{224.66931291990429, 242.68230996517576, 199.42454687115037}$0.223 \pm 0.010$ & \cellcolor[RGB]{241.4345999136516, 246.3452470486649, 234.2323174489654}$0.065 \pm 0.003$ & \cellcolor[RGB]{247.01519655423613, 246.9417465420949, 246.9797379276852}$3.647 \pm 0.408$ & \cellcolor[RGB]{247.4140121936798, 245.41295325756073, 246.44798374176025}$0.552 \pm 0.056$ \\
    CLOVE (k1 decomp.) & \cellcolor[RGB]{203.8023907740769, 233.60973511916387, 165.85601994090635}$0.314 \pm 0.002$ & \cellcolor[RGB]{157.59079754435268, 207.8571843708956, 102.03096544842693}$0.493 \pm 0.001$ & \cellcolor[RGB]{48.016700087747545, 110.91495273779967, 26.898252650052115}$0.953 \pm 0.001$ & \cellcolor[RGB]{200.0066325062906, 231.95940543751766, 159.74980011881533}$0.330 \pm 0.010$ & \cellcolor[RGB]{221.19337621567237, 241.17103313724886, 193.8328226078208}$0.238 \pm 0.006$ & \cellcolor[RGB]{240.9744371468836, 246.29111025257453, 233.17664992520352}$0.071 \pm 0.002$ & \cellcolor[RGB]{247.01302131635416, 246.9500849539757, 246.9826382448611}$3.125 \pm 0.333$ & \cellcolor[RGB]{247.62495183944702, 244.60435128211975, 246.1667308807373}$0.833 \pm 0.359$ \\
    HMCS & \cellcolor[RGB]{214.482286123981, 238.25316787999174, 183.03672115596942}$0.267 \pm 0.020$ & \cellcolor[RGB]{235.99274761493766, 245.70502913116914, 221.74806805779812}$0.129 \pm 0.002$ & \cellcolor[RGB]{48.50235297536935, 111.50284833860499, 27.00049536323565}$0.950 \pm 0.000$ & \cellcolor[RGB]{236.2745216862781, 245.73817902191507, 222.3944909273439}$0.126 \pm 0.005$ & \cellcolor[RGB]{238.95008124287497, 246.0529507344559, 228.5325393218896}$0.095 \pm 0.003$ & \cellcolor[RGB]{244.58428486646596, 246.71579821958423, 241.4580652818925}$0.028 \pm 0.001$ & \cellcolor[RGB]{247.0055085457118, 246.97888390810476, 246.99265527238427}$1.322 \pm 0.080$ & \cellcolor[RGB]{247.4288730621338, 245.3559865951538, 246.42816925048828}$0.572 \pm 0.068$ \\
    \thickhline
\end{tabular}
\caption{Average quality scores of the considered hyperbolic embedding algorithms for the 'dimacs10-cond-mat-2005' network. We show the results for the mapping accuracy, MA ($1^{\rm st}$ column), the edge prediction precision, EPP ($2^{\rm nd}$ column), the area under the receiver operating characteristic curve, AUC ($3^{\rm rd}$ column), the greedy routing score, GR ($4^{\rm th}$ column), the greedy success rate, GS ($5^{\rm th}$ column) and the greedy routing efficiency, GE ($6^{\rm th}$ column averaged over 2 different realizations. Beside the quality scores, we also display the real time in minutes ($7^{\rm th}$ column), the user time in minutes ($8^{\rm th}$ column) and the peak memory usage in GB ($9^{\rm th}$ column). In the top part of the table, we list the scores obtained for CLOVE with default settings ($1^{\rm st}$ row), for CLOVE with simulated annealing optimisation during the solution of the TSP problem ($2^{\rm nd}$ row) and for CLOVE with Louvain communities ($3^{\rm rd}$ row). For comparison, in the  $4^{\rm th}$ row we give the results for HMCS.}
\end{table}
\begin{table}
\begin{tabular}{|p{0.19\textwidth}|*{8}{>{\centering\arraybackslash}p{0.11\textwidth}|}}
    \thickhline
    & \makecell{MA} & \makecell{EPP} & \makecell{AUC} & \makecell{GR} & \makecell{GS} & \makecell{GE} & \makecell{Running \\ Time \\ (min)} & \makecell{Peak \\ Mem. \\ (GB)} \\
    \thickhline
    CLOVE & \cellcolor[RGB]{218.11406086972076, 239.83220037813948, 188.879141399116}$0.252 \pm 0.005$ & \cellcolor[RGB]{159.28350068777988, 208.9559565868045, 104.08002714836512}$0.487 \pm 0.026$ & \cellcolor[RGB]{39.06292856940742, 100.07617668928268, 25.013248119875247}$1.000 \pm 0.000$ & \cellcolor[RGB]{244.31694622237566, 246.68434661439713, 240.84475898074416}$0.032 \pm 0.001$ & \cellcolor[RGB]{244.53062638056306, 246.70948545653684, 241.33496640246818}$0.029 \pm 0.001$ & \cellcolor[RGB]{246.19881004136522, 246.90574235780767, 245.1619759772496}$0.009 \pm 0.000$ & \cellcolor[RGB]{219.10951831409145, 108.36302739585648, 168.33465589561922}$4486.492 \pm 75.595$ & \cellcolor[RGB]{246.26869678497314, 200.440438747406, 227.220219373703}$12.488 \pm 0.001$ \\
    CLOVE (with SA) & \cellcolor[RGB]{216.66310173563883, 239.20134858071253, 186.5449897486364}$0.258 \pm 0.017$ & \cellcolor[RGB]{150.86118460164548, 203.4888391273839, 93.88459188620243}$0.516 \pm 0.007$ & \cellcolor[RGB]{39.05703882832047, 100.06904700270374, 25.012008174383258}$1.000 \pm 0.000$ & \cellcolor[RGB]{244.56392755089556, 246.71340324128184, 241.4113632049957}$0.029 \pm 0.002$ & \cellcolor[RGB]{244.7577276071667, 246.73620324790195, 241.8559633340883}$0.026 \pm 0.002$ & \cellcolor[RGB]{246.26973354191992, 246.91408629904942, 245.32468283146335}$0.009 \pm 0.001$ & \cellcolor[RGB]{216.23877057132523, 97.7986757024768, 162.70799031979743}$4651.847 \pm 4.432$ & \cellcolor[RGB]{246.26728057861328, 200.43548202514648, 227.21774101257324}$12.488 \pm 0.001$ \\
    CLOVE (dendr.) & \cellcolor[RGB]{215.07002770402931, 238.50870769740405, 183.98221848039498}$0.265 \pm 0.010$ & \cellcolor[RGB]{198.69539441267875, 231.389301918556, 157.64041709865714}$0.336 \pm 0.007$ & \cellcolor[RGB]{39.09738357698077, 100.11788538266093, 25.020501805680162}$0.999 \pm 0.000$ & \cellcolor[RGB]{244.86873589341306, 246.7492630462839, 242.11062940253584}$0.025 \pm 0.000$ & \cellcolor[RGB]{245.02802679159393, 246.76800315195223, 242.47606146306842}$0.023 \pm 0.001$ & \cellcolor[RGB]{246.34421757529148, 246.92284912650487, 245.49555796684515}$0.008 \pm 0.000$ & \cellcolor[RGB]{249.13484320506944, 238.81643438056713, 244.15354239324074}$512.362 \pm 20.722$ & \cellcolor[RGB]{242.7406463623047, 188.0922622680664, 221.0461311340332}$14.840 \pm 1.640$ \\
    CLOVE (Louvain) & \cellcolor[RGB]{216.95253414709373, 239.32718875960597, 187.0105984105421}$0.257 \pm 0.004$ & \cellcolor[RGB]{155.44981431091796, 206.46742332463097, 99.43924890269017}$0.500 \pm 0.019$ & \cellcolor[RGB]{39.03283455911418, 100.03974709787505, 25.00691253876088}$1.000 \pm 0.000$ & \cellcolor[RGB]{244.22909248300726, 246.6740108803538, 240.64321216689902}$0.033 \pm 0.001$ & \cellcolor[RGB]{244.44409064736823, 246.69930478204333, 241.13644324984475}$0.030 \pm 0.001$ & \cellcolor[RGB]{246.1744103670468, 246.90287180788786, 245.1060002538132}$0.010 \pm 0.000$ & \cellcolor[RGB]{224.85643267045717, 128.47132938098957, 180.6148967105324}$4103.512 \pm 998.860$ & \cellcolor[RGB]{242.54572200775146, 187.41002702713013, 220.70501351356506}$14.970 \pm 2.179$ \\
    CLOVE (k1 decomp.) & \cellcolor[RGB]{215.50703103067113, 238.69870914377006, 184.6852238319492}$0.263 \pm 0.006$ & \cellcolor[RGB]{225.14466815766147, 242.888986155505, 200.18924877536844}$0.221 \pm 0.050$ & \cellcolor[RGB]{39.06449180057149, 100.07806902174443, 25.013577221172945}$1.000 \pm 0.000$ & \cellcolor[RGB]{244.56234195150822, 246.71321670017744, 241.40772565346}$0.029 \pm 0.002$ & \cellcolor[RGB]{244.77401063895974, 246.73811889870115, 241.8933185246723}$0.026 \pm 0.002$ & \cellcolor[RGB]{246.25279197753477, 246.91209317382763, 245.2858168896386}$0.009 \pm 0.001$ & \cellcolor[RGB]{236.11630140793403, 165.80668361578125, 206.69038220784722}$3250.133 \pm 714.723$ & \cellcolor[RGB]{244.59333610534668, 194.57667636871338, 224.2883381843567}$13.604 \pm 1.904$ \\
    HMCS & \cellcolor[RGB]{211.8986359822608, 237.12984173141774, 178.88041440624562}$0.279 \pm 0.004$ & \cellcolor[RGB]{223.64391449926845, 242.2364845648993, 197.77499289012752}$0.228 \pm 0.002$ & \cellcolor[RGB]{39.06573651348173, 100.07957577947789, 25.013839265996154}$1.000 \pm 0.000$ & \cellcolor[RGB]{245.3221598498614, 246.80260704116017, 243.15083730262322}$0.020 \pm 0.001$ & \cellcolor[RGB]{245.45066388173606, 246.8177251625572, 243.44564066986507}$0.018 \pm 0.001$ & \cellcolor[RGB]{246.47401909947357, 246.93811989405572, 245.7933379340864}$0.006 \pm 0.000$ & \cellcolor[RGB]{250.66110707069444, 232.96575622900463, 242.11852390574074}$878.666 \pm 46.770$ & \cellcolor[RGB]{244.72913455963135, 195.05197095870972, 224.52598547935486}$13.514 \pm 2.615$ \\
    \thickhline
\end{tabular}
\caption{Average quality scores of the considered hyperbolic embedding algorithms for the 'dimacs10-in-2004' network. We show the results for the mapping accuracy, MA ($1^{\rm st}$ column), the edge prediction precision, EPP ($2^{\rm nd}$ column), the area under the receiver operating characteristic curve, AUC ($3^{\rm rd}$ column), the greedy routing score, GR ($4^{\rm th}$ column), the greedy success rate, GS ($5^{\rm th}$ column) and the greedy routing efficiency, GE ($6^{\rm th}$ column averaged over 2 different realizations. Beside the quality scores, we also display the real time in minutes ($7^{\rm th}$ column), the user time in minutes ($8^{\rm th}$ column) and the peak memory usage in GB ($9^{\rm th}$ column). In the top part of the table, we list the scores obtained for CLOVE with default settings ($1^{\rm st}$ row), for CLOVE with simulated annealing optimisation during the solution of the TSP problem ($2^{\rm nd}$ row) and for CLOVE with Louvain communities ($3^{\rm rd}$ row). For comparison, in the  $4^{\rm th}$ row we give the results for HMCS.}
\end{table}
\begin{table}
\begin{tabular}{|p{0.19\textwidth}|*{8}{>{\centering\arraybackslash}p{0.11\textwidth}|}}
    \thickhline
    & \makecell{MA} & \makecell{EPP} & \makecell{AUC} & \makecell{GR} & \makecell{GS} & \makecell{GE} & \makecell{Running \\ Time \\ (min)} & \makecell{Peak \\ Mem. \\ (GB)} \\
    \thickhline
    CLOVE & \cellcolor[RGB]{206.82802744924092, 234.92522932575693, 170.72334850530063}$0.301 \pm 0.004$ & \cellcolor[RGB]{230.03864593679248, 245.00454658079911, 208.08865832558277}$0.200 \pm 0.004$ & \cellcolor[RGB]{47.90272758709304, 110.77698602648104, 26.874258439388008}$0.953 \pm 0.001$ & \cellcolor[RGB]{204.85000757059112, 234.06522068286571, 167.5413165266031}$0.309 \pm 0.006$ & \cellcolor[RGB]{227.7924913243222, 244.04021361927053, 204.44879039130097}$0.210 \pm 0.004$ & \cellcolor[RGB]{241.8317127469549, 246.3919662055241, 235.1433410077201}$0.061 \pm 0.001$ & \cellcolor[RGB]{247.07646935845486, 246.70686745925636, 246.89804085539353}$18.353 \pm 2.265$ & \cellcolor[RGB]{248.23224544525146, 242.27639245986938, 245.35700607299805}$1.643 \pm 0.328$ \\
    CLOVE (with SA) & \cellcolor[RGB]{206.30071998696607, 234.69596521172437, 169.8750712833802}$0.303 \pm 0.006$ & \cellcolor[RGB]{229.89018885154468, 244.95225602241072, 207.8233472829197}$0.200 \pm 0.002$ & \cellcolor[RGB]{47.97503039546168, 110.86451047871677, 26.88948008325509}$0.953 \pm 0.001$ & \cellcolor[RGB]{202.8874762709693, 233.21194620476928, 164.38420095764627}$0.318 \pm 0.005$ & \cellcolor[RGB]{226.66633985792222, 243.5505825469227, 202.63715542361402}$0.214 \pm 0.004$ & \cellcolor[RGB]{241.71397410742443, 246.37811460087346, 234.8732347170325}$0.062 \pm 0.001$ & \cellcolor[RGB]{247.06553659597222, 246.74877638210648, 246.91261787203703}$15.729 \pm 3.652$ & \cellcolor[RGB]{248.01933884620667, 243.09253442287445, 245.6408815383911}$1.359 \pm 0.481$ \\
    CLOVE (dendr.) & \cellcolor[RGB]{207.68130901684538, 235.2962213116719, 172.09601885318602}$0.297 \pm 0.007$ & \cellcolor[RGB]{232.2497576236816, 245.26467736749194, 213.16120866609307}$0.174 \pm 0.004$ & \cellcolor[RGB]{48.28446584414267, 111.23909023238323, 26.95462438824056}$0.951 \pm 0.001$ & \cellcolor[RGB]{233.1960815786003, 245.37600959748238, 215.3321871509066}$0.162 \pm 0.005$ & \cellcolor[RGB]{237.66569733503508, 245.90184674529823, 225.58601153331574}$0.110 \pm 0.003$ & \cellcolor[RGB]{244.31633295426184, 246.68427446520727, 240.8433520715419}$0.032 \pm 0.001$ & \cellcolor[RGB]{247.31401984368054, 245.7962572658912, 246.5813068750926}$75.365 \pm 13.718$ & \cellcolor[RGB]{248.694162607193, 240.5057100057602, 244.74111652374268}$2.259 \pm 0.423$ \\
    CLOVE (Louvain) & \cellcolor[RGB]{208.10514230604576, 235.48049665480252, 172.77783762276928}$0.295 \pm 0.005$ & \cellcolor[RGB]{230.0158946269954, 245.0018699561171, 208.03646414428354}$0.200 \pm 0.005$ & \cellcolor[RGB]{48.41597020145038, 111.3982797175452, 26.982309516094816}$0.950 \pm 0.001$ & \cellcolor[RGB]{210.55024636965408, 236.54358537811046, 176.71126589900874}$0.285 \pm 0.011$ & \cellcolor[RGB]{230.6872268270314, 245.08085021494486, 209.57657919142497}$0.192 \pm 0.008$ & \cellcolor[RGB]{242.2379165057348, 246.43975488302763, 236.07522021903867}$0.056 \pm 0.002$ & \cellcolor[RGB]{247.09011976175347, 246.65454091327837, 246.87984031766203}$21.629 \pm 4.908$ & \cellcolor[RGB]{248.70502161979675, 240.4640837907791, 244.726637840271}$2.273 \pm 0.351$ \\
    CLOVE (k1 decomp.) & \cellcolor[RGB]{207.9029073924246, 235.39256843148894, 172.4525031965091}$0.296 \pm 0.005$ & \cellcolor[RGB]{246.99892730943967, 246.99987380111054, 246.99753912165573}$0.000 \pm 0.000$ & \cellcolor[RGB]{47.52522024265083, 110.32000345162996, 26.794783208979123}$0.955 \pm 0.002$ & \cellcolor[RGB]{194.71954803213487, 229.66067305744994, 151.2444903125648}$0.353 \pm 0.004$ & \cellcolor[RGB]{221.67131104355727, 241.37883088850316, 194.60167428746172}$0.236 \pm 0.003$ & \cellcolor[RGB]{240.2338838442826, 246.20398633462148, 231.47773352511888}$0.080 \pm 0.001$ & \cellcolor[RGB]{247.08206576782987, 246.6854145566522, 246.89057897622686}$19.696 \pm 9.904$ & \cellcolor[RGB]{247.83687567710876, 243.79197657108307, 245.88416576385498}$1.116 \pm 0.484$ \\
    HMCS & \cellcolor[RGB]{221.47721034143686, 241.2944392788856, 194.28942533187669}$0.237 \pm 0.007$ & \cellcolor[RGB]{243.76981448292898, 246.61997817446223, 239.5895744020135}$0.038 \pm 0.002$ & \cellcolor[RGB]{49.12645628716976, 112.25834182131075, 27.131885534141002}$0.947 \pm 0.002$ & \cellcolor[RGB]{241.74799264190514, 246.3821167814006, 234.95127723731179}$0.062 \pm 0.002$ & \cellcolor[RGB]{243.06529795320418, 246.53709387684756, 237.97333059852724}$0.046 \pm 0.001$ & \cellcolor[RGB]{245.87543151280198, 246.86769782503552, 244.4201075881928}$0.013 \pm 0.000$ & \cellcolor[RGB]{247.03416503899305, 246.8690340171933, 246.95444661467593}$8.200 \pm 0.760$ & \cellcolor[RGB]{248.24382042884827, 242.23202168941498, 245.34157276153564}$1.658 \pm 0.202$ \\
    \thickhline
\end{tabular}
\caption{Average quality scores of the considered hyperbolic embedding algorithms for the 'douban' network. We show the results for the mapping accuracy, MA ($1^{\rm st}$ column), the edge prediction precision, EPP ($2^{\rm nd}$ column), the area under the receiver operating characteristic curve, AUC ($3^{\rm rd}$ column), the greedy routing score, GR ($4^{\rm th}$ column), the greedy success rate, GS ($5^{\rm th}$ column) and the greedy routing efficiency, GE ($6^{\rm th}$ column averaged over 2 different realizations. Beside the quality scores, we also display the real time in minutes ($7^{\rm th}$ column), the user time in minutes ($8^{\rm th}$ column) and the peak memory usage in GB ($9^{\rm th}$ column). In the top part of the table, we list the scores obtained for CLOVE with default settings ($1^{\rm st}$ row), for CLOVE with simulated annealing optimisation during the solution of the TSP problem ($2^{\rm nd}$ row) and for CLOVE with Louvain communities ($3^{\rm rd}$ row). For comparison, in the  $4^{\rm th}$ row we give the results for HMCS.}
\end{table}
\begin{table}
\begin{tabular}{|p{0.19\textwidth}|*{8}{>{\centering\arraybackslash}p{0.11\textwidth}|}}
    \thickhline
    & \makecell{MA} & \makecell{EPP} & \makecell{AUC} & \makecell{GR} & \makecell{GS} & \makecell{GE} & \makecell{Running \\ Time \\ (min)} & \makecell{Peak \\ Mem. \\ (GB)} \\
    \thickhline
    CLOVE & \cellcolor[RGB]{181.19511847220144, 223.17928742932375, 130.60461709792807}$0.410 \pm 0.009$ & \cellcolor[RGB]{213.4990616469387, 237.82567897692988, 181.45501221464056}$0.272 \pm 0.001$ & \cellcolor[RGB]{52.56556150035436, 116.42146918463949, 27.85590768428513}$0.929 \pm 0.003$ & \cellcolor[RGB]{126.00523838427912, 187.16440024279447, 64.36335256593864}$0.604 \pm 0.002$ & \cellcolor[RGB]{177.48185790961696, 220.76892530975135, 126.10961746953632}$0.423 \pm 0.002$ & \cellcolor[RGB]{234.6707608359659, 245.5495012748195, 218.7152748589806}$0.145 \pm 0.001$ & \cellcolor[RGB]{247.04942255572917, 246.81054686970487, 246.93410325902778}$11.861 \pm 1.782$ & \cellcolor[RGB]{247.80208778381348, 243.92533016204834, 245.93054962158203}$1.069 \pm 0.077$ \\
    CLOVE (with SA) & \cellcolor[RGB]{182.5995622995026, 224.09094394879992, 132.3047333099242}$0.405 \pm 0.004$ & \cellcolor[RGB]{213.74898130091583, 237.93433969605036, 181.85705687538632}$0.271 \pm 0.001$ & \cellcolor[RGB]{52.86020461536045, 116.77814242912055, 27.917937813760094}$0.927 \pm 0.002$ & \cellcolor[RGB]{128.461148246525, 188.94846465125306, 66.76875840368814}$0.595 \pm 0.006$ & \cellcolor[RGB]{179.2285130005268, 221.90271896525422, 128.2239894216903}$0.417 \pm 0.005$ & \cellcolor[RGB]{234.82584475020988, 245.56774644120117, 219.07105560342268}$0.143 \pm 0.002$ & \cellcolor[RGB]{247.0440396418229, 246.83118137301216, 246.94128047756945}$10.570 \pm 1.329$ & \cellcolor[RGB]{247.82581996917725, 243.83435678482056, 245.89890670776367}$1.101 \pm 0.094$ \\
    CLOVE (dendr.) & \cellcolor[RGB]{183.03030993615909, 224.37055206382257, 132.826164659561}$0.403 \pm 0.003$ & \cellcolor[RGB]{236.02740815470938, 245.70910684173052, 221.82758341374506}$0.129 \pm 0.005$ & \cellcolor[RGB]{53.99911733189916, 118.15682624387793, 28.15770891197877}$0.921 \pm 0.001$ & \cellcolor[RGB]{238.71027887434792, 246.02473869109974, 227.98240447644525}$0.098 \pm 0.002$ & \cellcolor[RGB]{240.76964899242364, 246.26701752852043, 232.70684180614833}$0.073 \pm 0.001$ & \cellcolor[RGB]{244.73360938093512, 246.73336580952179, 241.80063328567468}$0.027 \pm 0.000$ & \cellcolor[RGB]{247.05305792654514, 246.796611281577, 246.92925609793983}$12.734 \pm 2.299$ & \cellcolor[RGB]{248.10949969291687, 242.74691784381866, 245.52066707611084}$1.479 \pm 0.250$ \\
    CLOVE (Louvain) & \cellcolor[RGB]{184.1150906340082, 225.05003940609052, 134.18514580253492}$0.399 \pm 0.012$ & \cellcolor[RGB]{213.83027444684367, 237.96968454210594, 181.98783280579198}$0.270 \pm 0.001$ & \cellcolor[RGB]{53.460028098803164, 117.50424454065646, 28.044216441853298}$0.924 \pm 0.003$ & \cellcolor[RGB]{134.5709910940473, 192.9145029908728, 74.16488395595199}$0.573 \pm 0.012$ & \cellcolor[RGB]{183.5375061158273, 224.69978467167738, 133.44013898231725}$0.402 \pm 0.006$ & \cellcolor[RGB]{235.23345945714485, 245.61570111260528, 220.0061716958029}$0.138 \pm 0.002$ & \cellcolor[RGB]{247.0594818739236, 246.7719861499595, 246.92069083476852}$14.276 \pm 0.999$ & \cellcolor[RGB]{247.8657877445221, 243.68114697933197, 245.8456163406372}$1.154 \pm 0.056$ \\
    CLOVE (k1 decomp.) & \cellcolor[RGB]{181.0052340745819, 223.0560291361321, 130.37475703765176}$0.411 \pm 0.011$ & \cellcolor[RGB]{214.94098486700565, 238.45260211608942, 183.77462782953083}$0.265 \pm 0.002$ & \cellcolor[RGB]{52.75210112208548, 116.64728030568241, 27.89517918359694}$0.928 \pm 0.003$ & \cellcolor[RGB]{121.14350164223796, 183.08054137947988, 61.2518410510323}$0.623 \pm 0.016$ & \cellcolor[RGB]{173.50811342044534, 218.18947713256978, 121.29929519317069}$0.437 \pm 0.010$ & \cellcolor[RGB]{233.96034649971773, 245.46592311761384, 217.08550079347006}$0.153 \pm 0.003$ & \cellcolor[RGB]{247.03944725819446, 246.84878551025463, 246.94740365574074}$9.467 \pm 0.729$ & \cellcolor[RGB]{247.78728413581848, 243.9820774793625, 245.9502878189087}$1.050 \pm 0.040$ \\
    HMCS & \cellcolor[RGB]{194.28250796554568, 229.47065563719377, 150.54142585761696}$0.355 \pm 0.011$ & \cellcolor[RGB]{243.2242406385166, 246.55579301629606, 238.33796381777333}$0.044 \pm 0.003$ & \cellcolor[RGB]{54.014758071611105, 118.17575977089766, 28.161001699286548}$0.921 \pm 0.002$ & \cellcolor[RGB]{237.1382586018837, 245.8397951296334, 224.37600502785088}$0.116 \pm 0.005$ & \cellcolor[RGB]{239.4796188498425, 246.11524927645206, 229.74736089081512}$0.088 \pm 0.003$ & \cellcolor[RGB]{244.2991760378737, 246.68225600445572, 240.80399208688672}$0.032 \pm 0.001$ & \cellcolor[RGB]{247.02324506276042, 246.91089392608507, 246.9690065829861}$5.579 \pm 0.573$ & \cellcolor[RGB]{247.81096696853638, 243.89129328727722, 245.91871070861816}$1.081 \pm 0.014$ \\
    \thickhline
\end{tabular}
\caption{Average quality scores of the considered hyperbolic embedding algorithms for the 'facebook-wosn-links' network. We show the results for the mapping accuracy, MA ($1^{\rm st}$ column), the edge prediction precision, EPP ($2^{\rm nd}$ column), the area under the receiver operating characteristic curve, AUC ($3^{\rm rd}$ column), the greedy routing score, GR ($4^{\rm th}$ column), the greedy success rate, GS ($5^{\rm th}$ column) and the greedy routing efficiency, GE ($6^{\rm th}$ column averaged over 2 different realizations. Beside the quality scores, we also display the real time in minutes ($7^{\rm th}$ column), the user time in minutes ($8^{\rm th}$ column) and the peak memory usage in GB ($9^{\rm th}$ column). In the top part of the table, we list the scores obtained for CLOVE with default settings ($1^{\rm st}$ row), for CLOVE with simulated annealing optimisation during the solution of the TSP problem ($2^{\rm nd}$ row) and for CLOVE with Louvain communities ($3^{\rm rd}$ row). For comparison, in the  $4^{\rm th}$ row we give the results for HMCS.}
\end{table}
\begin{table}
\begin{tabular}{|p{0.19\textwidth}|*{8}{>{\centering\arraybackslash}p{0.11\textwidth}|}}
    \thickhline
    & \makecell{MA} & \makecell{EPP} & \makecell{AUC} & \makecell{GR} & \makecell{GS} & \makecell{GE} & \makecell{Running \\ Time \\ (min)} & \makecell{Peak \\ Mem. \\ (GB)} \\
    \thickhline
    CLOVE & \cellcolor[RGB]{234.54711110806198, 245.5349542480073, 218.43160783614218}$0.147 \pm 0.010$ & \cellcolor[RGB]{219.75314904835696, 240.54484741232912, 191.51593542561773}$0.245 \pm 0.006$ & \cellcolor[RGB]{46.964258555528545, 109.64094456721877, 26.676686011690222}$0.958 \pm 0.001$ & \cellcolor[RGB]{192.16750619564235, 228.55108965027927, 147.13903170603334}$0.364 \pm 0.007$ & \cellcolor[RGB]{221.04065889535693, 241.1046343023291, 193.58714691861763}$0.239 \pm 0.004$ & \cellcolor[RGB]{240.2247784677857, 246.20291511385713, 231.45684472021426}$0.080 \pm 0.001$ & \cellcolor[RGB]{247.2298101217014, 246.11906120014467, 246.69358650439816}$55.154 \pm 3.191$ & \cellcolor[RGB]{249.29185819625854, 238.2145435810089, 243.94418907165527}$3.056 \pm 0.464$ \\
    CLOVE (with SA) & \cellcolor[RGB]{235.8344778958333, 245.6864091642157, 221.38497870220576}$0.131 \pm 0.014$ & \cellcolor[RGB]{219.17214827435598, 240.29223838015477, 190.58128200657265}$0.247 \pm 0.003$ & \cellcolor[RGB]{47.08580451736263, 109.78807915259686, 26.70227463523424}$0.957 \pm 0.001$ & \cellcolor[RGB]{191.60026594286603, 228.30446345342, 146.22651477765405}$0.367 \pm 0.012$ & \cellcolor[RGB]{220.73727388938448, 240.97272777799327, 193.09909277857506}$0.240 \pm 0.009$ & \cellcolor[RGB]{240.1716180597167, 246.19666094820198, 231.33488848993835}$0.080 \pm 0.002$ & \cellcolor[RGB]{247.25680276309026, 246.01558940815394, 246.65759631587963}$61.633 \pm 18.897$ & \cellcolor[RGB]{249.50132942199707, 237.41157054901123, 243.6648941040039}$3.335 \pm 0.545$ \\
    CLOVE (dendr.) & \cellcolor[RGB]{236.4842104034306, 245.76284828275655, 222.87554151375255}$0.124 \pm 0.027$ & \cellcolor[RGB]{237.52573725062854, 245.8853808530151, 225.26492663379486}$0.111 \pm 0.014$ & \cellcolor[RGB]{50.610865459371354, 114.05525818766006, 27.444392728288705}$0.939 \pm 0.011$ & \cellcolor[RGB]{242.1458276932852, 246.42892090509238, 235.86395764930134}$0.057 \pm 0.007$ & \cellcolor[RGB]{243.22038613963537, 246.55533954583944, 238.3291211438694}$0.044 \pm 0.005$ & \cellcolor[RGB]{245.66474592108298, 246.8429112848333, 243.93677005424917}$0.016 \pm 0.002$ & \cellcolor[RGB]{247.05336243652778, 246.7954439933102, 246.92885008462963}$12.807 \pm 3.920$ & \cellcolor[RGB]{249.32181239128113, 238.09971916675568, 243.9042501449585}$3.096 \pm 0.615$ \\
    CLOVE (Louvain) & \cellcolor[RGB]{235.6816841915531, 245.66843343430037, 221.03445196885716}$0.133 \pm 0.003$ & \cellcolor[RGB]{219.8398428116345, 240.58254035288456, 191.65539930567286}$0.244 \pm 0.005$ & \cellcolor[RGB]{47.462236704103326, 110.24376022075666, 26.78152351665333}$0.955 \pm 0.001$ & \cellcolor[RGB]{196.89609540991881, 230.60699800431252, 154.74589261595634}$0.344 \pm 0.002$ & \cellcolor[RGB]{223.90142619487818, 242.34844617168616, 198.1892508352388}$0.227 \pm 0.002$ & \cellcolor[RGB]{240.54605333637133, 246.24071215722014, 232.19388706579306}$0.076 \pm 0.000$ & \cellcolor[RGB]{247.37344172637154, 245.56847338224247, 246.5020776981713}$89.626 \pm 14.368$ & \cellcolor[RGB]{249.3816499710083, 237.8703417778015, 243.82446670532227}$3.176 \pm 0.190$ \\
    CLOVE (k1 decomp.) & \cellcolor[RGB]{235.41443569648675, 245.6369924348808, 220.42135248017544}$0.136 \pm 0.006$ & \cellcolor[RGB]{219.32900350940005, 240.3604363084348, 190.83361434120877}$0.246 \pm 0.004$ & \cellcolor[RGB]{47.17702540308774, 109.89850443531674, 26.721479032229}$0.957 \pm 0.002$ & \cellcolor[RGB]{191.9822378183059, 228.47053818187214, 146.84099127292689}$0.365 \pm 0.004$ & \cellcolor[RGB]{221.10822872953486, 241.1340124911021, 193.69584621707781}$0.239 \pm 0.004$ & \cellcolor[RGB]{240.22155563224263, 246.20253595673444, 231.44945115632135}$0.080 \pm 0.001$ & \cellcolor[RGB]{247.25150445762154, 246.03589957911748, 246.6646607231713}$60.361 \pm 9.104$ & \cellcolor[RGB]{249.07506775856018, 239.04557359218597, 244.23324298858643}$2.767 \pm 0.359$ \\
    HMCS & \cellcolor[RGB]{230.3884502811271, 245.04570003307379, 208.89115064493865}$0.195 \pm 0.034$ & \cellcolor[RGB]{243.8123789133464, 246.62498575451136, 239.6872222129712}$0.038 \pm 0.003$ & \cellcolor[RGB]{62.19384635705608, 128.0767613795942, 29.88291502253812}$0.878 \pm 0.005$ & \cellcolor[RGB]{243.873059758081, 246.6321246774213, 239.82643120971528}$0.037 \pm 0.004$ & \cellcolor[RGB]{244.39715362948752, 246.6937827799397, 241.0287642088243}$0.031 \pm 0.003$ & \cellcolor[RGB]{246.0334852086923, 246.88629237749322, 244.7827013611176}$0.011 \pm 0.001$ & \cellcolor[RGB]{247.08390147534723, 246.67837767783564, 246.8881313662037}$20.136 \pm 2.708$ & \cellcolor[RGB]{249.35943484306335, 237.95549976825714, 243.85408687591553}$3.146 \pm 0.658$ \\
    \thickhline
\end{tabular}
\caption{Average quality scores of the considered hyperbolic embedding algorithms for the 'flickrEdges' network. We show the results for the mapping accuracy, MA ($1^{\rm st}$ column), the edge prediction precision, EPP ($2^{\rm nd}$ column), the area under the receiver operating characteristic curve, AUC ($3^{\rm rd}$ column), the greedy routing score, GR ($4^{\rm th}$ column), the greedy success rate, GS ($5^{\rm th}$ column) and the greedy routing efficiency, GE ($6^{\rm th}$ column averaged over 2 different realizations. Beside the quality scores, we also display the real time in minutes ($7^{\rm th}$ column), the user time in minutes ($8^{\rm th}$ column) and the peak memory usage in GB ($9^{\rm th}$ column). In the top part of the table, we list the scores obtained for CLOVE with default settings ($1^{\rm st}$ row), for CLOVE with simulated annealing optimisation during the solution of the TSP problem ($2^{\rm nd}$ row) and for CLOVE with Louvain communities ($3^{\rm rd}$ row). For comparison, in the  $4^{\rm th}$ row we give the results for HMCS.}
\end{table}
\begin{table}
\begin{tabular}{|p{0.19\textwidth}|*{8}{>{\centering\arraybackslash}p{0.11\textwidth}|}}
    \thickhline
    & \makecell{MA} & \makecell{EPP} & \makecell{AUC} & \makecell{GR} & \makecell{GS} & \makecell{GE} & \makecell{Running \\ Time \\ (min)} & \makecell{Peak \\ Mem. \\ (GB)} \\
    \thickhline
    CLOVE & \cellcolor[RGB]{201.16929772833367, 232.46491205579724, 161.6201746064498}$0.325 \pm 0.009$ & \cellcolor[RGB]{240.89822253038687, 246.28214382710433, 233.00180462853456}$0.072 \pm 0.001$ & \cellcolor[RGB]{62.473950752839194, 128.41583512185798, 29.941884369018776}$0.876 \pm 0.003$ & \cellcolor[RGB]{92.79729863883784, 159.2697308566238, 43.11027112885621}$0.737 \pm 0.007$ & \cellcolor[RGB]{156.43738020959393, 207.1084748728943, 100.6347234116137}$0.497 \pm 0.005$ & \cellcolor[RGB]{227.3149918581498, 243.8326051557173, 203.68063907615402}$0.212 \pm 0.002$ & \cellcolor[RGB]{247.2026927186632, 246.2230112451244, 246.7297430417824}$48.646 \pm 9.569$ & \cellcolor[RGB]{248.80933666229248, 240.06420946121216, 244.58755111694336}$2.412 \pm 0.265$ \\
    CLOVE (with SA) & \cellcolor[RGB]{200.46983102123394, 232.16079609618868, 160.49494555589808}$0.328 \pm 0.009$ & \cellcolor[RGB]{240.83396456345568, 246.2745840662889, 232.85438929263358}$0.073 \pm 0.000$ & \cellcolor[RGB]{62.30562017072299, 128.21206652245414, 29.906446351731155}$0.877 \pm 0.004$ & \cellcolor[RGB]{91.39341783767429, 158.0904709836464, 42.21178741611155}$0.742 \pm 0.007$ & \cellcolor[RGB]{155.25393493596826, 206.34027355492677, 99.20213176459316}$0.501 \pm 0.005$ & \cellcolor[RGB]{226.88780098070666, 243.6468699916116, 202.99341896896286}$0.214 \pm 0.002$ & \cellcolor[RGB]{247.2500592142014, 246.04143967889468, 246.66658771439816}$60.014 \pm 18.787$ & \cellcolor[RGB]{249.0720386505127, 239.05718517303467, 244.2372817993164}$2.763 \pm 0.363$ \\
    CLOVE (dendr.) & \cellcolor[RGB]{203.31599489902908, 233.39825865175177, 165.07355701148154}$0.316 \pm 0.005$ & \cellcolor[RGB]{245.1585436021601, 246.78335807084235, 242.77548238142612}$0.022 \pm 0.002$ & \cellcolor[RGB]{61.75550595753907, 127.54613879070519, 29.79063283316612}$0.880 \pm 0.002$ & \cellcolor[RGB]{241.8487078719666, 246.39396563199608, 235.18232982392337}$0.061 \pm 0.004$ & \cellcolor[RGB]{242.71202239256024, 246.49553204618357, 237.1628749005794}$0.050 \pm 0.003$ & \cellcolor[RGB]{245.1394181236461, 246.7811080145466, 242.7316062836587}$0.022 \pm 0.001$ & \cellcolor[RGB]{247.06490777050348, 246.75118687973668, 246.91345630599537}$15.578 \pm 2.134$ & \cellcolor[RGB]{249.52077221870422, 237.33703982830048, 243.63897037506104}$3.361 \pm 0.455$ \\
    CLOVE (Louvain) & \cellcolor[RGB]{201.54736076166094, 232.62928728767866, 162.22836296441108}$0.324 \pm 0.010$ & \cellcolor[RGB]{240.97876827630571, 246.29161979721243, 233.18658604564254}$0.071 \pm 0.001$ & \cellcolor[RGB]{62.65965133587449, 128.64063056447964, 29.980979228605158}$0.875 \pm 0.003$ & \cellcolor[RGB]{100.07934764372621, 165.38665202073003, 47.770782491984775}$0.708 \pm 0.010$ & \cellcolor[RGB]{163.12240422037428, 211.44787642375172, 108.72712089834783}$0.473 \pm 0.007$ & \cellcolor[RGB]{229.3736177486687, 244.7276598907255, 206.99234159568442}$0.203 \pm 0.003$ & \cellcolor[RGB]{247.29531000559027, 245.86797831190393, 246.60625332587963}$70.874 \pm 13.084$ & \cellcolor[RGB]{249.1283836364746, 238.84119606018066, 244.1621551513672}$2.838 \pm 0.243$ \\
    CLOVE (k1 decomp.) & \cellcolor[RGB]{201.18158499983494, 232.47025434775432, 161.639941086691}$0.325 \pm 0.010$ & \cellcolor[RGB]{240.8669180722811, 246.27846094968012, 232.92998851876249}$0.072 \pm 0.001$ & \cellcolor[RGB]{62.78031447065911, 128.78669646448208, 30.00638199382297}$0.875 \pm 0.007$ & \cellcolor[RGB]{90.75815306139312, 157.5568485715702, 41.805217959291596}$0.745 \pm 0.005$ & \cellcolor[RGB]{155.04941349646242, 206.20751402401947, 98.95455317992818}$0.502 \pm 0.004$ & \cellcolor[RGB]{225.4744686803918, 243.03237768712688, 200.7197974423694}$0.220 \pm 0.002$ & \cellcolor[RGB]{247.23366961592015, 246.10426647230614, 246.68844051210647}$56.081 \pm 8.887$ & \cellcolor[RGB]{249.121022939682, 238.8694120645523, 244.17196941375732}$2.828 \pm 0.415$ \\
    HMCS & \cellcolor[RGB]{207.0209123576431, 235.00909232941007, 171.0336416188172}$0.300 \pm 0.010$ & \cellcolor[RGB]{246.44236547587826, 246.9343959383386, 245.72072079760306}$0.007 \pm 0.000$ & \cellcolor[RGB]{64.94265117330089, 131.4042619466274, 30.461610773326502}$0.863 \pm 0.005$ & \cellcolor[RGB]{241.02445663226925, 246.29699489791403, 233.29140050932355}$0.070 \pm 0.003$ & \cellcolor[RGB]{242.05732011869208, 246.41850824925788, 235.66091086052887}$0.058 \pm 0.003$ & \cellcolor[RGB]{244.82429010396794, 246.74403412987857, 242.0086655326323}$0.026 \pm 0.001$ & \cellcolor[RGB]{247.0726179743403, 246.72163109836225, 246.90317603421295}$17.428 \pm 5.562$ & \cellcolor[RGB]{248.83164477348328, 239.97869503498077, 244.55780696868896}$2.442 \pm 0.210$ \\
    \thickhline
\end{tabular}
\caption{Average quality scores of the considered hyperbolic embedding algorithms for the 'livemocha' network. We show the results for the mapping accuracy, MA ($1^{\rm st}$ column), the edge prediction precision, EPP ($2^{\rm nd}$ column), the area under the receiver operating characteristic curve, AUC ($3^{\rm rd}$ column), the greedy routing score, GR ($4^{\rm th}$ column), the greedy success rate, GS ($5^{\rm th}$ column) and the greedy routing efficiency, GE ($6^{\rm th}$ column averaged over 2 different realizations. Beside the quality scores, we also display the real time in minutes ($7^{\rm th}$ column), the user time in minutes ($8^{\rm th}$ column) and the peak memory usage in GB ($9^{\rm th}$ column). In the top part of the table, we list the scores obtained for CLOVE with default settings ($1^{\rm st}$ row), for CLOVE with simulated annealing optimisation during the solution of the TSP problem ($2^{\rm nd}$ row) and for CLOVE with Louvain communities ($3^{\rm rd}$ row). For comparison, in the  $4^{\rm th}$ row we give the results for HMCS.}
\end{table}
\begin{table}
\begin{tabular}{|p{0.19\textwidth}|*{8}{>{\centering\arraybackslash}p{0.11\textwidth}|}}
    \thickhline
    & \makecell{MA} & \makecell{EPP} & \makecell{AUC} & \makecell{GR} & \makecell{GS} & \makecell{GE} & \makecell{Running \\ Time \\ (min)} & \makecell{Peak \\ Mem. \\ (GB)} \\
    \thickhline
    CLOVE & \cellcolor[RGB]{184.72279776548973, 225.314259898039, 135.16276162274437}$0.397 \pm 0.003$ & \cellcolor[RGB]{187.05725741181703, 226.32924235296392, 138.91819670596652}$0.387 \pm 0.004$ & \cellcolor[RGB]{44.552818939890656, 106.72183345355185, 26.16901451366119}$0.971 \pm 0.001$ & \cellcolor[RGB]{142.87617592213405, 198.30558787928, 84.2185287478465}$0.544 \pm 0.006$ & \cellcolor[RGB]{173.47869855085239, 218.17038326985156, 121.26368771945289}$0.437 \pm 0.005$ & \cellcolor[RGB]{234.35126883174937, 245.5119139802058, 217.98232261401324}$0.149 \pm 0.002$ & \cellcolor[RGB]{247.02866622616318, 246.89011279970777, 246.96177836511575}$6.880 \pm 1.251$ & \cellcolor[RGB]{247.51403069496155, 245.0295490026474, 246.31462574005127}$0.685 \pm 0.167$ \\
    CLOVE (with SA) & \cellcolor[RGB]{184.43318303967538, 225.18834045203278, 134.69685967252127}$0.398 \pm 0.003$ & \cellcolor[RGB]{187.89979511951913, 226.6955630954431, 140.27358345313945}$0.383 \pm 0.008$ & \cellcolor[RGB]{44.56161318893806, 106.73247912345134, 26.170865934513277}$0.971 \pm 0.001$ & \cellcolor[RGB]{141.11704154754284, 197.16369363612432, 82.08905029439399}$0.550 \pm 0.011$ & \cellcolor[RGB]{172.4628170360641, 217.51095140937494, 120.03393641207761}$0.440 \pm 0.008$ & \cellcolor[RGB]{234.24179854384957, 245.49903512280582, 217.73118489471375}$0.150 \pm 0.002$ & \cellcolor[RGB]{247.03134803998265, 246.8798325133999, 246.9582026133565}$7.524 \pm 0.250$ & \cellcolor[RGB]{247.46964120864868, 245.19970870018005, 246.37381172180176}$0.626 \pm 0.043$ \\
    CLOVE (dendr.) & \cellcolor[RGB]{184.41267953903895, 225.17942588653867, 134.6638757801931}$0.398 \pm 0.002$ & \cellcolor[RGB]{220.45044703418534, 240.8480204496458, 192.63767566368946}$0.242 \pm 0.006$ & \cellcolor[RGB]{45.260889780210114, 107.57897183920173, 26.318082058991603}$0.967 \pm 0.000$ & \cellcolor[RGB]{234.57172944834, 245.53785052333413, 218.48808520501527}$0.146 \pm 0.007$ & \cellcolor[RGB]{236.5847594960227, 245.77467758776737, 223.10621296146388}$0.123 \pm 0.005$ & \cellcolor[RGB]{243.3395834457283, 246.56936275832098, 238.60257378725902}$0.043 \pm 0.002$ & \cellcolor[RGB]{247.07792697612848, 246.70127992484086, 246.89609736516203}$18.702 \pm 1.792$ & \cellcolor[RGB]{247.8179178237915, 243.86464834213257, 245.90944290161133}$1.091 \pm 0.046$ \\
    CLOVE (Louvain) & \cellcolor[RGB]{184.31369299279822, 225.13638825773836, 134.50463655363194}$0.399 \pm 0.004$ & \cellcolor[RGB]{188.02464964860573, 226.74984767330685, 140.47443639123532}$0.383 \pm 0.002$ & \cellcolor[RGB]{44.745310397763866, 106.95484942887205, 26.209539031108182}$0.970 \pm 0.001$ & \cellcolor[RGB]{148.97892376672175, 202.26702069067903, 91.60606561234738}$0.523 \pm 0.007$ & \cellcolor[RGB]{178.4368161619962, 221.38881049112035, 127.26561956452173}$0.420 \pm 0.006$ & \cellcolor[RGB]{234.818700475445, 245.56690593828765, 219.05466579660919}$0.143 \pm 0.002$ & \cellcolor[RGB]{247.0333895969618, 246.87200654497974, 246.95548053738426}$8.014 \pm 1.032$ & \cellcolor[RGB]{247.53619861602783, 244.9445719718933, 246.2850685119629}$0.715 \pm 0.167$ \\
    CLOVE (k1 decomp.) & \cellcolor[RGB]{185.96305124308083, 225.85350054046992, 137.15795199973874}$0.391 \pm 0.004$ & \cellcolor[RGB]{196.20866813085388, 230.30811657863214, 153.6400313409389}$0.347 \pm 0.002$ & \cellcolor[RGB]{44.56276814880158, 106.73387723275981, 26.17110908395823}$0.971 \pm 0.001$ & \cellcolor[RGB]{136.08406774994037, 193.89667555697883, 75.99650306571729}$0.568 \pm 0.007$ & \cellcolor[RGB]{167.91556696232658, 214.55922767729973, 114.52937053334271}$0.456 \pm 0.006$ & \cellcolor[RGB]{232.3678847169974, 245.27857467258792, 213.43220611546462}$0.172 \pm 0.002$ & \cellcolor[RGB]{247.01640557411457, 246.9371119658941, 246.97812590118056}$3.937 \pm 0.439$ & \cellcolor[RGB]{247.43545269966125, 245.33076465129852, 246.41939640045166}$0.581 \pm 0.057$ \\
    HMCS & \cellcolor[RGB]{185.55571867317872, 225.67639942312118, 136.50267786554838}$0.393 \pm 0.001$ & \cellcolor[RGB]{239.75201265187812, 246.1472956061033, 230.37226431901453}$0.085 \pm 0.001$ & \cellcolor[RGB]{45.1195397847896, 107.40786395000848, 26.288324165218864}$0.968 \pm 0.000$ & \cellcolor[RGB]{235.88165622635114, 245.6919595560413, 221.49321134280555}$0.131 \pm 0.005$ & \cellcolor[RGB]{237.56812839817232, 245.8903680468438, 225.36217691345416}$0.111 \pm 0.004$ & \cellcolor[RGB]{243.68086290096613, 246.6095132824666, 239.38550900809878}$0.039 \pm 0.001$ & \cellcolor[RGB]{247.01431762652777, 246.94511576497686, 246.9809098312963}$3.436 \pm 0.206$ & \cellcolor[RGB]{247.4331681728363, 245.3395220041275, 246.42244243621826}$0.578 \pm 0.047$ \\
    \thickhline
\end{tabular}
\caption{Average quality scores of the considered hyperbolic embedding algorithms for the 'loc-brightkite\_edges' network. We show the results for the mapping accuracy, MA ($1^{\rm st}$ column), the edge prediction precision, EPP ($2^{\rm nd}$ column), the area under the receiver operating characteristic curve, AUC ($3^{\rm rd}$ column), the greedy routing score, GR ($4^{\rm th}$ column), the greedy success rate, GS ($5^{\rm th}$ column) and the greedy routing efficiency, GE ($6^{\rm th}$ column averaged over 2 different realizations. Beside the quality scores, we also display the real time in minutes ($7^{\rm th}$ column), the user time in minutes ($8^{\rm th}$ column) and the peak memory usage in GB ($9^{\rm th}$ column). In the top part of the table, we list the scores obtained for CLOVE with default settings ($1^{\rm st}$ row), for CLOVE with simulated annealing optimisation during the solution of the TSP problem ($2^{\rm nd}$ row) and for CLOVE with Louvain communities ($3^{\rm rd}$ row). For comparison, in the  $4^{\rm th}$ row we give the results for HMCS.}
\end{table}
\begin{table}
\begin{tabular}{|p{0.19\textwidth}|*{8}{>{\centering\arraybackslash}p{0.11\textwidth}|}}
    \thickhline
    & \makecell{MA} & \makecell{EPP} & \makecell{AUC} & \makecell{GR} & \makecell{GS} & \makecell{GE} & \makecell{Running \\ Time \\ (min)} & \makecell{Peak \\ Mem. \\ (GB)} \\
    \thickhline
    CLOVE & \cellcolor[RGB]{184.86056556190064, 225.3741589399568, 135.38438807784019}$0.396 \pm 0.002$ & \cellcolor[RGB]{191.18788609715745, 228.12516786832933, 145.56312111281852}$0.369 \pm 0.005$ & \cellcolor[RGB]{42.398948885410185, 104.11451707181233, 25.715568186402145}$0.982 \pm 0.001$ & \cellcolor[RGB]{127.65488154026777, 188.4250985436826, 65.79275133821888}$0.598 \pm 0.016$ & \cellcolor[RGB]{154.49618205691993, 205.8483988790533, 98.28485196363992}$0.504 \pm 0.013$ & \cellcolor[RGB]{228.72229938819, 244.44447799486522, 205.9445685810013}$0.206 \pm 0.005$ & \cellcolor[RGB]{248.55325303998265, 241.0458633467332, 244.9289959466898}$372.781 \pm 64.022$ & \cellcolor[RGB]{248.4369022846222, 241.49187457561493, 245.0841302871704}$1.916 \pm 0.334$ \\
    CLOVE (with SA) & \cellcolor[RGB]{186.1485837089283, 225.93416682996883, 137.45641727088469}$0.391 \pm 0.005$ & \cellcolor[RGB]{191.16936947374796, 228.1171171624991, 145.53333350124672}$0.369 \pm 0.006$ & \cellcolor[RGB]{42.36871325595853, 104.07791604668665, 25.709202790728114}$0.982 \pm 0.001$ & \cellcolor[RGB]{127.98512396741022, 188.6394664349856, 66.192518486865}$0.597 \pm 0.011$ & \cellcolor[RGB]{154.6307171923005, 205.935728703774, 98.4477102854164}$0.503 \pm 0.009$ & \cellcolor[RGB]{228.73502580021815, 244.45001121748615, 205.96504150469877}$0.205 \pm 0.004$ & \cellcolor[RGB]{248.39635079776042, 241.6473219419184, 245.13819893631944}$335.124 \pm 37.406$ & \cellcolor[RGB]{248.77564239501953, 240.1933708190918, 244.63247680664062}$2.368 \pm 0.299$ \\
    CLOVE (dendr.) & \cellcolor[RGB]{186.97411164334733, 226.29309201884666, 138.78444046973266}$0.387 \pm 0.001$ & \cellcolor[RGB]{233.10375782533785, 245.3651479794515, 215.12038559930448}$0.163 \pm 0.003$ & \cellcolor[RGB]{43.23862323120456, 105.13096496408974, 25.89234173288517}$0.978 \pm 0.001$ & \cellcolor[RGB]{220.76707145128023, 240.98568323968706, 193.14702798684212}$0.240 \pm 0.007$ & \cellcolor[RGB]{227.59406114285903, 243.95393962733002, 204.12957662112103}$0.210 \pm 0.007$ & \cellcolor[RGB]{239.51122146641808, 246.1189672313433, 229.81986101119443}$0.088 \pm 0.002$ & \cellcolor[RGB]{252.58265730270833, 225.59981367295137, 239.55645692972223}$1339.838 \pm 381.388$ & \cellcolor[RGB]{251.01708793640137, 231.60116291046143, 241.64388275146484}$5.356 \pm 0.422$ \\
    CLOVE (Louvain) & \cellcolor[RGB]{186.6280122042634, 226.14261400185364, 138.2276718068585}$0.389 \pm 0.004$ & \cellcolor[RGB]{193.3997125404569, 229.0868315393291, 149.12127669551765}$0.359 \pm 0.004$ & \cellcolor[RGB]{42.62509634285148, 104.38827452029389, 25.763178177442416}$0.981 \pm 0.001$ & \cellcolor[RGB]{130.0572816769419, 189.98455126397982, 68.70091992471916}$0.589 \pm 0.007$ & \cellcolor[RGB]{156.29131573447881, 207.01366109080203, 100.45790852068487}$0.497 \pm 0.006$ & \cellcolor[RGB]{229.30287473523978, 244.6969020587999, 206.87853761755966}$0.203 \pm 0.003$ & \cellcolor[RGB]{248.72887664975696, 240.37263950926504, 244.6948311336574}$414.930 \pm 37.695$ & \cellcolor[RGB]{248.58112144470215, 240.9390344619751, 244.89183807373047}$2.108 \pm 0.320$ \\
    CLOVE (k1 decomp.) & \cellcolor[RGB]{184.64736107174866, 225.2814613355429, 135.04140694150874}$0.397 \pm 0.001$ & \cellcolor[RGB]{245.45724036467786, 246.8184988664327, 243.4607278954374}$0.018 \pm 0.014$ & \cellcolor[RGB]{42.33418754468439, 104.03612176461795, 25.701934219933555}$0.982 \pm 0.001$ & \cellcolor[RGB]{123.04720204473834, 184.67964971758022, 62.47020930863254}$0.616 \pm 0.013$ & \cellcolor[RGB]{149.39644490101918, 202.53804318136332, 92.1114859328127}$0.521 \pm 0.010$ & \cellcolor[RGB]{223.02694482898636, 241.968236882168, 196.78247646402153}$0.230 \pm 0.004$ & \cellcolor[RGB]{247.86945522177083, 243.66708831654515, 245.84072637097222}$208.669 \pm 65.358$ & \cellcolor[RGB]{248.72118639945984, 240.40211880207062, 244.70508480072021}$2.295 \pm 0.490$ \\
    HMCS & \cellcolor[RGB]{189.62038092145144, 227.44364387889192, 143.04148235190013}$0.376 \pm 0.009$ & \cellcolor[RGB]{239.7068662479028, 246.14198426445915, 230.26869315695347}$0.086 \pm 0.004$ & \cellcolor[RGB]{42.98840991069113, 104.82807515504716, 25.839665244356027}$0.979 \pm 0.002$ & \cellcolor[RGB]{233.84781808124995, 245.45268448014704, 216.8273473628675}$0.155 \pm 0.006$ & \cellcolor[RGB]{235.36786630089787, 245.63151368245857, 220.31451680794217}$0.137 \pm 0.005$ & \cellcolor[RGB]{241.88837598986984, 246.39863246939646, 235.2733331532308}$0.060 \pm 0.002$ & \cellcolor[RGB]{247.2401711731771, 246.07934383615452, 246.67977176909721}$57.641 \pm 2.108$ & \cellcolor[RGB]{248.68184518814087, 240.55292677879333, 244.7575397491455}$2.242 \pm 0.465$ \\
    \thickhline
\end{tabular}
\caption{Average quality scores of the considered hyperbolic embedding algorithms for the 'loc-gowalla\_edges' network. We show the results for the mapping accuracy, MA ($1^{\rm st}$ column), the edge prediction precision, EPP ($2^{\rm nd}$ column), the area under the receiver operating characteristic curve, AUC ($3^{\rm rd}$ column), the greedy routing score, GR ($4^{\rm th}$ column), the greedy success rate, GS ($5^{\rm th}$ column) and the greedy routing efficiency, GE ($6^{\rm th}$ column averaged over 2 different realizations. Beside the quality scores, we also display the real time in minutes ($7^{\rm th}$ column), the user time in minutes ($8^{\rm th}$ column) and the peak memory usage in GB ($9^{\rm th}$ column). In the top part of the table, we list the scores obtained for CLOVE with default settings ($1^{\rm st}$ row), for CLOVE with simulated annealing optimisation during the solution of the TSP problem ($2^{\rm nd}$ row) and for CLOVE with Louvain communities ($3^{\rm rd}$ row). For comparison, in the  $4^{\rm th}$ row we give the results for HMCS.}
\end{table}
\begin{table}
\begin{tabular}{|p{0.19\textwidth}|*{8}{>{\centering\arraybackslash}p{0.11\textwidth}|}}
    \thickhline
    & \makecell{MA} & \makecell{EPP} & \makecell{AUC} & \makecell{GR} & \makecell{GS} & \makecell{GE} & \makecell{Running \\ Time \\ (min)} & \makecell{Peak \\ Mem. \\ (GB)} \\
    \thickhline
    CLOVE & \cellcolor[RGB]{200.63148054399127, 232.2310784973875, 160.7549904403338}$0.328 \pm 0.004$ & \cellcolor[RGB]{145.74240677242952, 200.16612369438408, 87.68817661925678}$0.534 \pm 0.005$ & \cellcolor[RGB]{42.678560957119295, 104.45299484282862, 25.774433885709325}$0.981 \pm 0.001$ & \cellcolor[RGB]{216.27529798558953, 239.03273825460414, 185.9211315420353}$0.260 \pm 0.012$ & \cellcolor[RGB]{229.53405071658815, 244.79741335503834, 207.25042941364183}$0.202 \pm 0.009$ & \cellcolor[RGB]{241.4408963393223, 246.34598780462616, 234.24676219021}$0.065 \pm 0.003$ & \cellcolor[RGB]{247.63263367657987, 244.5749042397772, 246.15648843122685}$151.832 \pm 27.578$ & \cellcolor[RGB]{248.76074934005737, 240.2504608631134, 244.65233421325684}$2.348 \pm 0.329$ \\
    CLOVE (with SA) & \cellcolor[RGB]{200.8077215559991, 232.30770502434743, 161.0385085900855}$0.327 \pm 0.009$ & \cellcolor[RGB]{145.50530402410254, 200.01221489283847, 87.40115750286095}$0.535 \pm 0.003$ & \cellcolor[RGB]{42.6304855062658, 104.39479824442702, 25.764312738161223}$0.981 \pm 0.001$ & \cellcolor[RGB]{214.69714769003815, 238.3465859521905, 183.38236802310485}$0.267 \pm 0.008$ & \cellcolor[RGB]{228.34648812838222, 244.2810817949488, 205.34000264131052}$0.207 \pm 0.006$ & \cellcolor[RGB]{241.3015652523914, 246.32959591204605, 233.92712028489797}$0.067 \pm 0.002$ & \cellcolor[RGB]{247.58855817482637, 244.7438603298322, 246.21525576689814}$141.254 \pm 26.627$ & \cellcolor[RGB]{248.44098138809204, 241.47623801231384, 245.07869148254395}$1.921 \pm 0.972$ \\
    CLOVE (dendr.) & \cellcolor[RGB]{201.41228937287792, 232.57056059690345, 162.01107420854274}$0.324 \pm 0.006$ & \cellcolor[RGB]{176.72334277379568, 220.27655583562176, 125.19141493670004}$0.426 \pm 0.038$ & \cellcolor[RGB]{42.91915094026032, 104.74423534873618, 25.82508440847586}$0.979 \pm 0.001$ & \cellcolor[RGB]{233.89082764224935, 245.45774442849992, 216.92601635574852}$0.154 \pm 0.010$ & \cellcolor[RGB]{236.54965429825938, 245.7705475645011, 223.02567750777152}$0.123 \pm 0.008$ & \cellcolor[RGB]{243.5642271861702, 246.59579143366707, 239.1179329565081}$0.040 \pm 0.003$ & \cellcolor[RGB]{247.66262087394097, 244.4599533165596, 246.11650550141204}$159.029 \pm 107.831$ & \cellcolor[RGB]{249.37876391410828, 237.88140499591827, 243.82831478118896}$3.172 \pm 0.547$ \\
    CLOVE (Louvain) & \cellcolor[RGB]{201.77171407406857, 232.72683220611677, 162.58927916263204}$0.323 \pm 0.006$ & \cellcolor[RGB]{145.63548598082537, 200.09671897000945, 87.55874618731492}$0.535 \pm 0.005$ & \cellcolor[RGB]{42.98868193109107, 104.82840444289971, 25.839722511808645}$0.979 \pm 0.001$ & \cellcolor[RGB]{218.07374436028033, 239.81467146099146, 188.81428440566836}$0.252 \pm 0.007$ & \cellcolor[RGB]{230.30885346807543, 245.03633570212654, 208.70854619146718}$0.196 \pm 0.005$ & \cellcolor[RGB]{241.59002695259232, 246.3635325826579, 234.58888536182945}$0.064 \pm 0.002$ & \cellcolor[RGB]{247.60325041185763, 244.68754008787906, 246.19566611752316}$144.780 \pm 22.495$ & \cellcolor[RGB]{248.57319140434265, 240.96943295001984, 244.90241146087646}$2.098 \pm 0.293$ \\
    CLOVE (k1 decomp.) & \cellcolor[RGB]{200.703166474949, 232.2622462934561, 160.87031128578752}$0.327 \pm 0.008$ & \cellcolor[RGB]{150.67433416918058, 203.36755025016987, 93.65840452058703}$0.517 \pm 0.003$ & \cellcolor[RGB]{42.6521166474151, 104.4209833100288, 25.768866662613704}$0.981 \pm 0.001$ & \cellcolor[RGB]{214.07328797198915, 238.07534259651703, 182.37876760711296}$0.269 \pm 0.013$ & \cellcolor[RGB]{227.87923254770052, 244.0779271946524, 204.5883306202139}$0.209 \pm 0.010$ & \cellcolor[RGB]{241.14796554591612, 246.31152535834306, 233.57474448768994}$0.069 \pm 0.003$ & \cellcolor[RGB]{247.4842280702257, 245.14379239746816, 246.3543625730324}$116.215 \pm 24.656$ & \cellcolor[RGB]{248.26110672950745, 242.16575753688812, 245.31852436065674}$1.681 \pm 0.672$ \\
    HMCS & \cellcolor[RGB]{203.0778734116335, 233.29472757027543, 164.69049201001909}$0.317 \pm 0.007$ & \cellcolor[RGB]{235.06404039442046, 245.59576945816713, 219.6175044342587}$0.140 \pm 0.006$ & \cellcolor[RGB]{42.87080259567057, 104.68570840528542, 25.814905809614856}$0.980 \pm 0.001$ & \cellcolor[RGB]{237.31040995834223, 245.8600482303932, 224.77094049266745}$0.114 \pm 0.005$ & \cellcolor[RGB]{239.15918431945528, 246.0775510964065, 229.01224637992684}$0.092 \pm 0.003$ & \cellcolor[RGB]{244.44259925225978, 246.69912932379526, 241.13302181400772}$0.030 \pm 0.001$ & \cellcolor[RGB]{247.04783478425347, 246.81663332702837, 246.93622028766202}$11.480 \pm 1.493$ & \cellcolor[RGB]{248.2853889465332, 242.07267570495605, 245.28614807128906}$1.714 \pm 0.513$ \\
    \thickhline
\end{tabular}
\caption{Average quality scores of the considered hyperbolic embedding algorithms for the 'wordnet-words' network. We show the results for the mapping accuracy, MA ($1^{\rm st}$ column), the edge prediction precision, EPP ($2^{\rm nd}$ column), the area under the receiver operating characteristic curve, AUC ($3^{\rm rd}$ column), the greedy routing score, GR ($4^{\rm th}$ column), the greedy success rate, GS ($5^{\rm th}$ column) and the greedy routing efficiency, GE ($6^{\rm th}$ column averaged over 2 different realizations. Beside the quality scores, we also display the real time in minutes ($7^{\rm th}$ column), the user time in minutes ($8^{\rm th}$ column) and the peak memory usage in GB ($9^{\rm th}$ column). In the top part of the table, we list the scores obtained for CLOVE with default settings ($1^{\rm st}$ row), for CLOVE with simulated annealing optimisation during the solution of the TSP problem ($2^{\rm nd}$ row) and for CLOVE with Louvain communities ($3^{\rm rd}$ row). For comparison, in the  $4^{\rm th}$ row we give the results for HMCS.}
\end{table}
\end{landscape}

\end{document}